\documentclass[12pt]{amsart}
\usepackage{geometry}                
\usepackage{graphicx}
\usepackage{amssymb}
\usepackage{epstopdf}
\usepackage{enumerate}
\usepackage[mathrm,colour,cntbysection]{czt}

\CountDefstrue

\title{The Specification of Sugarscape}
\author{Joseph Kehoe}
\date{9/03/2015}                                             

\begin{document}
\maketitle
\begin{abstract}
Sugarscape is a well known and influential Agent Based Social Simulation (ABSS).  Various parts of Sugarscape are supplied as examples in almost all Agent Based Model (ABM) toolkits. It has been used for demonstrating the applicability of different approaches to ABM.  However a lack of agreement on the precise definition of the rules within Sugarscape has curtailed its usefulness.  We provide a formal specification of Sugarscape using the Z specification language. This demonstrates the ability of formal specification to capture the definition of an ABM in a precise manner. It shows that formal specifications could be used as an approach to tackle the replication problem in the field of ABM.  It also provides the first clear interpretation of Sugarscape identifying areas where information is missing and/or ambiguous. This enables researchers to make proper comparisons between different implementations of this model.

\end{abstract}
\section{Introduction}
\subsection{Overview}
First we give an short informal summary of Sugarscape. We follow with a brief introduction to formal specification. We then specify the single resource simulation. Following the standard Z patterns of development we list the basic types and constants first, followed by the specification of the basic state attributes and invariant properties.  Then the rules are presented in order.

After this we present the extended specification, that is the specification with two resources. Here we highlight the differences between the single resource and two resource specifications by presenting the specifications in $\mathbf{bold~face}$  whenever the specification changes from the original. 

\section{Sugarscape}
\subsection{Agent Based social Simulations}
Sugarscape was the first large scale Agent Based Social Simulation (ABSS).  It was developed by Epstein and Axtell and presented in their book \emph{Growing Artificial Societies} \cite{Epstein1}.  The release of this simulation is considered an important event in the emerging field of Agent Based Social Simulation.

The Sugarscape ABSS was used to investigate how individual behaviour can influence and cause different social dynamics within large populations. It has been used to show how, for example, inheritance of wealth affects resource distribution in populations and how disease can spread through a population.  It remains influential today and every major simulation toolkit (Swarm, Repast, Mason and NetLogo) \cite{Railsback3,berryman2008review,Inchiosa14052002} comes with a partial implementation of Sugarscape that demonstrates that toolkit's approach to simulation. Since Sugarscape first appeared ABSSs have been applied to fields as diverse as Anthropology\cite{RubioCampillo2012347}, Biomedical Science, Ecology, Social Science \cite{Axtell00}, Epidemic modelling and Market Analysis\cite{macal1,Troitzsch1,gilbert2004agent}.

ABSS's employs a bottom-up approach to modelling populations.  Instead of precomputing the overall population behaviour, as done in equation based models, individual agents and their local interactions within the population are modelled. The behaviour of the overall population is left to emerge from these local interactions. This approach allows us to address failings in the top-down approach and demonstrates the causal factors behind the emergence of group dynamics. In cases where we do not know what the overall behaviour will be or where we are trying to find out the causes of this behaviour, bottom-up based ABSSs are the only possible approach.

\subsection{Issues with Sugarscape}
Currently, social science simulations are starting to embrace concurrency in an effort to allow for bigger, more complete and faster implementations of ABMs. Different concurrency researchers  have used the Sugarscape model as a testbed for benchmarking different approaches to parallelising ABMs \cite{lysenko2008,Perumalla3,Richmond1}. However although the rules of Sugarscape have been defined there is no general agreement on their exact meaning. These difficulties hamper the ability of researchers both to properly compare their approaches, provide complete implementations of Sugarscape or replicate their results.

Most of the rules require some form of conflict resolution. We have specified the rules in a manner consistent with the original intention (agents acting concurrently) but independent of any particular approach to how this concurrency is implemented.  That is, we have refrained from imposing any specific conflict resolution rules.

By formalising Sugarscape and providing a single precisely defined reference for the rules we can produce a standard definition of Sugarscape.  Compliance with this single reference will allow proper comparisons to be made between different approaches. It also leaves it open to the implementer to decide what approach to conflict resolution they wish to take. We detected ambiguities present in the current rule definitions, provided precise interpretations, where possible, and flagged irresolvable problems where not.
 
 We made the decision to restrict the initial specification to one pollution type and one resource type in an effort to guarantee clarity.  While the rules were designed so that they could be extended to arbitrary numbers of resources and pollutants, explicitly specifying for an arbitrary number of resources and pollutants would make the specification even more difficult to understand and thus more likely to either contain or cause mistakes. 
 
 Once we had a specification for the single resource scenario we extended the specification to a two resource situation. This allowed us to specify the final rule, $Trade$, as that rule requires two resources to function.
 
 This allowed for:
\begin{enumerate}
\item A simpler and easier to understand specification of the rules that use only one resource (trading clarity against completeness);
\item A complete (but separate) specification for simulations that use two resources.
\end{enumerate}
 We do not provide specifications for multiple pollutants as multiple pollutants were never actually implemented in Sugarscape\footnote{We leave this open as an exercise for the reader.}.
 
Similarly we did not provide a specification for more than two resources as we deem the benefits of doing so counterbalanced by both the complexity of the resulting specification and the lack of any requirement to use such a complex simulation for benchmarking purposes. Sugarscape has only ever been implemented with two resources types, known respectively as \emph{sugar} and \emph{spice}. Anyone wishing to extend Sugarscape further can use the two resource specification for guidance.
\subsection{Synchronous and Asynchronous Updating}
 Originally the rules were stated with an explicit assumption that the underlying implementation would be sequential. Concurrency was simulated through randomisation of the order of each rule application on the individual agents, and models that follow this regime are termed asynchronous.
\begin{quotation}
All results reported here have been produced by running the model on a serial computer; therefore only one agent is``active'' at any instant. In principle the model could be run on parallel hardware, permitting agents to move simultaneously (although \textbf{M} would have to be supplemented with a conflict resolution rule to handle cases in which two or more agents simultaneously decide to inhabit the same site).

[Footnote 12, Chapter II]
\end{quotation}

The alternative to asynchronous updating is synchronous updating. Synchronous updating assumes that all updates occur concurrently.  While it is clear that the original authors have no objection to employing synchronous updating on sugarscape it is well known that asynchronous and synchronous updating produce different results.  What is not known is how divergent these results are in the case of complex ABMs such as Sugarscape or indeed how to apply synchronous updating to all the complex interaction types in Sugarscape.

In order to answer these questions we present initially a specification that assumes a synchronous updating regime, as this is the most novel approach.  Following this we give the equivalent Asynchronous updating version for comparison.

\section{Single Resource Sugarscape}
Sugarscape is a discrete turn based simulation composed of a set of interacting agents that move across an environment.  The environment, or simulation space, is modelled as a two dimensional $M$ by $M$ grid or matrix of discrete locations known as the \emph{lattice}.  This lattice is toroidal in nature, that is, it wraps around on all four edges. Every lattice location has a position denoted by its x and y coordinates. For any lattice location [i, j] there are four direct (von Neumann) neighbours (up, down, left and right) at positions $[i, (j+1) \% M],[i, (j-1) \%M], [(i-1) \% M, j]$ and $[(i+1) \% M, j]$. We denote this set of von Neumann neighbours as $N_{1}(i, j)$, and further use $N_{k}(i,j)$ to denote the set of  von Neumann neighbours where each is a maximum distance of k locations from the location [i,j]. 

Each location can hold a number of resource and pollution types. While there is no limit placed on how many resource or pollution types can exist in a Sugarscape simulation we are unaware of any Sugarscape derived simulation that uses more than two resource types and one pollution type. When there is only one resource type it is called \emph{sugar} and if there are two then the second resource is known as \emph{spice}. These amounts are measured as natural numbers ($\geq 0$). Each individual location has limits placed on the maximum amount of resources of any type it may carry at any one time.  These limits are defined at simulation startup and remain fixed during a simulation run.  Agents consume the resources at their current location. Locations replenish their resources by some defined amount during each time step. Each location can also hold at most one agent at a time.

Agents reside at locations within the lattice but are mobile and can change location at most once per step.  At a minimum each agent has the following attributes:
\begin{description}
\item [Metabolism Rate (one per resource type)] The rate at which an agents resource stores decrease during each simulation step.  Different resource types have independent metabolism rates. Once an agent runs out of resources it \emph{dies} (is removed from the simulation);
\item [Age] The number of steps that the agent has been present in the simulation;
\item [Maximum Age] The maximum number of steps that an agent is allowed to exist during the simulation run. Once an agent reaches its maximum age it is removed from the simulation;
\item [Resource Store (one for each resource type)] The amount of each resource that the agent currently has;
\item [Vision] How far in each of the cardinal directions that the agent can see. An agent can only interact with locations and agents that are in its neighbourhood $N_{vision}$. To ensure locality all agent values for vision will be less than some predefined maximum and this maximum will be much smaller than the lattice dimension size ($M$).
\end{description}

In the more complex versions of Sugarscape agents can also have a ``culture'' identifier (identifying which tribe the agent belongs to), a set of outstanding loans of resources that the agent has given to (or received from) other agents, a set of diseases that the agent has contracted and, an immunity system that gives each agent immunity from certain diseases.
 
A simulation run consist of a series of turns or steps during which certain rules are applied to each location and agent.  Each rule is applied concurrently and instantaneously to each agent and/or location.  The rules are generally fairly simple and the only information that an agent (or location) can use when deciding how to apply a rule is local information, that is an agent or location at position [i,j] can only access information from locations and/or agents that are within the set $N_{k}$, where k$\leq$ vision (in most cases k=1).

The rules for locations decide how resources are replenished and how pollution is created or spread.  The rules for agents are more varied and determine agent movement and interaction.  Agent interaction can range from spreading disease, trading, entering financial agreements and even combat. There are a large number of rules but not all rules need to be (or indeed can be) applied in the same simulation run.  The rules are chosen based on what we wish to model. A simulation that wishes to see the effect of trading on wealth distribution would have no need for the combat or culture rules while one modelling disease transmission would only require the movement and disease transmission rules.

\subsection{Basic Types and Constants}
First we identify the basic types and any required constants. Many are self explanatory or will become clear when their associated rules are specified. A simulation is defined by the values given to these constants and the combination of rules employed.

\begin{axdef}
M: \nat_1 \hfill(1)\\
CULTURECOUNT:\nat_1\hfill(2)\\
MAXVISION :\nat_1\hfill(3)\\
MINMETABOLISM, MAXMETABOLISM:\nat\hfill(4)\\
SUGARGROWTH: \nat_1\hfill(5)\\
MAXAGE,MINAGE:\nat_1\hfill(6)\\
MAXSUGAR:\nat_1\hfill(7)\\
DURATION:\nat_1\hfill(8a)\\
RATE:\arithmos\hfill(8b)\\
INITIALSUGARMIN, INITIALSUGARMAX:\nat\hfill(9)\\
WINTERRATE,SEASONLENGTH: \nat_1\hfill(10)\\
PRODUCTION,CONSUMPTION: \nat\hfill(11)\\
COMBATLIMIT:\nat\hfill(12)\\
IMMUNITYLENGTH:\nat\hfill(13)\\
INITIALPOPULATIONSIZE:\nat\hfill(14)\\
POLLUTIONRATE:\nat\hfill(15)\\
CHILDAMT:\nat\hfill(16)\\
\where 

CULTURECOUNT \mod 2=1\\
MINMETABOLISM < MAXMETABOLISM\\
MAXAGE < MINAGE\\
MAXVISION < M\\
INITIALSUGARMIN < INITIALSUGARMAX\\
INITIALPOPULATIONSIZE\leq M * M
\end{axdef}

\begin{enumerate}
\item The simulation space is represented by a two dimensional $M$ by $M$ matrix of locations. Each location in the simulation space is referenced by two indices representing its position in this matrix;
\item $CULTURECOUNT$ determines the size of the bit sequence used to represent cultural allegiances. This is always equal to an odd number so that the number of 1's in the sequence is never equal to the number of 0's;
\item Agents can only ``see'' in the four cardinal directions, that is the locations to the north, south, east and west. Agents are endowed with a random vision strength that indicates how many locations the can ``see'' in each direction.  This endowment is always less than $MAXVISION$ and $MAXVISION$ is always less than $M$;
\item Agents consume an amount of sugar (resources) during each turn.  This sugar represents the amount of energy required to live.  Each agent is endowed, on creation, with a random metabolism between $MINMETABOLISM$ and $MAXMETABOLISM$;
\item Agents consume sugar (resources) from the location they occupy. Each location can renew its sugar at a rate determined by $SUGARGROWTH$. After each turn up to a maximum of $SUGARGROWTH$ units of sugar are added to each location (in accordance with the $Growback$ rule);
\item $MAXAGE$ and $MINAGE$ are, respectively,  the maximum and minimum allowable lifespan for any agent;
\item  $MAXSUGAR$ is the maximum amount of sugar that any location can possibly hold. This is known as the carrying capacity of a location;
\item $RATE$ and $DURATION$ are used for determining the rate of interest charged for loans and the duration of a loan;
\item $INITIALSUGARMIN$ and $INITIALSUGARMAX$ are the lower and upper limits for initial endowment of sugar given to a newly created agent;
\item If seasons are enabled then two seasons, winter and summer are allowed with a duration of $SEASONLENGTH$ turns (ticks) and a new separate lower seasonal grow back rate calculated using $WINTERRATE$ (as determined by the $SeasonalGrowback$ rule);
\item Pollution can occur at a rate determined by the production and consumption of resources determined by the $PRODUCTION$ and $CONSUMPTION$ constants respectively;
\item The combat rule posits the maximum reward $COMBATLIMIT$ that can be given to an agent through killing another agent;
\item Immunity in agents is represented using a fixed size sequence of bits of length $IMMUNITYLENGTH$;

\item We have some predetermined initial population size $INITIALPOPULATIONSIZE$ that is used to initialise the simulation;
\item $POLLUTIONRATE$ determines the number of steps that elapse before pollution levels diffuse to their neighbours;
\item A certain amount of sugar reserves, $CHILDAMT$, are required for an agent to have children.
\end{enumerate}

\begin{zed}
[ AGENT ] \hfill (1)\\
POSITION == 0 \upto M-1 \cross 0 \upto M-1~~~~~\hfill(2)\\
SEX ::= male | female\hfill(3)\\
BIT ::= 0| 1\hfill(4)\\
affiliation::= red | blue\hfill(5)\\
boolean::= true | false\\
\end{zed}

\begin{enumerate}
\item $AGENT$ is used as a unique identifier for agents;
\item $POSITION$ is also used to make specifying indices within the grid so as to make the schemas easier to read and more compact;
\item  All agents have a sex attribute;
\item $BIT$s are used to encode both culture preferences and diseases of agents;
\item Every agent has a cultural affiliation of either belonging to the blue tribe or red tribe.
 \end{enumerate}

Agents can, using the Mating rule, have offspring if they are fertile.  Fertility is determined by the age of the agent, where fertility starts at some predefined age and ends at another. These boundaries are defined for all agents. The numbers are set out by Epstein and Axtelland although there appears to be no special significance attached to these numbers we will stick with the originals. Male fertility ends 10 turns later than female fertility.
\begin{axdef}
  FEMALEFERTILITYSTART,FEMALEFERTILITYEND: \nat\\
 MALEFERTILITYSTART,MALEFERTILITYEND: \nat\\
\where
  12\leq FEMALEFERTILITYSTART \leq 15\\
  40 \leq FEMALEFERTILITYEND \leq 50\\ 
  12\leq MALEFERTILITYSTART \leq 15\\
 50 \leq MALEFERTILITYEND \leq 60 \\
 MALEFERTILITYEND=FEMALEFERTILITYEND+10\\
 
\end{axdef}

The replacement rule requires a sugar allocation be given to new agents set between 5 and 25. Again there appears to be no special significance attached to these numbers.
\begin{axdef}
 STARTSUGARMIN,STARTSUGARMAX:\nat
  \where
  STARTSUGARMIN=5\\
  STARTSUGARMAX=25\\
 \end{axdef}

\subsection{The Sugarspace Lattice}
The simulation space in Sugarscape consists of a finite discrete two-dimensional array of locations. Each location is identified its row and column value.  Each location contains a number of resources.   While only two resources are ever used it is clear that the intention of the original authors was that the simulation could be extended so that any number of different resources can be present. 

Similarly each location can contain a number of pollutant levels.  In practice, although the rule is explicitly defined for an arbitrary number of pollution types only one is ever used. Again, in line with actual Sugarscape usage and to make the specification more readable we assume only one pollution type. Pollution fluxes are used in the rules to help calculate how pollution levels change over time. Although explicitly referenced in the Pollution rule these do not need to be explicitly modelled in the specification.

\begin{schema}{Lattice}
\\
sugar: POSITION \pfun \nat\hfill(1)\\
maxSugar:POSITION \pfun \nat\hfill(2)\\
pollution:POSITION\pfun\nat\hfill(3)\\
\where

\dom sugar=\dom maxSugar=\dom pollution=POSITION\hfill(4)\\
\forall x:POSITION\spot sugar(x)\leq maxSugar(x)\leq MAXSUGAR \hfill(5)\\
\end{schema}

\begin{enumerate}
\item $sugar$ is a mapping that stores the amount of sugar stored at each position;
\item $maxSugar$ is a mapping that records the maximum amount of sugar that can be stored (carried) in each position;
\item $pollution$ records the amount of pollution at each location;
\item Every position has a sugar level, a maximum allowed sugar level (or carrying load) and a pollution level;
\item Every position's sugar level is less than or equal to the maximum allowed amount for that position which is in turn less than or equal to the $MAXSUGAR$ constant;

\end{enumerate}

 We need to track the number of turns that have occurred in the simulation. Each turn consists of the application of all rules that form part of the simulation.  
 \begin{schema}{Step}
 step : \nat\\
 \end{schema}

\subsection{Agents}
Every agent is situated on a location within the grid and each location is capable of containing only one agent at a time (putting an upper limit on the number of possible agents). Agents are mobile, that is they can move to a new location if a suitable unoccupied location is available. Movement is both discrete and instantaneous, it is possible for an agent to move to a new location instantly while skipping over all intermediate locations. The attributes that every agent has are:
\begin{description}
\item[ Vision] How far in the four cardinal directions that an agent can see;
\item[ Age] Number of turns  of the simulation that an agent has been alive;
\item[ Maximum Age] Age at which an agent dies;
\item[ Sex] Agents are either male or female;
\item[ Sugar Level] The amount of sugar that an agent currently holds. There is no limit to how much sugar an agent can hold;
\item[ Initial Sugar] The amount of sugar the agent was initialised with on creation;
\item[ Metabolism] The amount of energy, defined by sugar (or resource) consumption, used during every turn of the simulation;
\item[ Culture Tags] A sequence of bits that represents the culture of an agent;
\item[ Children]  For each agent we track its children (if any). To apply the Inheritance rule the full list of an agents children is required. 
\item[ Loans] Under the credit rule agents are allowed lend and/or borrow sugar for set durations and interest rates so we need to track these loans. For each loan we need to know the lender, the borrower, the loan principal and the due date (represented as the step number);
\item[ Diseases] Diseases are sequences of bits that can be passed between agents. An agent may carry more than one disease;
\item [ Immunity] Each agent has an associated bit sequence that confers immunity against certain diseases. If the bit sequence representing a disease is a subsequence of an agents immunity bit sequence then that agent is considered immune to that disease.

\end{description}
 
\begin{schema}{Agents}
population : \power AGENT\\
position : AGENT  \inj POSITION\\
sex: AGENT \pfun SEX\\
vision: AGENT \pfun \nat_1\\
age: AGENT \pfun \nat\\
maxAge: AGENT \pfun \nat_1\\
metabolism:AGENT\pfun\nat\\
agentSugar: AGENT \pfun \nat\\
initialSugar: AGENT \pfun \nat\\
agentCulture:AGENT \pfun \seq BIT\\
children: AGENT \pfun\power AGENT\\
loanBook: AGENT\rel(AGENT\cross(\nat,\nat))\\
agentImmunity: AGENT\pfun \seq BIT\\
diseases: AGENT\pfun\power \seq BIT\\
\where
population=\\
\t1\dom position=\dom sex=\dom vision\\
\t1=\dom maxAge=\dom agentSugar=\dom children\\
\t1=\dom agentCulture=\dom metabolism=\dom age\\
\t1=\dom agentImmunity=\dom diseases\hfill(1)\\
\\
\dom loanBook\subseteq population\hfill(2)\\
\dom (\ran loanBook)\subseteq population\hfill(3)\\
\forall x,y:AGENT; d: \seq BIT\spot \\
x,y \in population \land x\neq y\implies\hfill(4)\\
\t1((age(x)\leq maxAge(x)\land MINAGE\leq maxAge(x) \leq MAXAGE\\
\t1\land \#agentCulture(x)=CULTURECOUNT\\
\t1\land \#agentImmunity(x)=IMMUNITYLENGTH\\
\t1\land vision(x)\leq MAXVISION\\
\t1\land MINMETABOLISM\leq metabolism(x)\leq MAXMETABOLISM\\
\t1\land position(x)=position(y)\iff x=y)\\
d \in \ran diseases(x)\implies \#d < IMUNITYLENGTH\\
\end{schema}

\begin{enumerate}
\item Every existing agent has an associated age, sex, vision, etc. Note that the population holds only the currently existing agent IDs;
\item Only current members of the population can be lenders;
\item Only current members of the population can be borrowers
\item Every agent in the population is guaranteed to have a current age less than the maximum allowed age for that agent, a maximum age less than or equal to the global $MAXAGE$, a metabolism between the allowed limits and vision less than or equal to the maximum vision. The sequence of bits representing its culture tags is $CULTURECOUNT$ in size while those representing immunity is $IMMUNITYLENGTH$ in size.  All diseases are represented by sequences of bits that are shorter than the immunity sequence.
\end{enumerate}

 The entire simulation consists of locations, agents and a counter holding the tick count. We combine them all in the schema $SugarScape$.

 \begin{schema}{SugarScape}
 Agents\\
 Lattice\\
 Step\\
 \end{schema} 
 
 The initial state of the schema when the simulation begins must also be stated.
  \begin{schema}{InitialSugarScape}
Sugarscape'
 \where
 step'=0\hfill(1)\\
  \# population'=INITIALPOPULATIONSIZE\hfill(2)\\
  loanBook'=\emptyset\hfill(3)\\
 \forall a: AGENT\spot\hfill(4)\\
 a\in population'\implies\\
 \t1(age(a)=0\land diseases'(a)=\emptyset\land children'(a)=\emptyset\\
  \t1\land INITIALSUGARMIN\leq agentSugar'(a)\leq INITIALSUGARMAX)\\
  \t1\land initialSugar'(a)=agentSugar'(a)\\
 \end{schema}
 \begin{enumerate}
\item $step$ is set to zero;
\item The population is set to some initial size;
\item There are no loans as yet;
\item Every agent in the starting population has an age of zero, no diseases or children and some initial sugar level within the agreed limits. The other attributes have random values restricted only by the invariants;
\end{enumerate}

 \subsection{Rules}
 
There are a number of rules that can be employed in different combinations to give different simulations.  We will quote each rule as laid out in the appendix of \cite{Epstein1} and follow, where necessary, with a more detailed explanation of the rule. In many cases the simple rule definitions are not complete. Extra information, embedded in the original text, has been extracted where necessary to help complete these rules. The majority of rule definitions assume only one resource (sugar) and it is these that are specified in this section. 

The simulation is discrete with each time interval representing one complete set of rule applications. We use the $step$ variable in the $SugarScape$ schema to keep track of the current time interval number.

Where there exist ambiguities in the rule definitions we will identify them and propose one or more possible interpretations consistent with what we believe to be the authors intentions.  Throughout the rule definitions constants such as $\alpha,\beta$ are used but they have different meanings in each rule.  For the sake of clarity we will give each constant a meaningful and globally unique name.

\subsection{Tracking Steps}
 While not defined explicitly as a rule, we must ensure that we record the current step number. We increment the $Step$ variable before every sequence of rule applications that compose a single turn of the simulation.
 
There is an issue with metabolism in that every turn of the simulation requires that agents use up their sugar reserves at a rate determined by their metabolism.  It is not explicitly stated when or where this sugar deduction occurs within the rules.  It could be placed, for example, in the movement rule but it can also be placed, just as validly, within any rule that is guaranteed to be applied during every turn. Since there is no obvious  reason why one is superior to the other, as long as it is consistently applied, we choose to place the metabolism deduction within the $Tick$ schema. This new rule can be stated simply as follows:
\begin{description}
\item[ Tick] At the start of every time interval increase every agents age by one and decrease every agents sugar level by their metabolism rate.
\end{description}

 \begin{schema}{Tick}
 \Delta Agents\\
 \Delta Step
 \where
 population'=population\\
position'=position\\
sex'=sex\\
vision'=vision\\
maxAge'=maxAge\\
metabolism'=metabolism\\
initialSugar'=initialSugar\\

agentCulture'=agentCulture\\
children'=children\\
loanBook'=loanBook\\
agentImmunity'=agentImmunity\\
diseases'=diseases\\
 step'=step+1\hfill(1)\\
 \forall x:AGENT\spot x\in population\implies\\
 \t1age'(x)=age(x)+1\hfill(2)\\
 \t1\land agentSugar'(x)=agentSugar(x)-metabolism(x)\hfill(3)\\
 \end{schema}
  
 \begin{enumerate}
\item Add one to the step count;
\item Increase everyone's age by one;
\item Decrease everyone's agentSugar by their metabolism.
\end{enumerate}

\subsection{Sugarscape Growback$_{\alpha}$}

\begin{description}
\item[ Sugarscape Growback$_{\alpha}$] At each Lattice position, sugar grows back at a rate of $\alpha$ units per time interval up to the capacity at that position.
\end{description}

Growback determines the rate at which location resources are replenished. The integer constant $\alpha$ indicates the amount by which resources grow during a single step or time interval. 
If $\alpha=\infty$ then each resource returns to its maximum value during each turn, i.e. it is instantly fully replenished after each step.
The rule only refers to a single resource, \emph{sugar}, but the book explicitly defines one other resource \emph{spice}  and it is clear that generalisations allowing an arbitrary number of resource types to be held at each Lattice position are acceptable.

 Since we are dealing only with one resource, sugar, we only need to define $\alpha$ for this resource . The constant $SUGARGROWTH$ represents $alpha$ in this rule and we use this to update the sugar level of each position.

Since the  maximum carrying level of each resource cannot be exceeded we will set the resource levels to its maximum value if application of the replenishment rate would result in a value greater than this maximum.  With these definitions we can express the $Growback$ rule in a simple manner. The last line in the schema (see below) does the work of updating the resource levels of every location.
  
\begin{schema}{Growback}
 \Delta Lattice\\
 \where\\
\\
 pollution'=pollution\\
 maxSugar'=maxSugar\\
sugar'=\{x:POSITION\spot\\
\t1 x\mapsto min(\{sugar(x)+SUGARGROWTH,maxSugar(x)\})\}\hfill(1)\\
 \end{schema}
 
\begin{enumerate}
\item The new sugar levels are calculated using a simple formula to either, the maximum possible level for that location or the old level plus the $SUGARGROWTH$ whichever is the smaller. 
\end{enumerate}

\subsection{Seasonal Growback $S_{\alpha,\beta,\gamma}$}
\begin{description}
\item[ Seasonal Growback $S_{\alpha,\beta,\gamma}$] Initially it is \emph{summer} in the top, half of the Sugarscape and \emph{winter} in the bottom half.  Then every $\gamma$ time periods the seasons flip - in the region where it was summer it becomes winter and vice versa. For each site, if the season is summer then sugar grows back at a rate of $\alpha$ units per time interval; if the season is winter then the grow back rate is $\alpha$ units per $\beta$ time intervals.
\end{description}

Seasonal growback is an alternative to the previous grow back rule. Which rule is chosen will depend on what the simulation is trying to demonstrate. Seasonal grow back allow us to introduce seasonal factors into the original $Growback$ rule. There are two seasons (representing summer and winter) and each lasts $\gamma$ turns before switching. We rename $\gamma$ to $SEASONLENGTH$.  $\alpha$ is the summer season $SUGARGROWTH$ rate  and $\alpha/\beta$ is the winter season rate.  We use the existing $SUGARGROWTH$ to hold the summer rate and introduce $WINTERRATE$ as $\beta$.

Determining what season it is during a turn is fairly trivial. When $seasonLength$ divides into the $Step$ variable evenly it is summer in the top half and winter in the bottom half (and vice versa).  

\begin{schema}{SeasonalGrowback}
\Delta Lattice\\
\Xi Step\\
 \where
 pollution'=pollution\\
\\
 maxSugar'=maxSugar\\
\forall x: POSITION \spot\\
(step\div SEASONLENGTH)\mod 2 = 0 \implies sugar'=\hfill(1)\\
 \t1 \{ x: POSITION | first(x)<M\div 2 \spot\\
\t1  x\mapsto min(\{sugar(x)+SUGARGROWTH,maxSugar(x)\}) \} \hfill(1a)\\
 \t1\cup \\
 \t1 \{x: POSITION | first(x)\geq M\div 2 \spot \hfill(1b)\\
 \t1 x\mapsto min(\{sugar(x)+SUGARGROWTH\div WINTERRATE,maxSugar(x)\})\}\\

  (step\div SEASONLENGTH)\mod 2 \neq 0 \implies sugar'=\hfill(2)\\
 \t1\{x: POSITION | first(x)<M\div 2\spot \\
\t1 x\mapsto min(\{sugar(x)+SUGARGROWTH\div WINTERRATE,maxSugar(x)\})\}\hfill(2a)\\
 \t1\cup\\
 \t1 \{x: POSITION; y:\nat | first(x)\geq M\div 2\spot \\
 \t1 x\mapsto min(\{sugar(x)+SUGARGROWTH,maxSugar(x)\}) \}\hfill(2b)\\
 \end{schema}

\begin{enumerate}

\item If the season is summer then:

\begin{enumerate}[a)] 
\item Top half of grid is updated as normal;
\item Bottom half is updated at winter rate.
\end{enumerate}
\item Otherwise if it is winter:
\begin{enumerate}[a)] 
\item Top half of grid is updated at winter rate;
\item Bottom half is updated as normal.
\end{enumerate}

\end{enumerate}

\subsection{Movement - $M$}

\begin{description}
\item[ Movement - $M$] ~
\begin{itemize}
\item Look out as far as vision permits in each of the four lattice directions, north, south, east and west;
\item Considering only unoccupied lattice positions, find the nearest position producing maximum welfare;
\item Move to the new position
\item Collect all resources at that location
\end{itemize}

\end{description}

The previous rules affected only the locations but the remaining rules affect agents as well as locations.  The Movement rule determines how agents select their next location.  There are a number of different versions of this rule. We will specify the simplest rule first as it is the only movement rule explicitly defined in the appendix but we will also specify the other movement rules defined in the text.  We add a subscript to the rule title ($M_{basic}$) to distinguish between the different movement rule specifications.

Not explicitly stated within the rule but stated as a footnote to the rule is the restriction that the order in which the lattice directions are searched should be random. This comes into play when two or more available sites exist with the same welfare score.

This rule does not guarantee that an agent will move to the best location. To see why this is the case consider what happens if two agents both try to move to the same location. Only one can succeed and the other will have to move to a less advantageous location. How we decide which agent succeeds is not defined. We assume that either a conflict resolution or conflict avoidance rule is available to make this decision but it is not stated what this rule should be.  The original implementation is sequential with agents assumed to be moving in a random order thus enforcing collision avoidance. No guidance is provided for concurrent implementations.

To help make the specification clear we define some simple helper functions. The distance between two positions is only defined for positions that are directly horizontal or vertical to each other. This function must take into account the torus-like (wrap around) structure of the simulation.
 
  \begin{axdef}
distance :POSITION\\
\cross POSITION\\
\fun\nat\\
 \where
 \forall x1,x2,y1,y2:\nat \spot\\
distance((x1,y1),(x1,y2))=\hfill(1)\\
\t1min(\{|y2-y1|,M-|y2-y1|\})\\
distance((x1,y1),(x2,y2))=\hfill(2)\\
\t1min(\{|x1-x2|,M-|x1-x2|\})\\

distance((x1,y1),(x2,y2))=\infty\iff\\
\t1 x1\neq x2\land y1\neq y2\hfill(3)\\
 \end{axdef}
 
\begin{enumerate}
\item If two agents are vertically aligned we calculate distance based on the horizontal distance;
\item If two agents are horizontally aligned we calculate distance based on the vertical distance;
\item Otherwise the distance is defined as infinity.
\end{enumerate}
We use this to define the $adjacent$ function that lets us know if two agents are directly beside each other.

\begin{axdef}
adjacent: POSITION\\
\cross POSITION\\
 \fun boolean\\
 \where
 \forall a,b:POSITION\spot\\
adjacent(a,b)\iff distance(a,b)=1
 \end{axdef}
 
$visibleAgents$ takes an agent, a function mapping agents to positions  and the vision range of the agent and returns the set of agents that are within that agent's neighbourhood.

\begin{axdef}
visibleAgents: AGENT\\
\cross (AGENT \pinj POSITION)\\
\cross \nat\\
\pfun \finset AGENT
 \where
 \forall agent:AGENT; pos:AGENT\pinj POSITION;range:\nat\spot\\
visibleAgents(agent,pos,range)=\\
\t1\{ag:AGENT | ag\in \dom pos\land 1\leq distance(pos(ag),pos(agent))\leq range\}\\
 \end{axdef}

\begin{schema}{Movement_{basic}}
\Delta SugarScape\\
 \where
 step'=step\\
population'=population\\
maxSugar'=maxSugar\\
 pollution'=pollution\\
 sex'=sex\\
 vision'=vision\\
 age'=age\\
 maxAge'=maxAge\\
 agentCulture'=agentCulture\\
 loanBook'=loanBook\\
 diseases'=diseases\\
 agentImmunity'=agentImmunity\\
 children'=children\\
 metabolism'=metabolism\\
 initialSugar'=initialSugar\\
\forall a:AGENT; l:POSITION \spot\\
 a\in population' \implies\hfill(1)\\
\t1 distance(position'(a), position(a))\leq vision(a)\\
~\\
 (distance(position(a),l) \leq vision(a) \land (l\notin \ran position'))\implies \hfill(2)\\
\t1 sugar(l) \leq sugar(position'(a)) \hfill(2a)\\
\t1 \land(distance(l,position(a))<distance(position'(a),position(a)))\hfill(2b)\\
\t2 \implies sugar(l)<sugar(position'(a))\\
~\\
agentSugar'=\{\forall a:AGENT | a\in population'\spot\\
\t1  a\mapsto agentSugar(a)+sugar(position'(a))\}\hfill(3)\\
sugar'=sugar\oplus\{\forall l:POSITION| l \in \ran position' \spot l \mapsto 0\}\hfill(4)\\
\end{schema}

After the rule is applied the following will be the case for every agent:
\begin{enumerate}
\item They will be located within one of the locations in their original neighbourhood (possibly the same position as before);
\item After every agent has moved:
\begin{enumerate}[a)] 
\item There will exist no remaining available locations from the original neighbourhood of an agent that would have given a better welfare score than the location that agent now inhabits (we picked the maximum reward);
\item If there was more than one location with maximum reward then the agent moved to the closest location.
\end{enumerate} 
\item Agent sugar levels increase because they consume all the sugar at their new location (even if the new location is the same as their old location);
\item Location sugar levels are set to zero everywhere there is an agent present.
\end{enumerate}

The specification states what is true after the application of the rule but not how we achieve that state. In any implementation some conflict resolution strategy will be needed but in this specification we remain agnostic as to what it should be. 

The rule is well stated but requires that we precisely define $welfare$. For a single resource simulation welfare is precisely equal to the amount of sugar available at a location.  We will define welfare for multiple resource simulations later.

 \subsection{Pollution Formation $P_{\Pi,\chi}$} 
 \begin{description}
\item[ Pollution Formation $P_{\alpha,\beta}$] 
When sugar quantity $s$ is gathered from the Sugarscape, an amount of production pollution is gathered in quantity $\alpha s$. When sugar amount $m$ is consumed (metabolised), consumption pollution is generated according to $\beta m$. The total pollution on a site at time $t$, $p^{t}$, is the sum of the pollution present at the previous time, plus the pollution resulting from production and consumption activities, that is, $p^{t}=p^{t-1}+\alpha s+\beta m$.
\end{description}
This single resource pollution rule is easiest to understand and the most common form of the pollution rule.  When pollution is incorporated into the Sugarscape the movement rule is changed so that the welfare of a location is now defined using the sugar to pollution ratio - the greater the ratio the greater the welfare.  This ratio is defined as $sugar/(1+pollution)$ where the ``plus one'' prevents division by zero.

As the pollution rule requires that we know both the sugar consumed and sugar metabolised during the last move of an agent to that location it is simpler to incorporate the $Pollution Formation$ rule into the movement rule. The alternative is to track the sugar consumed during each move which would require another attribute defined in the $Agent$ schema.

\begin{schema}{Movement_{pollution}}
\Delta SugarScape\\
 \where
 step'=step\\
\\
 maxSugar'=maxSugar\\
 sex'=sex\\
 population'=population\\
 vision'=vision\\
 age'=age\\
 maxAge'=maxAge\\
 agentCulture'=agentCulture\\
 loanBook'=loanBook\\
 children'=children\\
 agentImmunity'=agentImmunity\\
 diseases'=diseases\\
 metabolism'=metabolism\\
 initialSugar'=initialSugar\\
\forall a:AGENT; l:POSITION\spot\\
a\in population' \implies distance(position'(a),position(a))\leq vision(a)\\
~\\
 (distance(position(a),l)\leq vision(a) \land (l\notin \ran position'))\\
\t1 \implies [sugar(l)/(1+pollution(l))] \\
\t2\leq [sugar(position'(a))/(1+pollution(position'(a)))] \hfill(1)\\
\land(distance(l,position(a))<distance(position'(a),position(a)))\\
\t1 \implies sugar(l)/(1+pollution(l)) \\
\t2< sugar(position'(a))/(1+pollution(position'(a)))\\ 

~\\
 sugar'=sugar\oplus \{\forall l:POSITION | l \in \ran position' \spot l\mapsto 0 \}\\
agentSugar'=\{\forall a:AGENT| a\in population \spot\\
 \t1\ a\mapsto agentSugar(a)+sugar(position'(a)) \}\\

~\\
pollution'=pollution\oplus\{\forall l:POSITION; x:AGENT| position'(x)=l\spot\\
\t1  l\mapsto (PRODUCTION*sugar(l)+CONSUMPTION*metabolism(x) ) \}\hfill(2)\\
\end{schema}

\begin{enumerate}
\item We use our new formula to calculate the desirability of a location;
\item The new pollution value for any location that an agent is present at is calculated as per rule definition.
\end{enumerate}

The rule as stated in the appendix is the generalised rule defined for an arbitrary number of pollutants and resources. We have specified the simpler version as it is easier to grasp. The more complex version has not been used in any of the Sugarscape simulations. We state the generalised rule below for completeness but do not present a formal specification of it.
 \begin{description}
\item[ Pollution Formation $P_{\Pi,\chi}$] 
For $n$ resources and $m$ pollutants, when n-dimensional resource vector $\mathbf{r}$ is gathered from the Sugarscape the m-dimensional pollution production vector $\mathbf{p}$ is produced according to $\mathbf{p=\Pi r}$, where $\mathbf{\Pi}$ is an $m\times n$ matrix; when n-dimensional (metabolism) vector $\mathbf{m}$ is consumed then m-dimensional consumption pollution vector $\mathbf{c}$ is produced according to $\mathbf{c=\chi m}$, where $\mathbf{\chi}$ is an $m\times n$ matrix.
\end{description}

 \subsection{Pollution Diffusion $D_{\alpha}$} 
\begin{description}
\item[ Pollution Diffusion $D_{\alpha}$] ~
\begin{itemize}
\item Each $\alpha$ time periods and at each site, compute the pollution flux the average pollution level over all its von Neumann neighbouring sites;
\item Each site's flux becomes its new pollution level.
\end{itemize}

\end{description}

This rule determines how pollution diffuses over grid.  Pollution diffusion is calculated every $\alpha$ turns and is computed as the average pollution level of all the locations von Neumann neighbours. We use the constant $POLLUTIONRATE$ in place of $alpha$.

The von Neumann neighbours of a location are those immediately above, below, left and right of the current locations (aka north, south, east and west). We define the four cardinal directions taking into account to fact that the grid wraps around at its edges (i.e. it is a torus). 

\begin{axdef}
north: POSITION \bij POSITION\\
south: POSITION \bij POSITION\\
east: POSITION \bij POSITION\\
west: POSITION \bij POSITION\\
 \where
 \forall x,y:\nat\spot\\
 \t1west ((x, y)) = ((x-1)\mod M, y)\\
  \t1east ((x, y)) = ((x+1)\mod M, y)\\
 \t1south ((x, y)) = (x, (y-1)\mod M)\\
 \t1north ((x, y)) = (x, (y+1)\mod M)\\
 \end{axdef}
 
We use this to define a function that returns true if two agents are von Neumann neighbours.  It takes as parameters the two agents and a function that maps each agent onto their location in the simulation.

 \begin{axdef}
  vonNeumanNeighbour: (AGENT\cross AGENT\cross (AGENT\inj POSITION)) \fun boolean 
\where
\forall a,b:AGENT;position:AGENT\inj POSITION\spot\\
  vonNeumanNeighbour(a,b,position)\iff\\ 
  \t1position(a)=north(position(b))\\
  \t1\lor position(a)=south(position(b))\\
  \t1\lor position(a)=east(position(b))\\
  \t1\lor position(a)=west(position(b))
\end{axdef}

\begin{schema}{PollutionDiffusion}
\Delta Lattice\\
\Xi Step
 \where
\\
maxSugar'=maxSugar\\
sugar'=sugar\\
(step \mod POLLUTIONRATE \neq 0) \implies pollution'=pollution\\
\\
 (step \mod POLLUTIONRATE = 0) \implies  pollution'=\\
 \t1\{\forall l:POSITION  \spot l\mapsto (pollution(north(l))+pollution(south(l))\\
 \t2+pollution(east(l))+pollution(west(l)))\div 4\}\\\\
\end{schema}

The $Pollution~Diffusion$ Rule can be simplified slightly, by removing the redundant mention of $flux$. 
\begin{description}
\item[ Pollution Diffusion $D_{\alpha}$] 
 After every $\alpha$ time periods and at each location, the average pollution level over all a site's von Neumann neighbouring locations becomes its new pollution level.
\end{description}

\subsection{Replacement - $R_{[a,b]}$} 
\begin{description}
\item[ Replacement - $R_{[a,b]}$] 
When an agent dies it is replaced by an agent of age 0 having random genetic attributes, random position on the Sugarscape, random initial endowment, and a maximum age selected from the range [a,b].

\end{description}

The two constants $a$ and $b$ we have defined already as $LOWERAGELIMIT$ and $UPPERAGELIMIT$ and we assume that the range is inclusive.  It is not stated whether the new agents immediately consume the resources at the location they are placed in.  We assume they do not, but accept that the alternative interpretation is equally valid. Although not part of the rule definition in the appendix it is stated elsewhere in the book that new agents will have initial resource levels set between 5 and 25.  We have defined $STARTSUGARMIN$ and $STARTSUGARMAX$ for this purpose.

Although the simulation can be run without employing the replacement rule (in an effort, for example, to determine the total carrying load - maximum tolerable population of agents - of a simulation space) there is no stated separate death rule. We will first add a schema that defines ``death'' explicitly to ensure consistency.   

\begin{description}
\item[ Death ] 
When an agent reaches its maximum allowed age or runs out of resources it is removed from the simulation and all its associated loans (either as borrower or lender) are considered void.
\end{description}
\begin{schema}{Death}
\Delta Agents\\
 \where
population'=population\setminus\\
\t3\{a:AGENT  |age(a)=maxAge(a)\lor agentSugar(a)=0\}\hfill(1)\\
loanBook'=population' \dres loanBook\rres\hfill(2)\\
\t1 \{x:AGENT\cross(\nat\cross\nat)|first(x)\in population'\} \\
\forall a:AGENT\spot\hfill(3)\\
a \in population' \implies \\
\t1 (sex(a)=sex'(a)\land vision(a)=vision'(a)\\
\t1\land maxAge(a)=maxAge'(a)\land agentCulture(a)=agentCulture'(a)\\
\t1\land position(a)=position'(a)\land age(a)=age'(a)\\
\t1\land agentSugar(a)=agentSugar'(a)\\
\t1\land metabolism'(a)=metabolism(a)\\
\t1\land diseases'(a)=diseases(a)\\
\t1\land agentImmunity'(a)=agentImmunity(a)\\
\t1\land children'(a)=children(a))\\
\t1\land initialSugar'(a)=initialSugar(a)\\
\end{schema}

\begin{enumerate}
\item We remove from the population all agents who have reached their maximum age or who have no sugar reserves; 
\item We remove all loans owed by or owing to these dying agents;
\item Any agent not being removed still has all attributes completely unchanged.
\end{enumerate}

The replacement rule follows readily from this rule, the only addition being the generation of new agents to replace the agents being removed. In effect we have broken the replacement rule into two parts $Death$ followed by $Replacement$; although the $Death$ rule may be used in isolation the $Replacement$ rule must always be preceded by the application of the $Death$ rule.
\begin{schema}{Replacement}
\Delta Agents\\
 \where
\#population'=INITIALPOPULATIONSIZE\hfill(1)\\
loanBook'= loanBook\\
\forall a:AGENT\spot\\
 a \in population \implies \hfill(2)\\
\t1 (a\in population'\\
\t1 \land sex(a)=sex'(a)\land vision(a)=vision'(a)\\
\t1\land maxAge(a)=maxAge'(a)\land agentCulture(a)=agentCulture'(a)\\
\t1\land position(a)=position'(a)\land age(a)=age'(a)\\
\t1\land agentSugar'(a)=agentSugar(a)\\
\t1\land metabolism'(a)=metabolism(a)\\
\t1\land diseases'(a)=diseases(a)\\
\t1\land agentImmunity'(a)=agentImmunity(a)\\
\t1\land children'(a)=children(a))\\
\t1\land initialSugar'(a)=initialSugar(a)\\
a\in population'\setminus population\implies\\
\t1 (age'(a) = 0\hfill(3)\\
\t1\land STARTSUGARMIN \leq agentSugar'(a) \leq STARTSUGARMAX\\
\t1\land initialSugar'(a)=agentSugar'(a)\\
\t1\land diseases'(a)=\emptyset\land children'(a)=\emptyset)\\
\end{schema}

\begin{enumerate}
\item The new population has the correct number of members; 
\item The existing agents remain unchanged and part of the new population;
\item All new agents have new values initialised within the allowed limits (those not stated explicitly are random values within the ranges set by the specification invariants.
\end{enumerate}
We do not state the positions of any new agents because they are chosen randomly. Our schema invariants ensure that they are on the grid in a location not occupied by any other agent.

We need to add some extra information to this rule definition to ensure that:
\begin{enumerate}
\item Newly created agents have no diseases, children or loans;
\item Their initial endowment of resources is within a set range.
\end{enumerate}

\subsection{Agent Mating $S$} 
\begin{description}
\item[ Agent Mating $S$] ~ \\ 
\begin{itemize}
\item Select a neighboring agent at random;
\item If the neighboring agent is of the opposite sex and if both agents are fertile and at least one of the agents has an empty neighboring site then a newborn is produced by crossing over the parents' genetic and cultural characteristics;
\item Repeat for all neighbors.
\end{itemize}
\end{description}

This rule determines how mating takes place amongst agents to produce offspring. An agent is fertile if its age is within preset boundaries. This is represented by the simple $isFertile$ function below. 

\begin{axdef}
  isFertile: (\nat\cross SEX)\inj boolean
\where
\forall age:\nat\spot\\
  isFertile(age,male) \iff\\
  \t1 MALEFERTILITYSTART\leq age \leq MALEFERTILITYEND \\
~\\
  isFertile(age,female) \iff\\
  \t1FEMALEFERTILITYSTART\leq age \leq FEMALEFERTILITYEND \\ 
\end{axdef}

We define two functions that take in an agent and a mapping from parents to offspring and returns the father or mother of the agent. 
\begin{axdef}
father: AGENT \cross((AGENT\cross AGENT)\pinj AGENT)\pinj AGENT\\
mother: AGENT \cross((AGENT\cross AGENT)\pinj AGENT)\pinj AGENT
\where
\forall x,m,f:AGENT;Offspring:(AGENT\cross AGENT)\pinj AGENT \spot\\
father(x,Offspring)=m\iff Offspring((m,f))=x\\
mother(x,Offspring)=f\iff Offspring((m,f))=x
\end{axdef}

The issues encountered with the mating rule are similar to those with movement. If two sets of parent try to produce offspring in the same vacant location only one can succeed. As there is no preferred conflict resolution rule we cannot state any preference for which agents succeed in producing children and which do not.  All we can state is that the maximum number of offspring will be produced given the space constraints but we cannot always be sure which offspring make it into this set. Neighbours in this rule refers to \emph{von Neumann Neighbours} only.

Mating although proceeding concurrently throughout the population is an exclusive event. That is, if agent $A$ is mating with agent $B$ then $A$ cannot be mating with any other agent at the same time: you can only ate with one partner at a time.  The rule itself specifies that each agent will mate with all available partners so the execution of the rule can involve a sequence of mating events for specific agents. 

Although it is not stated in the rule definition the accompanying book mentions that each parent should gift half of its sugar to its offspring and will only mate if it has a sugar level equal to or greater than its initial sugar level (that is its sugar level on creation). This significantly complicates the rule and dramatically changes its definition and characteristics.  However we will assume that this information was inadvertently omitted from the rule definition as the rule makes more sense if we include these extra factors.

Since each individual agent can involve itself in a sequence  of up to four mating events during rule execution we require a specification that retains global concurrency while still imposing a sequential ordering based on these constraints.  We do this by collecting all possible potential mating partners into a set and then dividing this set into a sequence of maximally sized sets where each subset contains only mating events that can occur concurrently.  These sets are produced using a conflict resolution rule that ensures that only pairing that can occur simultaneously appear within each such subset.  The rule then proceeds by executing mating events within each subset concurrently while the sets are evaluated in sequence.

\begin{schema}{AgentMating}
\Xi Lattice\\
\Delta Agents\\
 \where
 loanBook'=loanBook\\
\exists potentialMatingPairs: \power (AGENT\cross AGENT)|\hfill(1)\\
potentialMatingPairs=\{(a:AGENT,b:AGENT)| sex(a)\neq sex(b)\\
\t1 \land isFertile(age(a),sex(a)) \land isFertile(age(head),sex(head))\\
\t1\land adjacent(position(a),position(head))\}\\
(population',position',vision',agentSugar',agentCulture',metabolism'\\
\t1,children',diseases',agentImmunity',age',sex',initialSugar')=\hfill(2)\\
concurrentMating(getConfictFreePairs(potentialMatingPairs),population,position,vision,\\
\t1agentSugar,agentCulture,metabolism,children,\\
\t1diseases,agentImmunity,age,maxAge, sex,initialSugar)
\end{schema}
\begin{enumerate}
\item Generate the set of all possible mating pairs;
\item Recursively proceed with concurrent mating within the conflict free subsets.
\end{enumerate}

\begin{axdef}
getConfictFreePairs: \power (AGENT\cross AGENT)\\
\fun\\
\seq ( \power (AGENT\cross AGENT))
\where
\forall AllPairs: \power (AGENT\cross AGENT); a:AGENT;\spot\\
\exists conflictFreeSet: \power (AGENT\cross AGENT)|\\
conflictFreeSet\subseteq AllPairs\hfill(1)\\
\land  a\in \ran conflictFreeSet \implies a \notin \dom conflictFreeSet\\
\land  a\in \dom conflictFreeSet \implies a \notin \ran conflictFreeSet\\
\forall otherSet:\power (AGENT\cross AGENT) | \hfill(2)\\
otherSet\subseteq AllPairs\land  a\in \dom otherSet \implies a \notin \ran AllPairs\\
\land  a\in \ran otherSet \implies a \notin \dom AllPairs\spot \#otherSet\leq\#conflistFreeSet\\

getConfictFreePairs(\emptyset)=\emptyset\\
getConfictFreePairs(AllPairs)=\hfill(3)\\
\langle conflictFreeSet\rangle\cat getConfictFreePairs(AllPairs\setminus conflictFreeSet)\\
\end{axdef}

\begin{enumerate}
\item Generate a collision free (conflict resolved) set where each agent can only once within the set;
\item Ensure this set is as large as possible;
\item Recurse through the remaining pairs dividing them into more conflict free sets. 
\end{enumerate}

\begin{axdef}
concurrentMating: \seq \power(AGENT\cross AGENT)\\
 \cross \power AGENT\\
 \cross  AGENT  \inj POSITION\\
 \cross AGENT \pfun \nat_1\\
 \cross AGENT \pfun \nat\\
\cross AGENT \pfun \seq BIT\\
 \cross AGENT\pfun\nat\\
 \cross AGENT \pfun\power AGENT\\
  \cross AGENT\pfun\power \seq BIT\\
\cross AGENT\pfun \seq BIT\\
 \cross AGENT \pfun \nat\\
 \cross AGENT \pfun \nat_1\\
  \cross AGENT \pfun SEX\\
  \cross AGENT \pfun \nat\\
\rel\\ 
\power AGENT\\
\cross  AGENT  \inj POSITION\\
 \cross AGENT \pfun \nat_1\\
\cross AGENT \pfun \nat\\
\cross AGENT \pfun \seq BIT \\
\cross AGENT\pfun\nat \\
\cross AGENT \pfun\power AGENT\\
 \cross AGENT\pfun\power \seq BIT\\
\cross AGENT\pfun \seq BIT\\
 \cross AGENT \pfun \nat\\
 \cross AGENT \pfun \nat_1\\
  \cross AGENT \pfun SEX\\
  \cross AGENT \pfun \nat\\

\where
\forall tail :\seq \power(AGENT\cross AGENT);\\
head: \power(AGENT\cross AGENT);\\
 population: \power AGENT;\\
 position:  AGENT  \inj POSITION;\\
 vision: AGENT \pfun \nat_1;\\
 agentSugar: AGENT \pfun \nat;\\
agentCulture: AGENT \pfun \seq BIT;\\
 metabolism: AGENT\pfun\nat;\\
 children: AGENT \pfun\power AGENT;\\
 diseases: AGENT\pfun\power \seq BIT;\\
agentImmunity: AGENT\pfun \seq BIT;\\
age: AGENT \pfun \nat;\\
maxAge: AGENT \pfun \nat_1;\\
sex: AGENT \pfun SEX;\\
 initialSugar: AGENT \pfun \nat;\\
 \zbreak
  \exists newpopulation: \power AGENT;\\
 newposition:  AGENT  \inj POSITION;\\
 newvision: AGENT \pfun \nat_1;\\
 newagentSugar: AGENT \pfun \nat;\\
newagentCulture: AGENT \pfun \seq BIT;\\
 newmetabolism: AGENT\pfun\nat;\\
 newchildren: AGENT \pfun\power AGENT;\\
 newdiseases: AGENT\pfun\power \seq BIT;\\
newagentImmunity: AGENT\pfun \seq BIT;\\
newage: AGENT \pfun \nat;\\
newmaxAge: AGENT \pfun \nat_1;\\
newsex: AGENT \pfun SEX;\\
 newinitialSugar: AGENT \pfun \nat;|\\
(newpopulation,newposition,newvision,newagentSugar,newagentCulture,\\
\t1newmetabolism,newchildren,newdiseases,newagentImmunity,newage,newmaxAge,\\
\t1 newsex,newinitialSugar)=\\
applyMating(asSeq(head),population,position,vision,\\
\t1agentSugar,agentCulture,metabolism,children,\\
\t1diseases,agentImmunity,age,maxAge, sex,initialSugar)\spot\\

concurrentMating(\langle\rangle, population,position,vision,\\
\t1agentSugar,agentCulture,metabolism,children,\\
\t1diseases,agentImmunity,age,maxAge, sex,initialSugar)=\\
(population,position,vision,agentSugar,agentCulture,metabolism,\\
\t1children,diseases,agentImmunity,age,maxAge, sex,initialSugar)\\

concurrentMating(\langle head\rangle\cat tail,population,position,vision,\\
\t1agentSugar,agentCulture,metabolism,children,\\
\t1diseases,agentImmunity,age,maxAge, sex,initialSugar)=\\
concurrentMating(tail,newpopulation,newposition,newvision,\\
\t1newagentSugar,newagentCulture,newmetabolism,newchildren,\\
\t1newdiseases,newagentImmunity,newage,newmaxAge, newsex,newinitialSugar)\\

\end{axdef}

\begin{axdef}
applyMating:\seq (AGENT\cross AGENT)\\
 \cross \power AGENT\\
 \cross  AGENT  \inj POSITION\\
 \cross AGENT \pfun \nat_1\\
 \cross AGENT \pfun \nat\\
\cross AGENT \pfun \seq BIT\\
 \cross AGENT\pfun\nat\\
 \cross AGENT \pfun\power AGENT\\
  \cross AGENT\pfun\power \seq BIT\\
\cross AGENT\pfun \seq BIT\\
 \cross AGENT \pfun \nat\\
 \cross AGENT \pfun \nat_1\\
  \cross AGENT \pfun SEX\\
  \cross AGENT \pfun \nat\\
\rel\\ 
\power AGENT\\
\cross  AGENT  \inj POSITION\\
 \cross AGENT \pfun \nat_1\\
\cross AGENT \pfun \nat\\
\cross AGENT \pfun \seq BIT \\
\cross AGENT\pfun\nat \\
\cross AGENT \pfun\power AGENT\\
 \cross AGENT\pfun\power \seq BIT\\
\cross AGENT\pfun \seq BIT\\
 \cross AGENT \pfun \nat\\
 \cross AGENT \pfun \nat_1\\
  \cross AGENT \pfun SEX\\
  \cross AGENT \pfun \nat\\

\where
\forall population : \power AGENT; \\
position : AGENT  \inj POSITION;\\
sex: AGENT \pfun SEX;\\
vision: AGENT \pfun \nat_1;\\
age: AGENT \pfun \nat;\\
initialSugar: AGENT \pfun \nat;\\
maxAge: AGENT \pfun \nat_1;\\
metabolism:AGENT\pfun\nat;\\
agentSugar: AGENT \pfun \nat;\\
agentCulture:AGENT \pfun \seq BIT;\\
children: AGENT \pfun\power AGENT;\\
agentImmunity: AGENT\pfun \seq BIT;\\
diseases: AGENT\pfun\power \seq BIT;\\
head : AGENT\cross AGENT;\\
tail :\seq (AGENT\cross AGENT);\spot\\
\zbreak
\exists offspring,a,b:AGENT;\\
newsex: AGENT \pfun SEX;\\
newvision: AGENT \pfun \nat_1;\\
newmetabolism,newagentSugar,newinitialSugar: AGENT \pfun \nat;\\
newmaxAge: AGENT \pfun \nat_1;\\
newagentCulture:AGENT \pfun \seq BIT;\\
newchildren: AGENT \pfun\power AGENT;\\
newagentImmunity: AGENT\pfun \seq BIT;\\
inheritedImmunity: \seq BIT;\\
inheritedCulture: \seq BIT;\\
| offspring\notin population\\
a=first(head)\land b=second(head)\\

newchildren: children\cup\{offspring\mapsto \emptyset\,a\mapsto children(a)\cup\{ offspring\},\\
\t3b\mapsto children(b)\cup\{ offspring\}\}\\

newsex = sex\cup\{offspring\mapsto male\}\\
 \t1\lor newsex=sex\cup\{offspring\mapsto female\}\\
newvision = vision\cup\{offspring\mapsto vision(a)\}\\
\t1 \lor newvision = vision\cup\{offspring\mapsto vision(b)\}\\

newmaxAge = maxAge\cup\{offspring\mapsto maxAge(a)\}\\
\t1 \lor newmaxAge = maxAge\cup\{offspring\mapsto maxAge(b)\}\\
newmetabolism =metabolism\cup\{offspring\mapsto metabolism(a)\}\\
\t1 \lor newmetabolism=metabolism\cup\{offspring\mapsto metabolism(b)\}\\

newinitialSugar =initialSugar\oplus\\
\t1\{offspring\mapsto initialSugar(a)/2+initialSugar(b)/2,a\mapsto initialSugar(a)/2,b\mapsto initialSugar(b)/2\}\\
newagentSugar = agentSugar\cup\{offspring\mapsto initialSugar\}\\

\t1\land \forall n:1\upto IMMUNITYLENGTH\spot\\
\t2 (inheritedImmunity(n)=agentImmunity(a)(n)\\
\t2\lor inheritedImmunity(n)=agentImmunity(b)(n))\\
newagentImmunity: agentImmunity\cup\{offspring\mapsto inheritedImmunity\}\\

\t1\land \forall n:1\upto CULTURECOUNT\spot\\
\t2 (inheritedCulture(n)=agentCulture(a)(n)\\
\t2\lor inheritedCulture(n)=agentCulture(b)(n))\\
newagentCulture:agentCulture\cup\{offspring\mapsto inheritedCulture\}\\
\zbreak
applyMating(\langle\rangle, population,position,vision,agentSugar,agentCulture,\\
\t1metabolism,children,diseases, agentImmunity,age,maxAge, sex,initialSugar)=\\
 (population,position,vision,agentSugar,agentCulture,metabolism,\\
\t1children,diseases, agentImmunity,age,maxAge, sex,initialSugar)\\
~\\
applyMating(\langle head\rangle\cat tail, population,position,vision,agentSugar,agentCulture,\\
\t1metabolism,children,diseases, agentImmunity,age,maxAge, sex,initialSugar)=\\
\IF ((\exists loc:POSITION|  (adjacent(loc, position(ag)))\\
\t2\lor adjacent(loc, position(head))\land loc\notin \dom position)\\
\t2\land (agentSugar(head)>initialSugar(head)) \land (agentSugar(ag)>initialSugar(ag))) \THEN\hfill(2a)\\

\t1applyMating(tail, population\cup\{offspring\},position\cup\{offspring\mapsto loc\},\\
\t1newvision,newagentSugar,newagentCulture,newmetabolism,\\
\t1 newchildren,diseases\cup\{offspring\mapsto \emptyset\}, newagentImmunity,\\
\t1 age\cup\{offspring\mapsto 0\},newmaxAge, newsex,initialSugar)\\
\ELSE\hfill(2b)\\
\t1applyMating(tail, population,position,vision,agentSugar,agentCulture,\\
 \t2metabolism,children,diseases, agentImmunity,age,maxAge, sex,initialSugar)

\end{axdef}

\subsection{Agent Inheritance $I$} 
\begin{description}
\item[ Agent Inheritance $I$] 
When an agent dies its wealth is equally distributed among all its living children.
\end{description}
The rule definition is deceptively simple but some assumptions must be made in order to give it a precise definition. These assumptions are required because of the discrete nature of the simulation. Only living children can inherit from a parent. If a child is alive but scheduled to die at the same time as their parent then (because all agents who are due to die will die simultaneously)
 this child should not inherit from their parent. If we were to allow them to inherit we would either have to impose an ordering on the allocation of inheritance making the rule more complex or accept than the ordering will sometimes result in part of an inheritance disappearing. This extra complexity brings no real benefit to the simulation so we discount it.
 
The second assumption is that we allow for rounding errors. Resources ($sugar$) come in discrete amounts so division between children requires integer division. This is also true of division of the loans amongst an agents children. We just accept any rounding errors as part of the discrete nature of the simulation.

Finally we note that inheritance is separate from the actual death or replacement rule, it reallocates the resources of agents due to die but it does not remove those agents from the simulation.  We leave that to the actual Replacement or Death rule and assume that one of these rules is applied \emph{after} the inheritance rule.  This simplifies the $Inheritance$ schema.

To enable inheritance to handle the loan book (when an agent dies its loans are passed on to its children) we  introduce some helper functions. The $asSeq$ function turns a set of items into a sequence of items.  It does not specify the ordering in the sequence.
\begin{gendef}[X]
asSeq:\power X \fun \seq X
 \where
 \forall x:\power X; y:\seq X\spot\\
asSeq(x)=y\iff (\ran y=x\land \#y=\#x)\\
 \end{gendef}
 
The second function $disperseLoans$ takes in the loan book, a sequence containing all the dying agents and the children of the agents and produces an updated loan book with the loans of the dying agents now dispersed amongst their children.  To do this it employs a third function $oneAgentLoans$ that takes in a single agent (who is marked for removal) the loans (in a sequence) held by that agent and the set containing its children. It outputs a new set of loans generated by dispersing all this agents loans amongst its children. In both cases we use sequences for the parameter we are recursing over as it makes the recursion easier to specify.

 \begin{axdef}
disperseLoans: (\power(AGENT\cross(AGENT\cross (\nat\cross \nat)))\cross \seq AGENT\\
\t2 \cross (AGENT\pfun\power AGENT))\\
\fun \power(AGENT\cross (AGENT\cross (\nat\cross \nat)))
 \where
 \forall Loans:\power(AGENT\cross(AGENT\cross(\nat\cross\nat)));a:AGENT;\\
 \t1tail:\seq AGENT;Children:AGENT\pfun\power AGENT\spot\\
 disperseLoans(Loans,\langle\rangle,Children)=Loans\\
 ~\\
 disperseLoans(Loans,\langle a \rangle\cat tail,Children)=\\
disperseLoans((\{a\}\ndres Loans)\cup oneAgentLoans(a, asSeq(\ran(\{a\}\dres Loans)),\\
\t2Children(a)),tail,Children)
 \end{axdef}

 \begin{axdef}
 oneAgentLoans:AGENT\cross \seq(AGENT\cross (\nat\cross\nat))\cross\power AGENT \\
 \t1 \inj\power(AGENT\cross(AGENT\cross(\nat\cross\nat)))
 \where
  \forall a,borrower,inheritor:AGENT;Children:\power AGENT;amt,dur,newAmt:\nat; \\
 \t2 tail:\seq (AGENT\cross (\nat\cross\nat))\spot\\
 oneAgentLoans(a,\langle\rangle,Children)=\emptyset\\
 ~\\
 oneAgentLoans(a,\langle (borrower,(amt,dur)) \rangle\cat tail,Children)=\\
 \t1\{x: AGENT |x\in Children\spot (x,(borrower,(amt\div\#Children,dur)))\}\\
 \t1\cup~oneAgentLoans(a,tail,Children)\\
  \end{axdef}

 The getMother and getFather functions simply take in an agent and the $children$ set and finds the mother (father) of the agent from this set.
 \begin{axdef}
 getMother:AGENT\cross (AGENT\pfun\power AGENT) \cross AGENT \pfun SEX \pfun AGENT\\
 getFather:AGENT\cross (AGENT\pfun\power AGENT) \cross AGENT \pfun SEX \pfun AGENT
 \where
 \forall child,parent:AGENT; children:AGENT\pfun\power AGENT\spot\\
 getMother(child,children,sex)=parent\iff child\in children(parent) \land sex(parent)=female\\
 getFather(child,children,sex)=parent\iff child\in children(parent) \land sex(parent)=male
 \end{axdef}

An agent can inherit from at most two different agents, one male and one female. We use this to facilitate the specification by treating each sex separately.

\begin{schema}{Inheritance}
\Delta Agents\\
 \where
 population'=population \land sex'=sex\\
 position'=position \land  vision'=vision\\
 age'=age \land maxAge'=maxAge\\
 agentCulture'=agentCulture \land children'=children\\
 metabolism'=metabolism \land  diseases'=diseases\\
 agentImmunity'=agentImmunity\\
initialSugar'=initialSugar\\
 ~\\
\exists dying: \power AGENT; inheritFromFemale,inheritFromMale:AGENT\pfun\nat;\\
\forall x:AGENT;\exists p:AGENT |x\in population \setminus dying \spot\hfill(1)\\
\dom inheritFromFemale=\dom inheritFromMale=population\setminus dying\\
 dying=\{x:AGENT | x\in population \land \\
 \t1\land (age(x)=maxAge(x)\lor agentSugar(x)=0)\}\\
 
 getMother(x,children,sex) \notin dying\implies\hfill(1a)\\
 \t1 inheritFromFemale(x)=0\\
p=getMother(x,children,sex) \land p\in dying\implies\\
 \t1 inheritFromFemale(x)=\\
 \t2agentSugar(p)\div\#(population\cap children(p)\setminus dying))\\
 ~\\
 getFather(x,children,sex) \notin dying\implies \hfill(1b)\\
 \t1 inheritFromMale(x)=0\\
 p=getFather(x,children,sex) \land p\in dying\implies\\
 \t1 inheritFromMale(x)=\\
 \t2agentSugar(p)\div\#(population\cap children(p)\setminus dying))\\
 
x\in dying\hfill(3)\\
\t1 \implies agentSugar'(x)=0\\
x\notin dying\hfill(4)\\
\t1 \implies agentSugar'(x)=agentSugar(x)\\
\t2+inheritFromMale(x)+inheritFromFemale(x)\\
 loanBook'=disperseLoans(loanBook,asSeq(dying),children)\hfill(5)\\
\end{schema}

\begin{enumerate}
\item First we construct the set of dying agents. Then using this set of dying agents we can construct two functions, one mapping amounts inherited from a female parent and one mapping amounts inherited from a male parent.  These sets are then used to update the $sugar$ of each agent;
\begin{enumerate}[a)] 
\item The function giving the amount each inheriting agent gets from its female parent is constructed by finding all healthy agents who have a dying mother and determining their share of their dying mother's resources;
\item The function listing amounts each agent gets from a male parent is constructed in an almost identical manner.
\end{enumerate}
\item If an agent is dying its $sugar$ level is set to zero (because it is being reallocated to its children);
\item Otherwise the agents sugar level is its old level plus whatever it inherits from both dying parents;
\item Finally we update the loanBook using our $disperseLoans$ function.
\end{enumerate}

\subsection{Agent Culture $K$} 
\begin{description}
\item[ Agent cultural transmission] ~
\begin{itemize}
\item Select a neighboring agent at random;
\item Select a tag randomly;
\item If the neighbor agrees with the agent  at that tag position, no change is made; if they disagree, the neighbor's tag is flipped to agree with the agent's tag;
\item Repeat for all neighbors.  
\end{itemize}
\item[ Group membership]
Agents are defined to be members of the Blue group when 0s outnumber 1s on their tag strings, and members of the Red group in the opposite case.
\item[ Agent Culture $K$]
Combination of the ``agent cultural transmission'' and ``agent group membership'' rules given immediately above.
\end{description}

Group membership is defined with the assumption that there are always an odd number of tags. $tribe$ returns the affiliation of an agent based on the number of bits of each type in its culture sequence. If the majority of bits in a sequence are 0 then it belongs to the $blue$ tribe, otherwise it belongs to the $red$ tribe.  This is used by the culture rule. 

 \begin{axdef}
tribe: \seq BIT\fun affiliation
 \where
 \forall aSeq:\seq BIT\spot\\
 tribe(aSeq)=blue\iff \#(aSeq\rres\{0\}) >\#(aSeq\rres\{1\}) \\
 tribe(aSeq)=red\iff \#(aSeq\rres\{0\}) <\#(aSeq\rres\{1\}) 
 \end{axdef}

 $flipTags$ is a recursive function that takes in a culture tag sequence belonging to an agent, a sequence of neighbouring agents and the mapping containing all agent's culture tag sequences. It returns a new tag sequence generated by each neighbouring agent flipping one bit chosen at random of the original agent's tag sequence.  It is aided in this by the function $flipBit$ that takes in two bit sequences and returns a new sequence equal to the first bit sequence with one bit changed at random to match the other sequence at that position.
  
\begin{axdef}
 flipBit: \seq BIT \cross \seq BIT \rel \seq BIT
 \where
 \forall original, other, new:\seq BIT\spot\\
 flipBit(original,other)=new\iff\\
 \t1 \#original=\#other=\#new \land\\
\t1 \exists i:0\upto \#original\spot \forall j:0\upto \#original\spot\\
 \t2 ( i\neq j\implies new(j)=original(j)) \land new(i)=other(i)\\
 \end{axdef}
 
 \begin{axdef}
flipTags: \seq BIT\cross \seq AGENT \cross (AGENT\pfun \seq BIT) \rel \seq BIT
 \where
 \forall aSeq:\seq BIT; ag:AGENT; tail:\seq AGENT;\\
 \t1culturalResources:AGENT \pfun\seq BIT\spot\\
 flipTags(aSeq,\langle\rangle,culturalResources)=aSeq\\
 
 flipTags(aSeq,\langle ag\rangle\cat tail,culturalResources)=\\
\t1 flipTags(flipBit(aSeq,culturalResources(ag)),tail,culturalResources)\\
  \end{axdef}

The sequence of neighbours is provided by the $Culture$ scheme which employs the $asSeq$ function to convert a set of neighbours into a sequence.

\begin{schema}{Culture}
\Delta Agents\\
 \where
 population'=population\\
 sex'=sex\\
 position'=position\\
 vision'=vision\\
 age'=age\\
 maxAge'=maxAge\\
 agentSugar'=agentSugar\\
 children'=children\\
 loanBook'=loanBook\\
 diseases'=diseases\\
 metabolism'=metabolism\\
 agentImmunity'=agentImmunity\\
initialSugar'=initialSugar\\
 \forall a:AGENT \spot a \in population\implies agentCulture'(a)=\\
\t1 flipTags(agentCulture(a),\hfill(1)\\
\t2asSeq(\{b:AGENT | adjacent(position(a),position(b))\}),agentCulture)\\
\end{schema}

\begin{enumerate}
\item For every agent $a$ in the population we allow each other agent that counts $a$ as a neighbour to flip one bit at random of $a$'s culture bit sequence. 
\end{enumerate}

\subsection{Combat $C_{\alpha}$} 
\begin{description}
\item[ Agent Combat $C_{\alpha}$] ~
\begin{itemize}
\item Look out as far as vision permits in the four principle lattice directions;
\item Throw out all sites occupied by members of the agent's own tribe;
\item Throw out all sites occupied by members of different tribes who are wealthier then the agent;
\item The reward of each remaining site is given by the resource level at the site plus, if it is occupied, the minimum of $\alpha$ and the occupant's wealth;
\item Throw out all sites that are vulnerable to retaliation;
\item Select the nearest position having maximum reward and go there;
\item Gather the resources at the site plus the minimum of $\alpha$ and the occupants wealth if the site was occupied;
\item If the site was occupied then the former occupant is considered ``killed'' - permanently removed from play.
\end{itemize}
\end{description}

 $reward$ is used by the combat rule and values a position based on its sugar content and the sugar reserves held by any agent at that position.  The combat rule is really an extension of the movement rule where we are now allowed to move to locations occupied by other agents under certain predefined conditions.

\begin{axdef}
reward: POSITION\cross (POSITION\pinj \nat)\cross (AGENT\pinj POSITION)\\
\t1\cross (AGENT \pfun\nat)\cross\nat\\
 \fun \nat\\
 \where
 \forall l:POSITION; sugar:POSITION\pinj \nat; agentSugar:AGENT\pinj \nat;\\
 \t1positions:AGENT\pinj POSITION\spot\\
  \IF l\in \ran positions~\THEN\\
 \t1reward(l,sugar,positions,agentSugar,COMBATLIMIT)=\\
 \t2 sugar(l)+min(\{COMBATLIMIT,agentSugar(positions\inv(l))\})\\
 
\ELSE\\
  \t1reward(l,sugar,positions,agentSugar,COMBATLIMIT)=\\
 \t2 sugar(l)\\
 \end{axdef}

$availMoves$ returns the set of all safe moves that an agent can make. 

\begin{axdef}
availMoves:AGENT\cross (AGENT\pinj POSITION)\cross (POSITION\pinj\nat)\cross\\
(AGENT\pinj\nat)\cross (AGENT\pinj\seq BIT)\cross \nat\\
\t1\pfun\power POSITION
 \where
 \forall agent:AGENT;positions:AGENT\pinj POSITION;vision:\nat;\\
 \t1 sugar:POSITION\pinj\nat;agentSugar:AGENT\pinj\nat;\\
 \t1 culture: AGENT\pinj \seq BIT\spot\\
 
availMoves(agent,positions,sugar,agentSugar,culture,vision)=\\
\{l:POSITION;x:AGENT|distance(l,positions(agent))\leq vision\hfill(1)\\
\land positions(x)=l\implies (agentSugar(x)<agentSugar(agent)\hfill(2)\\
\t1\land tribe(culture(x))\neq tribe(culture(agent)))\\
\land ((distance(positions(x),l)\leq vision)\hfill(3)\\
\t1\land tribe(culture(x))\neq tribe(culture(agent)))\implies \\
\t2agentSugar(x)<agentSugar(agent)\\
\t2+reward(l,sugar,positions,agentSugar,COMBATLIMIT))\spot l\}\\
 \end{axdef}
\begin{enumerate}
\item Only locations within an agents neighbourhood are considered;
\item If a location is occupied it must be occupied by an agent belonging to a different tribe who has lower sugar levels;
\item We only consider a position already containing an agent from another tribe if there are no other agents from a different tribe within the neighbourhood of that location who are stronger than we will be once we have consumed the resources of the new location (that is agents who may retaliate against us for killing an agent belonging to their own tribe).
\end{enumerate}
We note that the rule as stated means we consider retaliation under all conditions even if we are just moving to an empty location.
It is unclear from the definition given as to how exactly we check for retaliation.  Do we base our check on agents visible from our current position or from the proposed position. We have assumed that it is based on the proposed position but it could easily be otherwise. We also assume that the range used is based on the vision of the moving agent as this seems logical.

The synchronous version of the combat rule assumes that all combat occurs instantaneously (concurrently).  We note that it is simpler to specify in that we just state the before and after states and make no mention of orderings of combat.
\begin{schema}{Combat_{Synchronous}}
\Delta SugarScape\\
 \where
 step'=step\\
\\
 maxSugar'= maxSugar\\
 pollution'=pollution\\
 loanBook'= population'\dres loanBook\rres (population'\dres (\ran loanBook))\hfill(1)\\
    population'\subseteq population\hfill(2)\\
   sugar'=sugar\oplus\{p:POSITION | p\in\ran position'\spot p\mapsto 0\}\hfill(3)\\

  \forall ag:AGENT; l:POSITION  \spot\\

  ag\in population' \implies\hfill(4)\\
  \t1(sex'(ag)=sex(ag)\\
  \t1\land vision'(ag)=vision(ag)\\
  \t1\land age'(ag)=age(ag)\\
  \t1\land maxAge'(ag)=maxAge(ag)\\
  \t1\land children'(ag)=children(ag)\\
  \t1\land agentCulture'(ag)=agentCulture(ag)\\
  \t1\land agentImmunity'(ag)=agentImmunity(ag)\\
  \t1\land metabolism'(ag)=metabolism(ag)\\
  \t1\land diseases'(ag)=diseases(ag)\\
  \t1\land initialSugar'(a)=initialSugar(a)\\
  \t1\land agentSugar'(ag)=agentSugar(ag)\hfill(5)\\
\t2+reward(position'(ag),sugar,position,agentSugar,COMBATLIMIT)\\
\t1\land position'(ag)\in\hfill(6)\\
 \t2 availMoves(ag,position,sugar,agentSugar,agentCulture,vision(ag)))\\
  \t1)\\
 
ag\in population \setminus population'\implies \hfill(7)\\
\t1 \exists x:AGENT \spot position'(x)=position(ag)\land tribe(culture(x))\neq tribe(culture(ag))\\

 ~\\
 (l\in availMoves(ag,position,sugar,agentSugar,agentCulture,vision(ag))\hfill(8)\\
 \t1\land reward(l,sugar,position,agentSugar,COMBATLIMIT)\\
 \t2 \geq reward(position'(ag),sugar,position,agentSugar,COMBATLIMIT)\\
 \t1\land distance(position(ag),l) < distance(position(ag),position'(ag)))\\
 \t1\implies \exists x:AGENT \spot position'\inv(l)=x\land position(x)\neq l\\

\end{schema}

\begin{enumerate}
\item Every agent that is removed from the simulation is also removed from the $loanBook$;
\item No new agents are introduced;
 \item Location sugar levels are updated;
\item Every agent that remains in the population has all its attributes unchanged apart from (possibly) position and sugar;

  \item We update the sugar levels of each agent using the reward function;
 \item Every agent has moved somewhere within their old neighbourhood;
\item Every agent that is no longer part of the population was removed by combat, that is, there is another agent (the agent that killed them) now situated in their old position;

\item If a location available to an agent and the reward of that location is better or equal to that agent's new position and it was closer than that agents new position to its old position then it must be the case that some other agent has just moved to that location (otherwise we would have moved there);

\end{enumerate}

We have had to make some assumptions here.  It is not stated what happens when there are no available moves, for example if all sites are subject to retaliation. We have assumed that a move is preferable to staying still and that the only time that an agent stays in the same position is when there are no available moves. That is, if every site, including our current one, is subject to retaliation then we do not move anywhere. A more complex interpretation would be to for an agent that cannot escape retaliation to attack another agent anyway and hope for the best but purely in the interests of simplicity we have agents remain where they are.

\subsection{Credit $L_{d~r}$} 
\begin{description}
\item[ Credit $L_{d~r}$]~
\begin{itemize}
\item An agent is a potential lender if it is too old to have children, in which case the maximum amount it may lend is one-half of its current wealth;
\item An agent is a potential lender if it is of childbearing age and has wealth in excess of the amount necessary to have children, in which case the maximum amount it may lend is the excess wealth;
\item An agent is a potential borrower if it is of childbearing age and has insufficient wealth to have a child and has income (resources gathered, minus metabolism, minus other loan obligations) in the present period making it credit-worthy for a loan written at terms specified by the lender;
\item If a potential borrower and a potential lender are neighbors then a loan is originated with a duration of d years at the rate of r percent, and the face value of the loan is transferred from the lender to the borrower;
\item At the time of the loan due date, if the borrower has sufficient wealth to repay the loan then a transfer from the borrower to the lender is made; else the borrower is required to pay back half of its wealth and a new loan is originated for the remaining sum;
\item  If the borrower on an active loan dies before the due date then the lender simply takes a loss;
\item If the lender on an active loan dies before the due date then the borrower is not required to pay back the loan, unless inheritance rule $I$ is active, in which case the lender's children now become the borrower's creditors.
\end{itemize}
\end{description}

$totalOwed$ calculates the total amount owed from a given sequence of loans. We have assumed that interest is simple interest and not compound.
\begin{axdef}
totalOwed: \seq (AGENT\cross(\nat\cross\nat))\fun\nat
\where
\forall a:AGENT; amt,dur:\nat; tail:\seq(AGENT\cross(\nat\cross\nat))\spot\\
totalOwed(\langle\rangle)=0\\
totalOwed(\langle(a,(amt,dur))\rangle\cat tail)=(amt+amt*RATE*DURATION)\\
\t2+totalOwed(tail)\\
\end{axdef}

$canLend$ and $willBorrow$ are simple rules. The definition of what determines credit-worthiness is missing so we have assumed it means an agent has enough money to pay \emph{all} their outstanding loans.

\begin{axdef}
  canLend: \nat\cross SEX \cross \nat \fun boolean\\
\where
\forall age,sugar:\nat\spot\\
canLend(age,male,sugar) \iff \\
\t1age> MALEFERTILITYEND \\
\t1\lor (MALEFERTILITYSTART \leq age \leq MALEFERTILITYEND\\
\t2\land sugar>CHILDAMT)\\
canLend(age,female,sugar)\iff\\
\t1 age> FEMALEFERTILITYEND \\
\t1\lor (FEMALEFERTILITYSTART \leq age \leq FEMALEFERTILITYEND\\
\t2\land sugar>CHILDAMT)\\
\end{axdef}

\begin{axdef}
  willBorrow: \nat\cross SEX\cross \nat\cross \power(AGENT\cross(\nat\cross\nat)) \fun boolean\\
\where
\forall age,sugar:\nat;loans:\power(AGENT\cross(\nat\cross\nat))\spot\\
willBorrow(age,male,sugar,loans)\iff\\
\t1(MALEFERTILITYSTART \leq age \leq MALEFERTILITYEND\\
\t2\land sugar<CHILDAMT)\\
\t2\land sugar >totalOwed(asSeq(loans))\\
willBorrow(age,female,sugar,loans)\iff\\
\t1(FEMALEFERTILITYSTART \leq age \leq FEMALEFERTILITYEND\\
\t2\land sugar<CHILDAMT)\\
\t2\land sugar >totalOwed(asSeq(loans))\\
\end{axdef}

$amtAvail$ depends on whether an agent can still have children.  If they are no longer fertile then they can loan out half their available sugar. If that are still fertile then they have to retain enough sugar to have children.
\begin{axdef}
  amtAvail:\nat\cross SEX\cross \nat \fun \nat\\
\where
\forall age,sugar:\nat\spot\\
amtAvail(age,male,sugar)=\\
\IF (age> MALEFERTILITYEND)\THEN \\
\t1 sugar\div 2\\
\ELSE\IF (isFertile(age,male)\land sugar>CHILDAMT)\THEN\\
\t1sugar-CHILDAMT\\
\ELSE\\
\t1 0\\

amtAvail(age,female,sugar)=\\
\IF (age> FEMALEFERTILITYEND)\THEN \\
\t1 sugar\div 2\\
\ELSE\IF (isFertile(age,female)\land sugar>CHILDAMT)\THEN\\
\t1sugar-CHILDAMT\\
\ELSE\\
\t1 0\\
\end{axdef}

$amtReq$ is the amount that a lender requires.  This is not defined so we can only use a best guess as to what it is. We assume that the amount required is that which gives the borrower enough sugar to have children.  This is the simplest sensible definition we can think of. 
\begin{axdef}
  amtReq: \nat \fun \nat\\
  \where
  \forall sugar:\nat\spot\\
  amtReq(sugar)=CHILDAMT-sugar
\end{axdef}

We supply some simple helper functions that extract the borrower and lender from a loanBook entry, calculate the amount due from a loan, the principal and the due date (defined as the $step$ when payment is due).
\begin{axdef}
  lender: AGENT\cross(AGENT\cross(\nat\cross\nat))\inj AGENT\\
  borrower: AGENT\cross(AGENT\cross(\nat\cross\nat))\inj AGENT\\
  amtDue: AGENT\cross(AGENT\cross(\nat\cross\nat))\inj \nat\\
  principal:AGENT\cross(AGENT\cross(\nat\cross\nat))\inj \nat\\
  due: AGENT\cross(AGENT\cross(\nat\cross\nat))\inj \nat\\
\where
\forall l,b:AGENT;p,d:\nat\spot\\
lender(l,(b,(p,d)))=l\\
borrower(l,(b,(p,d)))=b\\
amtDue(l,(b,(p,d)))= p+p*RATE*DURATION\\
principal(l,(b,(p,d)))= p\\
due(l,(b,(p,d)))= d\\
\end{axdef}

Finally, using these functions we can present the $PayLoans$ schema.

\begin{schema}{PayLoans}
\Delta Agents\\
\Xi Step
 \where
 population'=population\\
 sex'=sex\\
 position'=position\\
 vision'=vision\\
 age'=age\\
 maxAge'=maxAge\\
 agentCulture'=agentCulture\\
 agentImmunity'=agentImmunity\\
 children'=children\\
 diseases'=diseases\\
 metabolism'=metabolism\\
 initialSugar'=initialSugar\\
 
\exists dueLoans,newLoans:(AGENT\rel(AGENT\cross(\nat\cross\nat)))\spot\\ 
 dueLoans= loanBook\rres (\ran(loanBook)\rres\{a:(\nat\cross\nat)| second(a)=step\})\\
 (loanBook',agentSugar')=\\
 \t1 payExclusiveLoans(chooseConflictFreeSets(dueLoans),agentSugar,loanBook)\\
\end{schema}

This schema is complicated by the fact that it is possible that an agent has a loan due and cannot pay this loan off.  In this case, according to the rule definition, the borrower must pay half of its sugar to the lender and renegotiate another loan to cover the remainder of its debt.  Under this rule some issues will arise if the borrower has more than one due loan and cannot pay these loans off.  The lender must pay each borrower in sequence the amount of half its sugar.  This cannot be performed simultaneously (for example if we owe three loans we cannot give each lender half our sugar as this would mean giving out more sugar than we actually have). In order to remain true to the rule definition we must, when we have more than one loan due, pay each loan in some sequence (defined using a conflict resolution rule e.g. pay biggest loan first). The helper function $chooseConflictFreeLoans$ returns a sequence of groups of loans that are conflict free (i.e. a borrower can only appear once in each group).

The function $payExclusiveLoans$ takes in this sequence of loan sets and processes each set concurrently in the same manner as the $Mating$ rule.

 \begin{axdef}
chooseConflictFreeLoans:(AGENT\rel(AGENT\cross(\nat\cross\nat)))\\
 \rel\\
  \seq (AGENT\rel(AGENT\cross(\nat\cross\nat)))
\where
\forall a:AGENT; dueLoans:(AGENT\rel(AGENT\cross(\nat\cross\nat)))\spot\\
chooseConflictFreeLoans(\emptyset) = \langle\rangle\\
chooseConflictFreeLoans(dueLoans)=\\
\t1\exists maxSet:(AGENT\rel(AGENT\cross(\nat\cross\nat))) | maxSet\subseteq dueLoans\\
\t1\land  \#(\{a\}\dres (\ran dueLoans))>0\implies \#(\{a\}\dres (\ran maxSet))=1\hfill(1)\\
\t1 \langle maxSet\rangle\cat chooseConflictFreeLoans(dueLoans\setminus maxSet)
\end{axdef}

\begin{enumerate}
\item We choose the largest convict free set possible where a set is deemed conflict free if all borrowers only appear in that set at most once.
\end{enumerate}

\begin{axdef}
payExclusiveLoans:\seq (AGENT\rel(AGENT\cross(\nat\cross\nat)))\\
 \cross AGENT \pfun \nat\\
 \cross (AGENT\rel(AGENT\cross(\nat\cross\nat))) \\
\rel\\
 ((AGENT\rel(AGENT\cross(\nat\cross\nat)))\\
 \cross AGENT \pfun \nat)
\where
\forall tail:\seq (AGENT\rel(AGENT\cross(\nat\cross\nat)));\\
 head,loanBook: (AGENT\cross(AGENT\cross(\nat\cross\nat)));\\
agentSugar: AGENT \pfun \nat\spot\\
payExclusiveLoans(\langle\rangle,agentSugar,loanBook)=\\
\t1(loanBook, agentSugar)\\
payExclusiveLoans(\langle head\rangle\cat tail,agentSugar,loanBook)=\\
\t1\exists newAgentSugar:AGENT \pfun \nat; newLoans:(AGENT\rel(AGENT\cross(\nat\cross\nat))) |\\

\t1(newLoans,newAgentSugar)=makePayments(asSeq(head),\emptyset,agentSugar)\spot\\
 \t1payExclusiveLoans(tail,newAgentSugar,(LoanBook\setminus head) \cup newLoans)
\end{axdef}

$makePayments$ is a recursive function that goes through a sequence of loans and makes the final payment on each one. It is used in the $PayLoans$ schema where it takes in a sequence of the due loans and the agents current sugar levels and returns a set of renegotiated loans, where payment is unable to be made, and the new agent sugar levels.

\begin{axdef}
  makePayments: \seq(AGENT\cross(AGENT\cross(\nat\cross\nat)))\cross\\
  \t1\power(AGENT\cross(AGENT\cross(\nat\cross\nat)))\cross(AGENT\pfun\nat)\\
  \fun\\
  (\power(AGENT\cross(AGENT\cross(\nat\cross\nat)))\cross (AGENT\pinj\nat))
\where
\forall renegotiatedLoans,new:\power(AGENT\cross(AGENT\cross(\nat\cross\nat)));\\
\t1updatedSugar,agentSugar:AGENT\pfun\nat;\\
\t1loan:(AGENT\cross(AGENT\cross(\nat\cross\nat)));\\
\t1tail:\seq(AGENT\cross(AGENT\cross(\nat\cross\nat)))\spot\\
~\\
  makePayments(\langle\rangle,renegotiatedLoans,updatedSugar)=\hfill(1)\\
  \t1(renegotiatedLoans,updatedSugar)\\
  makePayments(\langle loan\rangle\cat tail,new,agentSugar)=\hfill(2)\\
   \IF amtDue(loan)\leq agentSugar(borrower(loan))\THEN\hfill(2a)\\
 \t1 makePayments(tail,new,agentSugar\\
 \t2\oplus\{lender(loan)\mapsto agentSugar(lender(loan))+amtDue(loan), \\
 \t2borrower(loan)\mapsto agentSugar(borrower(loan))-amtDue(loan)\})\\
\ELSE\hfill(2b)\\
\t1makePayments(tail,new\cup\{(lender(loan),\\
 \t1(borrower(loan),(amtDue(loan)-agentSugar(borrower(loan))\div 2,\\
 \t2due(loan)+DURATION)))\},\\
 \t1agentSugar\oplus\{lender(loan)\mapsto agentSugar(lender(loan))\\
 \t2+agentSugar(borrower(loan))\div 2, \\
\t2borrower(loan)\mapsto agentSugar(borrower(loan))\div 2\})\\ 
\end{axdef}

For the final part of the $Credit$ rule we need to be able to work out the total owed by an agent over all loans. First we define two helper functions: $sumLoans$ and $totalOwed$.

\begin{axdef}
sumLoans: \seq(AGENT\cross(AGENT\cross(\nat \cross\nat)))\inj \nat
\where
\forall tail:\seq (AGENT\cross(AGENT\cross(\nat \cross\nat)));\\
\t1 top:AGENT\cross(AGENT\cross(\nat \cross\nat))\spot\\
sumLoans(\langle\rangle)=0\\
sumLoans(\langle top\rangle\cat tail)= sumLoans(tail)+amtDue(top)\\
\end{axdef}

\begin{axdef}
totalOwed: AGENT\cross(AGENT\cross(AGENT\cross(\nat \cross\nat)))\inj \nat\\
totalLoaned: AGENT\cross(AGENT\cross(AGENT\cross(\nat \cross\nat)))\inj \nat
\where
\forall  agent:AGENT; loans:AGENT\cross(AGENT\cross(\nat \cross\nat))\spot\\
totalOwed(agent,loans)=\\
\t1sumLoans(asSeq(loans\rres (\{agent\}\dres(\ran loans))))\\
totalLoaned(agent,loans)=\\
\t1sumLoans(asSeq(\{agent\}\dres loans))\\
\end{axdef}

\begin{schema}{MakeLoans}
\Delta Agents\\
\Xi Step
 \where
  population'=population\land sex'=sex\\
 position'=position\land vision'=vision\\
 age'=age\land maxAge'=maxAge\\
 initialSugar'=initialSugar\\
 agentCulture'=agentCulture\land agentImmunity'=agentImmunity\\
  diseases'=diseases\land children'=children\land metabolism'=metabolism\\
 
\exists newLoans: \power(AGENT\cross(AGENT\cross(\nat \cross\nat)));\\
\forall ag,lender,borrower:AGENT; amt,due:\nat\spot\\
loanBook'=loanBook\cup newLoans\hfill(1)\\
 ag\in\dom newLoans\implies\\
 \t1 agentSugar'(ag)=agentSugar(ag)-totalLoaned(ag,newLoans)\hfill(2a)\\
ag\in\dom(\ran newLoans)\implies\hfill(2b)\\\ 
\t1agentSugar'(ag)=agentSugar(ag)+totalOwed(ag,newLoans)\\
ag\notin\dom (newLoans)\cup \dom(\ran newLoans)\implies\hfill(2c)\\
\t1 agentSugar'(ag)=agentSugar(ag)\\
 willBorrow(age(ag),sex(ag),agentSugar'(ag),\\
 \t1\ran (loanBook'\cap\{a:AGENT\cross(AGENT\cross(\nat \cross\nat))\\
\t3|borrower(a)=borrower(loan)\}))\implies\hfill(2d)\\
 \t2\lnot\exists ag2:AGENT\spot canLend(age(ag2),sex(ag2),agentSugar'(ag2))\\
 \t3\land adjacent(position(ag2),position(ag))\\
 
 totalLoaned(ag,newLoans)\leq amtAvail(age(ag),sex(ag),agentSugar(ag))\hfill(3)\\
totalOwed(ag,newLoans)\leq amtReq(agentSugar(ag))\hfill(4)\\

(lender,(borrower,(amt,due)))\in newLoans \implies\hfill(5)\\
\t1 (canLend(age(lender),sex(lender),agentSugar(lender))\hfill(5a)\\
\t1\land willBorrow(age(borrower),sex(borrower),agentSugar(borrower),\\
\t2\{borrower\}\dres(\ran loanBook))\\

\t1\land amt\leq min(\{amtAvail(age(lender),sex(lender),agentSugar(lender)),\\
\t6amtReq(agentSugar(borrower))\})  \hfill(5b)\\
\t1\land due=step+DURATION\hfill(5c)\\
 \t1\land adjacent(position(lender),position(borrower)))\hfill(5d)\\
 
\end{schema}

\begin{enumerate}
\item The new loan book is the old book plus the new loans;
\item
The following properties ensure sugar is updated correctly and that the correct amount of borrowing has taken place:
\begin{enumerate}[a)]
\item If an agent is a lender then their new sugar levels decrease by the amount the have lent;
\item If an agent is a borrower then their sugar has increased by the amount they have borrowed;
\item Any agent that neither borrowed or lent has the same sugar levels as before;
\item If there remain any agents who still need to borrow then it is because there are no agents in their neighbourhood who are still in a position to borrow.
\end{enumerate}
\item The total amount loaned by any agent is no greater than the amount that agent had available;
\item The total amount borrowed is less than or equal to the amount required by the borrower;
\item Every loan in this set must have the following properties:
\begin{enumerate}[a)]
\item The lender must be in a position to lend;
\item The borrower must need to borrow;
\item The amount is less than or equal to the minimum of (i) the amount required by the borrower and (ii) the maximum amount available from the lender;
\item The due date of the loan is set by the $DURATION$ constant;
\item the borrower and lender must be neighbours.
\end{enumerate}

\end{enumerate}

\subsection{Agent Disease $E$} 
\begin{description}
\item[ Agent immune response]~
\begin{itemize}
\item If the disease is a substring of the immune system then end (the agent is immune), else (the agent is infected) go to the following step;
\item The substring in the agent immune system having the smallest Hamming distance from the disease is selected and the first bit at which it is different from the disease string is changed to match the disease.
\end{itemize}

\item[ Disease transmission] For each neighbor, a disease that currently afflicts the agent is selected at random and given to the neighbor.
\item[ Agent disease processes $E$]
Combination of ``agent immune response'' and ``agent disease transmission'' rules given immediately above
\end{description}

$subseq$ is a function for determining whether one sequence is a subsequence of another. $hammingDist$ determines the number of bit differences in two sequences of the same size.

\begin{axdef}
 subseq:\seq BIT\cross\seq BIT \fun boolean
\where
\forall mid,aSequence:\seq BIT\spot\\
subseq(mid,aSequence)\iff\\
\t1 \exists prefix,suffix:\seq BIT \spot prefix\cat mid\cat suffix=aSequence\\
\end{axdef}

\begin{axdef}
hammingDist:\seq BIT\cross\seq BIT\fun \nat
\where
\forall tail, rest:\seq BIT\spot\\
hammingDist(\langle\rangle,\langle\rangle)=0\\
hammingDist(\langle 1\rangle\cat tail,\langle 1\rangle\cat rest)= hammingDist(tail,rest)\\
hammingDist(\langle 0\rangle\cat tail,\langle 0\rangle\cat rest)= hammingDist(tail,rest)\\
hammingDist(\langle 0\rangle\cat tail,\langle 1\rangle\cat rest)= 1+hammingDist(tail,rest)\\
hammingDist(\langle 1\rangle\cat tail,\langle 0\rangle\cat rest)= 1+hammingDist(tail,rest)\\
\end{axdef}

$applyDiseases$ takes in a bit sequence representing the immunity of an agent and a list of diseases that affect the agent and produces a new immunity bit sequence that is updated by the disease list. More precisely, for every disease not in the immunity sequence a single bit in the closest subsequence that matches the disease is flipped to make the sequence more closely match the disease. It uses another function $processInfection$ to process each disease in the disease set.

\begin{axdef}
 applyDiseases: \seq BIT\cross \seq\seq BIT \fun \seq BIT
\where
\forall I,d:\seq BIT;tail:\seq\seq BIT\spot\\
 applyDiseases(I,\langle\rangle)=I\\
 applyDiseases(I,\langle d\rangle\cat tail)=\\
\t1 applyDiseases(processInfection(I,d), tail)\\
\end{axdef}

   \begin{axdef}
processInfection: \seq BIT\cross\seq BIT \fun \seq BIT
\where\forall I,d:\seq BIT\spot\\
\IF subseq(d,I)\THEN\\
\t1 processInfection(I,d)=I\\
\ELSE\\
\t1\exists a,b,c:\seq BIT;\forall x:\seq BIT\spot\\
\t1  a\cat b\cat c=I\\
\t1 (\#b=\#x\land subseq(x,I))\implies hammingDist(b,d)\leq hammingDist(x,d)\\
\t1\exists i:1\upto\#I \spot( y(i)\neq d(i)\land\forall j:\nat\spot j<i\implies d(j)=y(j))\\
\t1processInfection(I,d)=I\oplus\{(i+\#a)\mapsto b(i)\})\\

\end{axdef}

$ImmuneResponse$ is the simplest part of this rule to specify.  The recursive function $applyDiseases$ does all the work.
\begin{schema}{ImmuneResponse}
\Delta Agents\\
 \where
loanBook'=loanBook\\
 population'=population\\
 sex'=sex\\
 position'=position\\
 vision'=vision\\
 age'=age\\
 maxAge'=maxAge\\
 agentCulture'=agentCulture\\
 diseases'=diseases\\
  children'=children\\
  agentSugar'=agentSugar\\
  metabolism'=metabolism\\
  initialSugar'=initialSugar\\
 agentImmunity'=\{a:AGENT |a\in population\spot\\
 \t1 a\mapsto applyDiseases(agentImmunity(a),asSeq(diseases(a)))\}\\
 
   \forall x:AGENT\spot x\in population\implies agentSugar'(x)=agentSugar(x)-\\
 \t1\#\{d:\seq BIT| d\in diseases(a)\land \lnot subseq(d,agentImmunity(a)\}\hfill(1)\\
\end{schema}

Although not stated in the rule definition careful reading of the accompanying text \cite{Epstein1} shows that there is a penalty that is applied to each agent carrying diseases that it has no immunity to. The text states that for every disease carried by an agent that it has no immunity to, sugar metabolism is increased by one.  So if an agent carried two diseases that it has no immunity to then its metabolism rate increases by two.  This extra cost can equally be deducted by the metabolism rule or the disease rule. Purely for the sake of narrative it is placed in the \emph{ImmuneResponse} rule where it is first referenced in the original Sugarscape book. This is implemented by the final two lines (1) of the \emph{ImmuneResponse} schema.

The transmission of diseases is the more complex part of this rule. We will again use a recursive helper function $newDiseases$ to construct a set of diseases that an agent can catch from its neighbours. It takes the set of neighbours and their current diseases as input and constructs a set of diseases where one disease is chosen from each neighbour. 

\begin{axdef}
newDiseases: \seq AGENT\cross (AGENT\pfun \power(\seq BIT)) \fun \power\seq BIT
\where
\forall a:AGENT;tail:\seq AGENT; diseases:AGENT\pfun \power(\seq BIT)\spot\\
newDiseases(\langle\rangle,diseases)=\emptyset\\
newDiseases(\langle a\rangle\cat tail,diseases)=\\
\t1\IF diseases(a)=\emptyset \THEN\\
\t2newDiseases(tail,diseases)\\
\t1\ELSE\\
\t2\exists d:\seq BIT|d\in diseases(a)\spot\{d\}\cup newDiseases(tail,diseases))\\
\end{axdef} 

\begin{schema}{Transmission}
\Delta Agents\\
 \where
 loanBook'=loanBook\\
 population'=population\\
 sex'=sex\\
 position'=position\\
 vision'=vision\\
 age'=age\\
 maxAge'=maxAge\\
 agentCulture'=agentCulture\\
 agentImmunity'=agentImmunity\\
 children'=children\\
 agentSugar'=agentSugar\\
 metabolism'=metabolism\\
 initialSugar'=initialSugar\\
\forall a: AGENT\spot a\in population\implies\\
\t1diseases'(a)=diseases(a)\cup\hfill(1)\\
\t2 newDiseases(asSeq(visibleAgents(a,position,1)),diseases)
\end{schema}

 \begin{enumerate}
\item $visibleAgents$ returns the set of neighbours of an agent and this set is then passed to the $newDisease$ function which returns a set of diseases, one chosen from each agent in the neighbour set.
\end{enumerate}

\subsection{Rule Application Sequence}
Each tick of simulation time consists of the application of a set sequence of rules.  Not all rules can be used together so we identify the allowable sequences of rules.  We note that it is not stated in the book what order the rules are to be applied.  In the absence of this information we will pick one ordering and restrict ourselves to this ordering.

The display the different allowable combinations of rules in any given simulation we use the following terminology.
\begin{description}
\item[ $\{ Rule \}$] The indicates that $Rule$ is optional.  We can choose to include it or not in a simulation;
\item[ $RuleA|RuleB$] This indicates that there is a choice of which rule to apply - either one or the other but not both.
\end{description}
This rule ordering is for simulations using only a single resource and so omits the $Trade$ rule.

\begin{zed}
Tick\\
[\semi Growback|\semi SeasonalGrowback]\\
[\semi Movement_{basic}|(\semi Movement_{pollution}\semi PollutionDiffusion)|\semi Combat]\\
\{\semi\ Inheritance\}\{\semi Death\{[\semi Replacement|\semi AgentMating]\}\}\\
\{\semi Culture\}\{\semi PayLoans\semi MakeLoans\}\\
\{\semi Transmission\semi ImmuneResponse\}
\end{zed}

\section{Asynchronous Sugarscape Specification}
 AU is the sequential application of rules to agents during a simulation step.  If, for example, all agents move during a single step then a sequential ordering is imposed on all of the agents and they will move one at a time (that is, sequentially) based on that ordering.  This is in contrast to SU where all agents will attempt to move simultaneously (concurrently).  AU is easier to implement that SU as it maps directly onto the current standard sequential programming practice.  AU requires no collision detection and resolution (as for example when two agents try to simultaneously move to the same location) because concurrency is excluded - only one agent can move at any one time.  It is well know that the AU and SU approaches can deliver different simulation results.
 
 Although AU and SU are both commonly used in CA based simulations, where agent interactions are simple in nature, AU is prevelent in ABM. This is due to the lack of any good SU algorithms that can handle the complex interactions such as $Movement$, $Combat$ or $Trade$ that can appear in ABM based simulations.  
 
 We have provided a specification of Sugarscape that assumes SU.  For the sake of completeness and to allow us to make comparisons between synchronous and asynchronous updating in Sugarscape we will now present an AU based specification of the rules of Sugarscape.
\subsection{Variants of Asynchronous Updating}
There are a number of varieties of AU.  These variations differ in how they sequentially order agents for updating.  The best known variations are \cite{Schonfisch1999123}:
\begin{description}
\item[ Fixed Direction Line-By-Line] The locations in the lattice representing the simulation space are updated in the order they appear in the lattice (usually left to right, top-down);

\item[ Fixed Random Sweep] The order that is used is determined randomly at the start of the simulation and this order is used for every step in the simulation;

\item[ Random New Sweep] The order that the agents are updated in is determined randomly at the start of each step (each step uses a different order);

\item[ Uniform Choice] Each agent has an equal probability of being chosen.  If there are $n$ agents, then $n$ agents are chosen randomly during a step.  During any single step an agent may not be picked at all or may be picked more than once (in contrast Random New Sweep guarantees every agent is picked exactly once per step);

\item[ Exponential Waiting Time] This is a \emph{Time Driven} method, all the others are \emph{step driven}. Every agent has its own clock which rings when the agent is to be updated. The waiting times for the clock are exponentially distributed (with mean 1). The probability that an event occurs at time $t$ follows $e^{-t}$ where $t$ is a real number, $t\geq 0$. This is most similar to \emph{Uniform Choice}.
\end{description}

We will provide a specification for each variation in turn.

Fixed Direction Line-by-Line takes in a set of agents and their positions on the lattice.  It produces a sequence of agents where every agent appears once and only once in the sequence and the order of the sequence is determined by the agents position on the lattice.

\begin{axdef}
lineByLine:  AGENT\inj POSITION\\
  \fun seq AGENT
 \where
 \forall thePositions:AGENT\inj POSITION; theSequence:\seq AGENT\spot\\
lineByLine(theSet)=theSequence\\
 \t1\iff \ran theSequence=\dom thePositions \land \#theSequence=\#thePositions~~~\hfill(1)\\
 (n,a)\in theSequence \iff n=first(thePositions(a))*DIM+second(thePositions(a)) ~~~\hfill(2)
\end{axdef}

\begin{enumerate}
\item Each agent in the population appears in the sequence once and only once;
\item If one agent appears before another in the sequence then it also appears before that agent on the lattice.
\end{enumerate}

Fixed Random Sweep returns a sequence of the agents in some fixed random ordering.  This random ordering is chosen once at the start of the simulation and is fixed for the entire simulation run.

 \begin{axdef}
RANDOMORDER: \seq POSITION\\
\where
\#RANDOMORDER=\#POSITION\hfill(1)\\
\forall n,m:\nat \spot RANDOMORDER(n) =RANDOMORDER(m)~~~~\hfill(2)\\
\t1\iff n=m
\end{axdef}
\begin{enumerate}
\item $RANDOMORDER$ is a globally defined sequence containing an ordering of positions on the lattice;
\item Each position on the lattice appears once and only once in this sequence.
\end{enumerate}
Any ordering that satisfies these constrains is allowable according to our specification.  This introduces the randomness into the sequence.

 \begin{axdef}
fixedRandom:  AGENT  \inj POSITION\\
  \rel seq AGENT
 \where
 \forall thePositions:AGENT  \inj POSITION; theSequence:\seq AGENT\spot\\
 fixedRandom(thePositions)=theSequence\\
 \t1\iff \ran theSequence=\dom thePositions \land \#theSequence=\#thePositions~~\hfill(1)\\
 
  \forall i:0\upto \#theSequence-2; a_{1},a_{2}: AGENT\spot\\
  \t1 (i,a_{1})\in theSequence \land (i+1,a_{1})\in theSequence\implies\\
  \t1 (\exists x_{1},x_{2}:\nat | (x_{1},a_{1}) ,(x_{2},a_{2})\in RANDOMORDER \\
  \t2 \land x_{1}<x_{2}  \hfill(2)
\end{axdef}

\begin{enumerate}
\item Every agent in the population appears once and only once in the resulting sequence;
\item The ordering of agents in the sequence is based on the ordering defined in $RANDOMORDERING$.
\end{enumerate}

Random New sweep is simpler to specify.  We return a random ordering of agents after each call. We only need to ensure that every agent appears in this sequence exactly once.

\begin{axdef}
rndNewSweep: AGENT  \inj POSITION\\
 \rel seq AGENT
 \where
 \forall thePositions:AGENT  \inj POSITION; theSequence:\seq AGENT\spot\\
 rndNewSweep(thePositions)=theSequence\\
 \t1\iff \ran theSequence=\dom thePositions \land \#theSequence=\#thePositions\hfill(1)
\end{axdef}

 \begin{enumerate}
\item Every agent in the population appears once and only once in the resulting sequence;
\end{enumerate}

 Uniform Choice allows for an agent to be picked multiple times. The only constraints are that the sequence returned contains only agents in the population and that the size of the sequence equals the number of agents.
 
\begin{axdef}
uniformChoice: AGENT  \inj POSITION\\
 \rel \seq AGENT
 \where
 \forall thePositions:AGENT  \inj POSITION; n:\nat; theSequence:\seq AGENT| 0\leq n< \#theSequence \spot\\
 uniformChoice(thePositions)=theSequence\iff\\
 \t1theSequence(n)\in \dom thePositions\hfill(1)\\ 
 \t1\land \#theSequence=\#thePositions\hfill(2)
 \end{axdef}
 
 \begin{enumerate}
\item Every agent in the sequence is an agent from the simulation population;
\item the size of the sequence equals the total number of individual agents in the population.
\end{enumerate}

Each variation of asynchronous updating can now be covered by the simple matter of swapping in the appropriate ordering function within the specifications.
\subsection{Growback, Seasonal Growback and Replacement}
$Replacement, Growback$ and $Seasonal Growback$ belong to the category of rules we term $independent$.  This category includes all rules where the agent involved in the update (or rule execution) does not interact with any other agent - the update result is independent of any outside factor.  It follows then that the order in which these rules are executed will have no bearing on their outcome.  Given this we need make no changes to any of these rules.
\subsection{Pollution Diffusion}
$Pollution Diffusion$ is defined specifically as a synchronous rule.  There is no asynchronous alternative to this rule as imposing AU would redefine the rule entirely.  For this reason we do not produce a AU specification of this rule.

\subsection{Movement}

The specification of rules under an AU regime follows a standard pattern. First we impose an ordering on all the agents subject to the rule and then we recursively apply the update to each agent in the defined order.  Each individual agent update can affect the global state  and these changes must be passed forward to the next sequence of agent updates.  This is in contrast to SU where all updates occur simultaneously.

We always define the application of the rule to agents in a sequence recursively.  While the rules themselves can be quite simple the $Z$ notation forces us to pass to each update all parts of the global state that can be changed. This can result in large function signatures.

\begin{schema}{AsyncMovement_{basic}}
\Delta SugarScape\\
 \where
 step'=step\\
 loc'=loc\\
maxSugar'=maxSugar\\
 pollution'=pollution\\
 sex'=sex\\
 vision'=vision\\
 age'=age\\
 maxAge'=maxAge\\
 agentCulture'=agentCulture\\
 loanBook'=loanBook\\
 diseases'=diseases\\
 agentImmunity'=agentImmunity\\
 children'=children\\
 metabolism'=metabolism\\
 population'=population\\
 initialSugar'=initialSugar\\
 (sugar',agentSugar',position')=\\
 \t1applyMove(rndNewSweep(position),vision,sugar,agentSugar,position)
 \end{schema}
 Movement is a typical example of this structure. The main specification $AsyncMovement_{basic}$ simply passes the relevant state information alongside the ordering of agents (according to whatever AU variant we are using) to the recursive function $applyMove$. This recursive function applies the move rule to each agent in turn and returns the final updated agent position, agent sugar levels and lattice sugar levels.
  
 \begin{axdef}
applyMove: \seq AGENT \\
\cross AGENT\pfun\nat\\
 \cross (POSITION\pinj\nat)\\
  \cross AGENT\pinj\nat\\
\cross AGENT\pinj POSITION \\
\t1\rel\\
 ((POSITION\pinj\nat) \\
\cross AGENT\pinj\nat\\
 \cross AGENT\pinj POSITION)

\where

\forall head: AGENT; tail:\seq AGENT; population:\power AGENT;\\
positions:AGENT\pinj POSITION; sugar: POSITION\pinj\nat;\\
agentSugar: AGENT\pinj\nat;vision:AGENT\pfun\nat; \spot\\

applyMove(\langle\rangle,vision,sugar,agentSugar,positions) =\hfill(1)\\
\t1(sugar,agentSugar,positions)\\
~\\
applyMove(\langle head\rangle\cat tail,vision,sugar,agentSugar,positions) =\hfill(2)\\
\exists newLoc:POSITION | newLoc\in neighbourhood(position(head),vision(head)) \hfill(3)\\
\land \forall otherLoc:POSITION | otherLoc\in neighbourhood(position(head),vision(head))\\
\t1  \implies sugar(otherLoc)\leq sugar(newLoc) \spot\\
\t1 applyMove(tail,vision,sugar\oplus \{newLoc \mapsto 0\},\\
\t2agentSugar\oplus \{head \mapsto agentSugar(head)+sugar(newLoc)\},\\
\t2positions\oplus \{head \mapsto newLoc\})
 \end{axdef}
 
 \begin{enumerate}
\item The base case. If there are no agents left to update then we simply return the current state;
\item The recursive case.  If we have agents left to process then we move the first agent in the list and apply the rule to the remaining agents;
\item Find the best location for the agent to move to based on sugar levels at each location.
\end{enumerate}

\subsection{Pollution Diffusion}
The movement rule for pollution is almost identical to the simpler basic movement rule.  It only differs in that it takes pollution into account when selecting the best new position for an agent to move to.
\begin{schema}{AsyncMovement_{pollution}}
\Delta SugarScape\\
 \where
 step'=step\\
 loc'=loc\\
 maxSugar'=maxSugar\\
 sex'=sex\\
 pollution'=pollution\\
 vision'=vision\\
 age'=age\\
 maxAge'=maxAge\\
 agentCulture'=agentCulture\\
 loanBook'=loanBook\\
 children'=children\\
 agentImmunity'=agentImmunity\\
 diseases'=diseases\\
 metabolism'=metabolism\\
  population'=population\\
  (sugar',agentSugar',position')=\hfill(1)\\
  \t1applyMove_{pollution}(rndNewSweep(position),vision,sugar,\\
  \t5agentSugar,position,pollution)
  ~\\
pollution'=pollution\oplus\{\forall l:POSITION; x:AGENT | position'(x)=l\spot\hfill(2)\\
\t1  l\mapsto (PRODUCTION*sugar(l)+CONSUMPTION*metabolism(x) ) \}\\
  \end{schema}
  
 \begin{enumerate}
\item Call the recursive $applyMove_{pollution}$ to apply movement rule to each agent in turn;
\item Update location pollution levels based on agent movement.
\end{enumerate}  
  
\begin{axdef}
applyMove_{pollution}: \seq AGENT \\
\cross AGENT\pfun\nat \\
\cross (POSITION\pinj\nat)\\
\cross AGENT\pinj\nat\\
 \cross AGENT\pinj POSITION\\
  \cross POSITION\pinj\nat\\
\t1\rel\\
 ( (POSITION\pinj\nat)\\
  \cross AGENT\pinj\nat\\
   \cross AGENT\pinj POSITION)

\where

\forall head: AGENT; tail:\seq AGENT; population:\power AGENT;\\
positions:AGENT\pinj POSITION; sugar: POSITION\pinj\nat;\\
agentSugar: AGENT\pinj\nat;\\
 vision:AGENT\pfun\nat; pollution: POSITION\pinj\nat \spot\\

applyMove_{pollution}(\langle\rangle,vision,sugar,agentSugar,positions,pollution) =\\
\t1(sugar,agentSugar,positions)\\
~\\
applyMove_{pollution}(\langle head\rangle\cat tail,vision,sugar,agentSugar,positions,pollution) =\\
\exists newLoc:POSITION | newLoc\in neighbourhood(position(head),vision(head)) \\
\land \forall otherLoc:POSITION | otherLoc\in neighbourhood(position(head),vision(head))\\
\t1  \implies sugar(otherLoc)\div(1+pollution(otherLoc)) \leq \\
\t2sugar(newLoc)\div(1+pollution(position'(newLoc)))\spot\\
\t1 applyMove_{pollution}(tail,vision,sugar\oplus \{newLoc \mapsto 0\},\\
\t2agentSugar\oplus \{head \mapsto agentSugar(head)+sugar(newLoc)\},\\
\t2positions\oplus \{head \mapsto newLoc\},pollution)
 \end{axdef}

\subsection{Combat}
 Asynchronous Combat is undertaken with the $applyAllCombat$ function which applies the combat rule to each agent in a random order using the $singleFight$ function.  We note in passing that the synchronous specification seems to us to be simpler than the asynchronous one (even if the implementation is not).

\begin{schema}{Combat_{async}}
\Delta SugarScape\\
 \where
 step'=step\\
\\
 maxSugar'= maxSugar\\
 pollution'=pollution\\
 loanBook'= population'\dres loanBook\rres (population'\dres (\ran loanBook))\\
~\\
  \forall ag:AGENT; l:POSITION  \spot\\
  ag\in population' \implies\\
  \t1(sex'(ag)=sex(ag)\\
  \t1\land vision'(ag)=vision(ag)\\
  \t1\land age'(ag)=age(ag)\\
  \t1\land maxAge'(ag)=maxAge(ag)\\
  \t1\land children'(ag)=children(ag)\\
  \t1\land agentCulture'(ag)=agentCulture(ag)\\
  \t1\land agentImmunity'(ag)=agentImmunity(ag)\\
  \t1\land metabolism'(ag)=metabolism(ag)\\
  \t1\land diseases'(ag)=diseases(ag))\\
  \t1\land initialSugar'(ag)=initialSugar(ag)\\
   (population',position',sugar',agentSugar')=\\
   \t1applyAllCombat(rndNewSweep(position),population,position,sugar,\\
   \t4agentSugar,vision,agentCulture)\\
\end{schema}

\begin{axdef}
applyAllCombat: \seq AGENT\\
\cross \power AGENT\\
\cross (AGENT\pinj POSITION)\\
\cross (POSITION\pinj\nat)\\
\cross(AGENT\pinj\nat)\\
\cross (AGENT\pfun\nat)\\
\cross (AGENT\pfun \seq BIT)\\
 \fun\\
 (\power AGENT\\
 \cross (AGENT\pinj POSITION)\\
 \cross (POSITION\pinj\nat)\\
 \cross(AGENT\pinj\nat))\\
\where
\forall head: AGENT; tail:\seq AGENT; population:\power AGENT;\\
positions:AGENT\pinj POSITION; sugar: POSITION\pinj\nat; agentSugar: AGENT\pinj\nat;\\
 vision:AGENT\pfun\nat; culture: AGENT\pfun \seq BIT\spot\\

applyAllCombat(\langle\rangle,population,positions,sugar,agentSugar,vision,culture) =\\
\t1(population,positions,sugar,agentSugar,vision,culture)\\
~\\
applyAllCombat(\langle head\rangle\cat tail,population,positions,sugar,agentSugar,vision,culture) =\\

\IF  (head\in population) \THEN\\
\t1 applyAllCombat(tail,\\
\t2 singleFight(head,population,positions,sugar,agentSugar,vision,culture))\\
 \ELSE\\
\t1 applyAllCombat(tail,population,positions,sugar,agentSugar,vision,culture)
\end{axdef}

\begin{axdef}
singleFight:AGENT\cross \\
\power AGENT\\
\cross (AGENT\pinj POSITION)\\
\cross (POSITION\pinj\nat)\\
\cross(AGENT\pinj\nat)\\
\cross (AGENT\pfun\nat)\\
 \cross (AGENT\pfun \seq BIT)\\
 \fun\\
 (\power AGENT\\
 \cross (AGENT\pinj POSITION)\\
 \cross (POSITION\pinj\nat)\\
 \cross(AGENT\pinj\nat))\\
\where
\forall agent: AGENT; population:\power AGENT; positions:AGENT\pinj POSITION;\\
sugar: POSITION\pinj\nat agentSugar: AGENT\pinj\nat; vision:AGENT\pfun\nat;\\
 culture: AGENT\pfun \seq BIT\spot\\
singleFight(agent,population,positions,sugar,agentSugar,vision,culture)=\\
\IF~(availMoves(agent,positions,sugar,agentSugar,culture,vision(agent))=\emptyset)~\THEN\\
\t1 (population,positions,sugar,agentSugar)\\
\ELSE\\
\t1\exists loc:POSITION; available:\power POSITION;\\
\t1\forall otherLoc:POSITION|\\
\t1 loc,otherLoc\in availMoves(agent,positions,sugar,agentSugar,culture,vision(agent))\\
 \t1 \land otherLocation\neq location\spot\\
 \t2 reward(loc,sugar,position,agentSugar,COMBATLIMIT)\\
 \t2 \geq reward(otherLoc,sugar,position,agentSugar,COMBATLIMIT)\\
 \t1(distance(position(agent),loc) > distance(position(otherLoc),position'(agent))\implies\\
 \t2reward(loc,sugar,position,agentSugar,COMBATLIMIT)\\
 \t2 > reward(otherLoc,sugar,position,agentSugar,COMBATLIMIT))\spot\\
\t1(population\setminus\{positions\inv(loc)\},(positions\nrres \{loc\})\oplus\{agent\mapsto loc\},\\
\t1sugar\oplus\{loc\mapsto 0\},agentSugar\oplus\{agent\mapsto agentSugar(agent)+\\
\t2reward(position'(agent),sugar,position,agentSugar,COMBATLIMIT)\})\\
\end{axdef}

\subsection{Disease}
Disease is a simple rule that follows the standard pattern for AU specification. We place all agents into a sequence, ordered according to the variation of AU we are using, and apply the rule to each agent in turn updating the state as we go along.

\begin{schema}{Transmission}
\Delta Agents\\
 \where
 loanBook'=loanBook\\
 population'=population\\
 sex'=sex\\
 position'=position\\
 vision'=vision\\
 age'=age\\
 maxAge'=maxAge\\
 agentCulture'=agentCulture\\
 agentImmunity'=agentImmunity\\
 children'=children\\
 agentSugar'=agentSugar\\
 metabolism'=metabolism\\
 initialSugar'=initialSugar\\
diseases'=\\
\t1 applyTransmission(rndNewSweep(position),diseases,position)\hfill(1)
\end{schema}

\begin{enumerate}
\item Call recursive $applyTransmission$ on each agent in population in determined order.
\end{enumerate}

 \begin{axdef}
applyTransmission:\seq AGENT\\
 \cross AGENT\pfun\power \seq BIT\\
  \cross AGENT  \inj POSITION\\
\rel\\
 AGENT\pfun\power \seq BIT
\where
\forall head: AGENT; tail:\seq AGENT; diseases:AGENT\pfun\power \seq BIT;\\
 position:AGENT  \inj POSITION\\
\exists newInfections:\power\seq BIT | newInfections\hfill(1)\\
\t1=newDiseases(asSeq(visibleAgents(head,position,1)),diseases)\spot\\

applyTransmission(\langle\rangle,diseases,position) =\hfill(2)\\
\t1diseases\\

applyTransmission(\langle head\rangle\cat tail,diseases,position)=\hfill(3)\\
\t1applyTransmission(tail,diseases\oplus\{head\mapsto (diseases(head)\cup newInfections)\}\\
\t2,position)
\end{axdef}

\begin{enumerate}
\item Construct a set of new infections for an agent using the previously defined $newDiseases$;
\item Base case: Noting to do, return new disease mapping;
\item Recursive case: Add new diseases to the first agent in the list (according to the rule definition) and then recursively apply the rule to the rest of the list.
\end{enumerate}

\subsection{Culture}

$Culture$ is specified in an identical manner to $Disease$.
 \begin{schema}{AsyncCulture}
\Delta Agents\\
 \where
 population'=population\\
 sex'=sex\\
 position'=position\\
 vision'=vision\\
 age'=age\\
 maxAge'=maxAge\\
 agentSugar'=agentSugar\\
 children'=children\\
 loanBook'=loanBook\\
 diseases'=diseases\\
 metabolism'=metabolism\\
 agentImmunity'=agentImmunity\\
  initialSugar'=initialSugar\\
 agentCulture'=applyCulture(rndNewSweep(position),agentCulture,position)
\end{schema}
   \begin{enumerate}
\item Call recursive $applyCulture$ on each agent in population in determined order.
\end{enumerate}
 
 \begin{axdef}
applyCulture:\seq AGENT\\
 \cross AGENT \pfun\seq BIT\\
  \cross AGENT  \inj POSITION\\
\rel\\
 AGENT \pfun\seq BIT
\where
\forall head: AGENT; tail:\seq AGENT; culture:AGENT \pfun\seq BIT;\\
 position:AGENT  \inj POSITION;\\
 \exists  n:\power AGENT|n=neighbours(head,position(head),1)\spot\\
applyCulture(\langle\rangle,culture,position) =\hfill(1)\\
\t1culture\\
applyCulture(\langle head\rangle\cat tail,culture,position)=\hfill(2)\\
\t1applyCulture(tail,\\
\t1culture\oplus\{head\mapsto flipTags(culture(head), asSeq(n),culture)\}\\
\t1,position)
\end{axdef}
\begin{enumerate}
\item Base case: return new values for culture tags;
\item Recursive case: Flip the tags of the first agent in the list and repeat (recursively) for the remaining agents in the list (sequence). 
\end{enumerate}

\subsection{Inheritance}
  
$Inheritance$ also follows the same pattern as $Culture$ and $Disease$.

\begin{schema}{AsyncInheritance}
\Delta Agents\\
 \where
 population'=population\\
 sex'=sex\\
 position'=position\\
 vision'=vision\\
 age'=age\\
 maxAge'=maxAge\\
 agentCulture'=agentCulture\\
 children'=children\\
 metabolism'=metabolism\\
 diseases'=diseases\\
 agentImmunity'=agentImmunity\\
  initialSugar'=initialSugar\\

(loanBook',agentSugar') =\hfill(1)\\
\t1 applyInheritance(rndNewSweep(position),children,loanBook,agentSugar)\\
 \end{schema}
  \begin{enumerate}
\item Use recursive $applyInheritance$ function to calculate inheritance based on each agent in turn.
\end{enumerate}
 
   \begin{axdef}
applyInheritance:\seq AGENT \\
\cross AGENT \pfun\power AGENT \\
\cross (AGENT\rel (AGENT \cross (\nat \cross \nat))) \\
\cross AGENT \pfun \nat\\
\cross AGENT\pfun \nat\\
\cross AGENT\pfun \nat\\
\pfun \\
(AGENT\rel (AGENT \cross (\nat \cross \nat))) \\
\cross AGENT \pfun \nat
\where
\forall head: AGENT; tail:\seq AGENT; children:AGENT \pfun\power AGENT;\\
 loans:AGENT\rel (AGENT \cross (\nat \cross \nat));\\
agentSugar:AGENT \pfun \nat; age,maxAge:AGENT\pfun \nat\\
\exists newLoans:AGENT\rel (AGENT \cross (\nat \cross \nat));newAgentSugar:AGENT \pfun \nat\\
| newAgentSugar=agentSugar\oplus (\{head\mapsto 0\} \cup\hfill(1)\\
\t2\{(a,amt)|a\in children(head)\land amt=agentSugar(a)\\
\t3+agentSugar(head)/ \#children(head)\})\\
\land newLoans=(\{head\}\ndres loans)\cup oneAgentLoans(a, asSeq(\ran(\{a\}\dres loans)),\\
\t3Children(head))\spot\hfill(2)\\
applyInheritance(\langle\rangle,children,loans,agentSugar,age,maxAge) =\hfill(3)\\
\t1(loans,agentSugar)\\

applyInheritance(\langle head\rangle\cat tail,children,loans,agentSugar,age,maxAge)=\\
\IF(age(head)=maxAge(head) \lor agentSugar(head)=0) \THEN\hfill(4)\\
\t1applyInheritance(tail,children,newLoans,newAgentSugar,age,maxAge)\\
\ELSE\\
\t1applyInheritance(tail,children,loans,agentSugar,age,maxAge)\\
\end{axdef}

  \begin{enumerate}
\item Distribute the dying agents sugar equally amongst its children;
\item Distribute any loans where the dying agent is the lender equally amongst its children;
\item Base  case of recursion.  Nothing to do but return results;
\item Recursive case: If the first agent in the list is dying then handle that agents inheritance and recurse through the rest of the agents otherwise just ignore it and apply the rule to rest of agents.
\end{enumerate}

\subsection{Mating}
The AU specification of $Mating$ is simpler than the SU version as it does not have to construct conflict free sets. It just puts all of the potential pairs in a sequence ordered according to the variant of AU we are using and applies the rule to each in turn.
\begin{schema}{AsyncAgentMating}
\Xi Lattice\\
\Delta Agents\\
 \where
 loanBook'=loanBook\\
\exists potentialMatingPairs: \power (AGENT\cross AGENT)|\\
potentialMatingPairs=\{(a:AGENT,b:AGENT)| sex(a)\neq sex(b)\\
\t1 \land isFertile(age(a),sex(a)) \land isFertile(age(head),sex(head))\\
\t1\land adjacent(position(a),position(head))\}\\

(population',position',vision',agentSugar',agentCulture',\\
\t1metabolism',children',diseases',agentImmunity',age',sex',initialSugar')=\hfill(1)\\
applyMating(rndNewSweep(potentialMatingPairs),population,position,vision,\\
\t1agentSugar,agentCulture,metabolism,children,\\
\t1diseases,agentImmunity,age,maxAge, sex,initialSugar)\\
\end{schema}

\begin{enumerate}
\item Call $applyMating$ function on the agents in sequence.
\end{enumerate}

\subsection{Credit}

\begin{schema}{MakeLoans}
\Delta Agents\\
\Xi Step
 \where
 population'=population\\
 sex'=sex\\
 position'=position\\
 vision'=vision\\
 age'=age\\
 maxAge'=maxAge\\
 agentCulture'=agentCulture\\
 agentImmunity'=agentImmunity\\
 diseases'=diseases\\
 children'=children\\
 metabolism'=metabolism\\
  initialSugar'=initialSugar\\
 (loanBook',agentSugar')=\hfill(1)\\
 \t1applyLoans(rndNewSweep(position),population,position,agentSugar,age,sex,loanBook,step)\\
 \end{schema}
 
 \begin{enumerate}
\item Call $applyLoans$ on each agent in turn.
\end{enumerate}

 \begin{axdef}
applyLoans:\seq AGENT\\
 \cross \power AGENT\\
 \cross  AGENT  \inj POSITION\\
\cross AGENT \pfun \nat\\
 \cross AGENT \pfun \nat\\
 \cross AGENT \pfun SEX\\
\cross (AGENT\rel(AGENT\cross(\nat \cross\nat)))\\
\cross \nat\\
\fun\\
 (AGENT\rel(AGENT\cross(\nat \cross\nat)))\\
  \cross AGENT \pfun \nat

\where
\forall population : \power AGENT; \\
position : AGENT  \inj POSITION;\\
sex: AGENT \pfun SEX;\\
age: AGENT \pfun \nat;\\
agentSugar: AGENT \pfun \nat;\\
head,ag : AGENT;\\
loanBook,loans:(AGENT\rel(AGENT\cross(\nat \cross\nat)));\\
tail :\seq AGENT;\\
step:\nat;\\
\exists newAgentSugar: AGENT \pfun \nat;newLoans:(AGENT\rel(AGENT\cross(\nat \cross\nat)));\\
neighbours:\power AGENT |\\
\t1 neighbours=\{b:AGENT | vonNeumanNeighbour(lender,b,position)\}\hfill(1)\\
\t1 (newLoans,newAgentSugar)=\\
\t3singleLenderLoans(lender,asSeq(neighbours),\\
\t7agentSugar,age,sex,loanBook,step)\\
applyLoans(\langle\rangle, population,position,agentSugar,age,sex,loanBook,step) =\hfill(2)\\
\t1(loanBook,agentSugar)\\
applyLoans(\langle lender \rangle\cat tail, population,position,\\
\t4agentSugar,age,sex,loanBook,step) =\hfill(3)\\
\t1 applyLoans(tail, population,position,newAgentSugar,age,sex,newLoans,step)\\
\end{axdef}

\begin{enumerate}
\item Construct the set of neighbours of an agent $lender$, the updated loan book and the updated sugar levels gotten by the $lender$ giving loans to its neighbours;
\item Base Case: Nothing to do just return existing values;
\item Recursive case: Recursively call $applyLoans$ on the remainder of the agents (excluding the first agent $lender$) and the new loan and sugar levels gotten by $lender$ generating new loans.
\end{enumerate}

  \begin{axdef}
singleLenderLoans:AGENT\\
\cross \seq AGENT\\
 \cross \power AGENT\\
\cross AGENT \pfun \nat\\
 \cross AGENT \pfun \nat\\
 \cross AGENT \pfun SEX\\
\cross (AGENT\rel(AGENT\cross(\nat \cross\nat)))\\
\cross \nat\\
\fun\\
 (AGENT\rel(AGENT\cross(\nat \cross\nat)))\\
  \cross AGENT \pfun \nat

\where
\forall sex: AGENT \pfun SEX;\\
age: AGENT \pfun \nat;\\
agentSugar: AGENT \pfun \nat;\\
head,lender : AGENT;\\
loanBook:(AGENT\rel(AGENT\cross(\nat \cross\nat)));\\
tail :\seq AGENT;\\
step:\nat;\\

singleLenderLoans(lender,\langle\rangle,agentSugar,age,sex,loanBook,step) =\\
\t1(loanBook,agentSugar)\\

singleLenderLoans(lender,\langle head \rangle\cat tail, agentSugar,age,sex,loans,step) =\\
\IF canLend(age(lender),sex(lender),agentSugar(lender))\\
\t1\land willBorrow(age(head),sex(head),agentSugar(head),\{head\}\dres \ran (loanBook)))\\
\THEN\\
\t1\exists newAgentSugar: AGENT \pfun \nat;\\
\t1newLoans:(AGENT\cross(AGENT\cross(\nat \cross\nat))); amt:\nat |\\
\t2 amt= min(amtAvail(age(lender),sex(lender),agentSugar(lender)),\\
\t3amtReq(agentSugar(head)))\\
\t2 newAgentSugar=agentSugar\oplus\{lender\mapsto agentSugar(lender)-amt,\\
\t3head\mapsto agentSugar(head)+amt\}\\
\t2 newLoans=loanBook\cup\{(lender,(head,(amt,step+DURATION)\} \\
\t1 singleLenderLoans(lender,tail,newAgentSugar,age,sex,newLoans,step)\\
\ELSE\\
\t1 singleLenderLoans(lender,tail, agentSugar,age,sex,loanBook,step)
\end{axdef}
$singleLenderLoans$ calculates all loans that a particular agent can give to its neighbours.

\begin{enumerate}
\item If there are no loans in the sequence then just return the current loans and sugar levels as is;
\item If the loan sequence is not empty then apply the payment details to the first loan and make the payments on the rest:
\begin{enumerate}[a)]
\item If the first loan is capable of being paid by the borrower we simply move the correct amount of sugar from the borrower to the lender;
\item If the borrower cannot pay off the loan then they pay back half their sugar and the loan is renegotiated for the remainder.
\end{enumerate}

\end{enumerate}
Using these functions we can now specify the $PayLoans$ part of the $Credit$ rule.

\begin{schema}{PayLoans}
\Delta Agents\\
\Xi Step
 \where
 population'=population\\
 sex'=sex\\
 position'=position\\
 vision'=vision\\
 age'=age\\
 maxAge'=maxAge\\
 agentCulture'=agentCulture\\
 agentImmunity'=agentImmunity\\
 children'=children\\
 diseases'=diseases\\
 metabolism'=metabolism\\
  initialSugar'=initialSugar\\
 
\exists dueLoans,newLoans:\power(AGENT\cross(AGENT\cross(\nat\cross\nat)))\spot\\ 
 dueLoans= loanBook\rres (\ran(loanBook)\rres\{a:(\nat\cross\nat)| second(a)=step\})\hfill(1)\\
(newLoans,agentSugar')=makePayments(asSeq(dueLoans),\emptyset,agentSugar)\hfill(2)\\\\
loanBook'=(loanBook\setminus dueLoans)\cup newLoans\hfill(3)\\
\end{schema}

 \begin{enumerate}
\item We create the set of due loans;
\item We now create the set of renegotiated loans and update the agentSugar levels using the $makePayments$ function;
\item Finally we update the loan book by removing all loans that were due and adding any new renegotiated loans.
\end{enumerate}

\section{Added Spice}
\subsection{Introduction}
We have defined all the rules so far under the assumption that there is only one resource (known as $sugar$). The final rule, $Trade$, is only defined for simulations with at least two resources. In fact the rules are meant to be general enough that they will work with any number of resources although we know of no sugarscape based simulation that used more than two resources.  The second resource is known as $spice$.

We will show how to extend the rules to deal with two resources. In order to avoid unnecessary clutter and make the differences as clear as possible we will show the differences between the one and two resource schemas with $\mathbf{bold face}$. Any part of a schema that is not an exact copy of the previously defined version will be in $\mathbf{bold face}$.

\subsection{Basic Types}
The basic types are copies of those already defined for $sugar$.
\begin{axdef}
MAXSPICEMETABOLISM:\nat\hfill(1)\\
SPICEGROWTH:\nat\hfill(2)\\
MAXSPICE:\nat\hfill(3)\\
INITIALSPICEMIN,INITIALSPICEMAX:\nat\hfill(4)\\
SPICEPRODUCTION, SPICECONSUMPTION:\nat\hfill(5)\\
SPICECOMBATLIMIT:\nat\hfill(6)\\
SPICECHILDAMT:\nat\hfill(7)\\
\end{axdef}

\begin{enumerate}
\item Agents metabolise spice during each move at an individually set rate less than $MAXSPICEMETABOLISM$;
\item Spice grows back at a predefined rate;
\item Each location can hold a set maximum amount of spice;
\item Agents created after mating start with an initial spice endowment;
\item Pollution can be caused by production and consumption of spice;
\item $SPICECOMBATLIMIT$ is required to help determine the reward from attacking an agent using the combat rule.
\item We posit that a minimum amount of spice is needed for agent mating to occur.
\end{enumerate}

Note that these constants are replicas of their $sugar$ counterparts.

\subsection{The SpiceScape}
The spice grid contains everything in the $Lattice$ scheme and just adds information on the extra $spice$ resource. 
\begin{schema}{SpiceLattice}
Lattice\\
\mathbf{spice:POSITION\pfun\nat}\\
\mathbf{maxSpice:POSITION\pfun\nat}\\
\where
\mathbf{\dom spice=\dom maxSpice=POSITION}\hfill(1)\\
\mathbf{\forall x:POSITION\spot spice(x)\leq maxSpice(x)\leq MAXSPICE}\hfill(2)\\
\end{schema}
\begin{enumerate}
\item Every location has an associated amount of spice and maximum carrying capacity;
\item Every position's spice levels are within the acceptable levels.
\end{enumerate}

\subsection{Agents}

\begin{schema}{SpiceAgents}
Agents\\
\mathbf{agentSpice:AGENT\pfun\nat}\\
\mathbf{initialSpice:AGENT\pfun\nat}\\
\mathbf{spiceMetabolism:AGENT\pfun\nat}\\
\mathbf{spiceLoanBook: AGENT\rel(AGENT\cross(\nat\cross\nat))}\\
\where
\mathbf{\dom spiceMetabolism=\dom agentSpice==\dom initialtSpice population}\hfill(1)\\
\mathbf{\dom spiceLoanBook\subseteq population}\hfill(2)\\
\mathbf{\dom (\ran spiceLoanBook)\subseteq population}\\
\mathbf{\forall x:AGENT\spot x\in population\implies}\\
\mathbf{\t1 spiceMetabolism(x)\leq MAXSPICEMETABOLISM}\hfill(3)\\

\end{schema}
\begin{enumerate}
\item Every agent has a spice metabolism and a spice store;
\item The spiceLoanBook has the same invariants as the original loanBook;
\item Every agents metabolism is less than or equal to the defined maximum.

\end{enumerate}

Finally we combine them into an overall schema as before:
 \begin{schema}{SpiceScape}
 SpiceAgents\\
 SpiceLattice\\
 Step \\
 \end{schema}
 
 The initialisation scheme and tick schemas are also largely unchanged. 
  \begin{schema}{InitialSpiceScape}
 SpiceAgents~'\\
 SpiceLattice~'\\
 Step~' 
 \where
 step'=0\\
  \# population'=INITIALPOPULATIONSIZE\\
  loanBook'=\emptyset\\
   \mathbf{spiceLoanBook'=\emptyset}\\
 \forall a: AGENT\spot\\
 \t1a\in population'\implies\\
 \t2age'(a)=0\\
 \t2diseases'(a)=\emptyset\\
  \t2children'(a)=\emptyset\\
  \t2INITIALSUGARMIN\leq agentSugar'(a)\leq INITIALSUGARMAX\\
  initialSugar'(a)=agentSugar'(a)\\
\mathbf{\t2INITIALSPICEMIN\leq agentSpice'(a)\leq INITIALSPICEMAX}\\
\mathbf{initialSpice'(a)=agentSpice'(a)}\\

 \end{schema}

  \begin{schema}{Tick_{spice}}
 \Delta SpiceAgents\\
 \Delta Step\\
 \where
 population'=population\\
position'=position\\
sex'=sex\\
vision'=vision\\
maxAge'=maxAge\\
metabolism'=metabolism\\
 initialSugar'=initialSugar\\
 \mathbf{initialSpice'=initialSpice}\\
\mathbf{spiceMetabolism'=spiceMetabolism}\\
agentCulture'=agentCulture\\
children'=children\\
loanBook'=loanBook\\
\mathbf{spiceLoanBook'=spiceLoanBook}\\
agentImmunity'=agentImmunity\\
diseases'=diseases\\
 step'=step+1\\
 \forall x:AGENT\spot x\in population\implies\\
 \t1(age'(x)=age(x)+1\\
 \t1\land agentSugar'(x)=agentSugar(x)-metabolism(x)\\
\mathbf{\t1\land agentSpice'(x)=agentSpice(x)-spiceMetabolism(x))}\\
 \end{schema}
 
\subsection{Rules}
As well as defining the final rule, $Trade$, we will also expand the other rules to allow them to operate on a simulation with two resources. We define the new rule ($Trade$) first.

\subsection{Agent Trade $T$}
\begin{description}
\item[ Agent Trade $T$]~
\begin{itemize}
\item Agent and neighbour compute their MRSs; if these are equal then end, else continue;
\item The direction of exchange is as follows: spice flows from the agent with the higher MRS to the agent with the lower MRS while sugar goes in the opposite direction;
\item The geometric mean of the two MRSs is calculated-this will serve as the bargaining price, $p$;
\item The quantities to be exchanged are as follows: if p$>$1 the p units of spice for 1 unit of sugar; if p $<$ 1 the 1/p units of sugar for 1 unit of spice;
\item If this trade will (a) make both agents better off (increases the welfare of both agents), and (b) not cause the agents' MRSs to cross over one another, then the trade is made and return to start, else end.
\end{itemize}
\end{description}

MRS is calculated simply for an agent as the fraction obtained by dividing its spice level times its sugar metabolism by its spice metabolism times its sugar level, as set out below.
\begin{axdef}
  MRS:\nat\cross\nat\cross\nat\cross\nat \inj \arithmos
\where
\forall sugar,sugarMetabolism,spice,spiceMetabolism:\nat\spot\\
MRS(sugar,sugarMetabolism,spice,spiceMetabolism)=\\
\t1 (spice*sugarMetabolism)/(spiceMetabolism*sugar)
\end{axdef}

$tradePairs$ constructs a sequence of all possible trading partners based on the proximity of the agents to each other.
\begin{axdef}
  tradePairs: \seq AGENT\cross AGENT\pinj POSITION  \inj \seq (AGENT\cross AGENT)
\where
\forall tail: \seq AGENT; positions: AGENT\pinj POSITION; a:AGENT \spot\\
tradePairs(\langle \rangle,positions)=\langle\rangle\\
tradePairs(\langle a\rangle\cat tail,positions)=\\
\t1 trade(tail,positions\cat asSeq(\{b:agent | adjacent(position(a),position(b))\spot(a,b)\})
\end{axdef}

\begin{schema}{Trade}
\Delta SpiceAgents\\
\where
spiceLoanBook'=spiceLoanBook\\
loanBook'=loanBook\\ 
 population'=population\\
  initialSugar'=initialSugar\\
   \mathbf{initialSpice'=initialSpice}\\
 sex'=sex\\
 metabolism'=metabolism\\
 spiceMetabolism'=spiceMetabolism\\
 position'=position\\
 vision'=vision\\
 age'=age\\
 maxAge'=maxAge\\
 agentCulture'=agentCulture\\
 agentImmunity'=agentImmunity\\
 children'=children\\
 \exists allPairs:\seq (AGENT,AGENT)| allPairs=tradePairs(asSeq(population),position)\\
 
 (agentSugar',agentSpice') =\hfill(1)\\
 \t1 executeTrades(allTrades(chooseExclusiveTrades(allPairs),\\
 \t2(agentSugar,agentSpice),metabolism,spiceMetabolism))\\
\end{schema}

$Trade$ is similar to $Mating$ in that trading must be done in exclusive pairs.  An agent cannot carry out two simultaneous trades and the rule forces each agent to trade with all its neighbours in some sequence.  As with $Mating$ we construct conflict free sets of trading pairs that can proceed simultaneously and then order these sets.

\begin{axdef}
chooseExclusiveTrades:AGENT\cross AGENT \rel \seq \power(AGENT,AGENT)
\where
\forall a,b:AGENT; tradingPairs:AGENT\rel AGENT\spot\\
chooseExclusiveTrades(\emptyset) = \langle\rangle\\
chooseExclusiveTrades(tradingPairs)=\\
\t1\exists maxSet:AGENT\rel AGENT| maxSet\subseteq tradingPairs\\
\t1\land  ((a,b)\in tradingPairs \land (a,b)\notin maxSet) \iff\\
\t2 \exists c:AGENT | \{(a,c),(c,a),(b,c),(c,b)\}\cap maxSet\neq\emptyset)\\
\t1 \langle maxSet\rangle\cat chooseExclusiveTrades(tradingPairs\setminus maxSet)
\end{axdef}

\begin{axdef}
executeTrades:\seq (AGENT\cross AGENT) \cross AGENT \pfun \nat \cross AGENT \pfun \nat\\
\rel (AGENT \pfun \nat , AGENT \pfun \nat)
\where
\forall tail:\seq (AGENT\cross AGENT); head: (AGENT\cross AGENT);\\
agentSugar,agentSpice,metabolism,spiceMetabolism: AGENT \pfun \nat\spot\\
executeTrades(\langle\rangle,agentSugar,agentSpice,metabolism,spiceMetabolism)=\\
\t1(agentSugar,agentSpice)\\
executeTrades(\langle head\rangle\cat tail,agentSugar,sugar)=\\
\exists newAgentSugar,newAgentSpice:AGENT \pfun \nat |
(newAgentSugar,newAgentSpice)=\\
\t1allTrades(head,(agentSugar,agentSpice,metabolism,spiceMetabolism)\\
\spot executeTrades(tail,newAgentSugar,newAgentSpice,metabolism,spiceMetabolism)
\end{axdef}

$allTrades$ recursively goes through the sequence of trading partners and gets each trading pair to update the sugar and spice levels based on their trades. 
\begin{axdef}
  allTrades:\seq(AGENT\cross AGENT)\cross ((AGENT\pinj\nat)\cross (AGENT\pinj\nat))\cross\\
  \t2(AGENT\pinj\nat)\cross(AGENT\pinj\nat)\\
  \t1\rel ((AGENT\pinj\nat)\cross (AGENT\pinj\nat))
\where
\forall head:AGENT\cross AGENT;tail:\seq(AGENT\cross AGENT)\\
sugar,spice,sugarMetabolism,spiceMetabolism:AGENT\pinj\nat\spot\\
allTrades(\langle\rangle,(sugar,spice),sugarMetabolism,spiceMetabolism)=(sugar,spice)\\
allTrades(\langle head\rangle\cat tail,(sugar,spice),sugarMetabolism,spiceMetabolism)=\\
\t1allTrades(tail,pairTrade(head,(sugar,spice),sugarMetabolism,spiceMetabolism)\\
\t2,sugarMetabolism,spiceMetabolism)\\

\end{axdef}

Each trading partnership will execute a series of trades until their MRS scores cross over. $pairTrade$ is complicated by the fact that there are multiple options:
\begin{enumerate}
\item If their MRS scores are equal then they perform no trades;
\item If their MRS scores are not equal then the direction of trade will depend on which MRS score is higher;
\begin{enumerate}[a)]
\item Within a trade a value of $p$ (based on MRS scores) determines the price of the resources.
\end{enumerate}
\item Trades between the pair continue until their new and old MRS scores cross over.
\end{enumerate}

\begin{axdef}
  pairTrade: (AGENT\cross AGENT)\cross ((AGENT\pinj\nat)\cross (AGENT\pinj\nat))\cross\\
  \t2(AGENT\pinj\nat)\cross (AGENT\pinj\nat)\\
  \t1\rel ((AGENT\pinj\nat)\cross (AGENT\pinj\nat))
\where
\forall a,b:AGENT;sugar,spice:AGENT\pinj\nat;\\
metabolism,spiceMetabolism:AGENT\pinj\nat|\\
\exists p,mrsA,mrsB,newMrsA,newMrsB:\arithmos; newSugar,newSpice:AGENT\pinj\nat\\
 mrsA=MRS(sugar(a),metabolism(a),spice(a),spiceMetabolism(a))\\
 mrsB=MRS(sugar(b),metabolism(b),spice(b),spiceMetabolism(b))\\
 p= \sqrt{mrsA * mrsB}\\
(mrsA>mrsB \land p>1)\implies\\
\t1(newSugar=sugar\oplus \{(a,sugar(a)+1),(b,sugar(b)-1) \} \\    
\t1\land newSpice=spice\oplus\{a\mapsto spice(a)-p,b\mapsto spice(b)+p\})\\
(mrsA>mrsB \land p\leq 1)\implies\\
\t1(newSugar=sugar\oplus\{a\mapsto sugar(a)+(1\div p),b\mapsto sugar(b)-(1\div p)\}\\
\t1newSpice=spice\oplus\{a\mapsto spice(a)-1,b\mapsto spice(b)+1\})\\
(mrsA\leq mrsB \land p>1)\implies\\
\t1 (newSugar=sugar\oplus\{b\mapsto sugar(b)+1,a\mapsto sugar(a)-1\}\\
\t1 \land newSpice=spice\oplus\{b\mapsto spice(b)-p,a\mapsto spice(a)+p\})\\
(mrsA\leq mrsB \land p\leq 1)\implies\\
\t1 (newSugar=sugar\oplus\{b\mapsto sugar(b)+(1\div p),a\mapsto sugar(a)-(1\div p)\}\\
\t1 \land newSpice=spice\oplus\{b\mapsto spice(b)-1,a\mapsto spice(a)+1\})\\
 newMrsA=MRS(newSugar(a),metabolism(a),newSpice(a),spiceMetabolism(a))\\
 newMrsB=MRS(newSugar(b),metabolism(b),newSpice(b),spiceMetabolism(b))\\
\spot\\
pairTrade((a,b),(sugar,spice),metabolism,spiceMetabolism)=(sugar,spice)\\
\t1\iff mrsA=mrsB\\
pairTrade((a,b),(sugar,spice),metabolism,spiceMetabolism)=(newSugar,newSpice)\\
\t1\iff ((mrsA>mrsB\land newMrsA\leq newMrsB)\\
\t1\lor(mrsA<mrsB\land newMrsA\geq newMrsB))\\
pairTrade((a,b),(sugar,spice),metabolism,spiceMetabolism)=\\
\t1pairTrade((a,b),(newSugar,newSpice),metabolism,spiceMetabolism)\\
\t1\iff ((mrsA> mrsB\land newMrsA> newMrsB)\\
\t1\lor(mrsA< mrsB\land newMrsA< newMrsB))\\
\end{axdef}

\subsection{Asynchronous Trade}
 
\begin{schema}{Trade}
\Delta SpiceAgents\\
\where
spiceLoanBook'=spiceLoanBook\\
loanBook'=loanBook\\ 
 population'=population\\
 sex'=sex\\
 metabolism'=metabolism\\
 spiceMetabolism'=spiceMetabolism\\
  initialSugar'=initialSugar\\
   \mathbf{initialSpice'=initialSpice}\\
 position'=position\\
 vision'=vision\\
 age'=age\\
 maxAge'=maxAge\\
 agentCulture'=agentCulture\\
 agentImmunity'=agentImmunity\\
 children'=children\\
 
 \exists traders: \power(AGENT\cross AGENT)\spot\\
 (agentSugar',agentSpice') =\hfill(1)\\
 \t1 allTrades(tradePairs(rndNewSweep(position),positions),(agentSugar,agentSpice),\\
 \t2metabolism,spiceMetabolism)\\
\end{schema}

\begin{enumerate}
\item The new sugar and spice allocations are derived by conducting all possible trades using the recursive helper function $allTrades$.
\end{enumerate}

 \subsubsection{Growback}
\begin{schema}{Growback_{spice}}
 \Delta SpiceLattice\hfill(1)\\
 \where\\
 pollution'=pollution\\
 maxSugar'=maxSugar\\
 \mathbf{maxSpice'=maxSpice}\hfill(2)\\
 sugar'=sugar\oplus\{x:POSITION \spot\\
 \t1 x\mapsto min(\{sugar(x)+SUGARGROWTH,maxSugar(x)\})  \}\\
 \mathbf{spice'=spice\oplus\{x:POSITION \spot}\\
  \mathbf{\t1 x\mapsto min(\{spice(x)+SPICEGROWTH,maxSpice(x)\})  \}}\hfill(3)\\
 \end{schema} 
 
\begin{enumerate}
\item $Lattice$ is replaced with $SpiceLattice$. In all subsequent schemas $SpiceLattice$ replaces $Lattice$ and $SpiceAgent$ replaces $Agent$;
\item $maxSpice$ remains unchanged;
\item The new spice levels are calculated using the same simple formula used for sugar growback. 
\end{enumerate}

 \subsubsection{Seasonal Growback}

  \begin{schema}{SeasonalGrowback_{spice}}
\Delta SpiceLattice\\
\Xi Step\\
 \where
 pollution'=pollution\\
\\
 maxSugar'=maxSugar\\
 \mathbf{maxSpice'=maxSpice}\\
\forall x: POSITION \spot\\
(step\div SEASONLENGTH)\mod 2 = 0 \implies sugar'=\\
 \t1 \{ x: POSITION | first(x)<M\div 2 \spot\\
\t1 x\mapsto min(\{sugar(x)+SUGARGROWTH,maxSugar(x)\})\}\\
 \t1\cup \\
 \t1 \{x: POSITION | first(x)\geq M\div 2 \spot\\
 \t1 x\mapsto  min(\{sugar(x)+SUGARGROWTH\div WINTERRATE,maxSugar(x)\})\}\\
 
\forall x: POSITION \spot\\
  (step\div SEASONLENGTH)\mod 2 \neq 0 \implies sugar'=\\
 \t1\{x: POSITION | first(x)<M\div 2\spot\\
\t1 x\mapsto min(\{sugar(x)+SUGARGROWTH\div WINTERRATE,maxSugar(x)\}) \}\\
 \t1\cup\\
 \t1 \{x: POSITION; y:\nat | first(x)\geq M\div 2\spot\\
 \t1 x\mapsto  min(\{sugar(x)+SUGARGROWTH,maxSugar(x)\}) \}\\
 
 \mathbf{ \forall x: POSITION \spot}\hfill(1)\\
  \mathbf{(step\div SEASONLENGTH)\mod 2 = 0 \implies spice'=}\\
  \mathbf{\t1 \{ x: POSITION | first(x)<M\div 2 \spot}\\
 \mathbf{\t1 x \mapsto  min(\{spice(x)+SPICEGROWTH,maxSpice(x)\})\}}\\
  \mathbf{\t1\cup }\\
 \mathbf{ \t1 \{x: POSITION | first(x)\geq M\div 2\spot } \\
  \mathbf{\t1  x\mapsto  min(\{spice(x)+SPICEGROWTH\div WINTERRATE,maxSpice(x)\})\}}\\

 \mathbf{\forall x: POSITION \spot}\\
  \mathbf{ (step\div SEASONLENGTH)\mod 2 \neq 0 \implies spice'=}\\
 \mathbf{ \t1\{x: POSITION | first(x)<M\div 2\spot}\\
 \mathbf{\t1 x\mapsto min(\{spice(x)+SPICEGROWTH\div WINTERRATE,maxSpice(x)\})\}}\\
 \mathbf{ \t1\cup}\\
 \mathbf{ \t1 \{x: POSITION | first(x)\geq M\div 2\spot}\\
 \mathbf{ \t1 x\mapsto min(\{spice(x)+SPICEGROWTH,maxSpice(x)\})\}}\\
 \end{schema}

\begin{enumerate}
\item Seasonal growback  adds a rule for spice grow back that is an exact replica of the sugar rule.  We note that we only use the one $WINTERRATE$ instead of a separate rate for sugar and spice. In the absence of any explicit direction on this point this solution seems to be the most obvious.
\end{enumerate}

\subsection{Movement}
In order to update Movement we will need to implement a welfare function that can be used to measure the desirability of a location. With two resources the desirability of  any location becomes a subjective measure, what one agent may rate highly another may not.  Welfare is dependent on the agents current levels of spice and sugar, so an agent with low spice levels may consider a location containing spice more desirable than one containing sugar. Overall the desirability of a location is determined by the agents current resource levels (wealth) and the relative metabolism rates for each resource.

This is in contrast to the previous approach where welfare just equaled the amount of sugar in a location.  This welfare measure is precisely defined in the book and we follow this definition.

\begin{axdef}
  welfare: \nat\cross\nat\cross\nat\cross\nat\cross\nat\cross\nat \inj \arithmos
\where
\forall agentSugar,sugarMetabolism,\\
\t1agentSpice,spiceMetabolism,\\
\t1locationSugar,locationSpice:\nat\spot\\
welfare(agentSugar,sugarMetabolism,agentSpice,spiceMetabolism,\\
\t1locationSugar,locationSpice)=\\
\t1 (locationSugar+agentSugar)*(sugarMetabolism~\div\\
\t3(sugarMetabolism+spiceMetabolism))\\
\t1 * (locationSpice+agentSpice)* (spiceMetabolism\div\\
\t3(sugarMetabolism+spiceMetabolism))
\end{axdef}

Movement can now be restated by replacing the previous measure of a locations desirability (sugar level) with this new measure and the updating of spice levels.  In all other respects the schema remains unchanged.

\begin{schema}{Movement_{basicSpice}}
\Delta SpiceScape\\
 \where
 step'=step\\
population'=population\\
maxSugar'=maxSugar\\
 pollution'=pollution\\
 sex'=sex\\
 vision'=vision\\
 age'=age\\
  initialSugar'=initialSugar\\
   \mathbf{initialSpice'=initialSpice}\\
 metabolism'=metabolism\\
 maxAge'=maxAge\\
 agentCulture'=agentCulture\\
 loanBook'=loanBook\\
 diseases'=diseases\\
 agentImmunity'=agentImmunity\\
 children'=children\\
 \mathbf{maxSpice'=maxSpice}\\
 \mathbf{spiceLoanBook'=spiceLoanBook}\\
 \mathbf{spiceMetabolism'=spiceMetabolism}\\
\forall a:AGENT; l:POSITION \spot\\
\t1 a\in population\implies distance(position'(a),position(a))\leq vision(a)\\
\t1distance(position(a),l)\leq vision(a) \land (l\notin \ran position'))\\
\t2 \implies \mathbf{welfare(agentSugar(a),metabolism(a),}\hfill(1)\\
\t3\mathbf{agentSpice(a),spiceMetabolism(a),sugar(l),spice(l))}\\
\t2 \mathbf{<}\\
 \t2\mathbf{welfare(agentSugar(a),metabolism(a),}\\
\t3\mathbf{agentSpice(a),spiceMetabolism(a),}\\
\t4\mathbf{sugar(position'(a)),spice(position'(a)))}\\

agentSugar'=\{a:AGENT  | a\in population \spot\\
\t1 a\mapsto agentSugar(a)+sugar(position'(a))\}\\
sugar'=sugar\oplus \{loc:POSITION| loc \in \ran position'\spot loc \mapsto 0\}\\

\mathbf{agentSpice'=\{a:AGENT | a\in population \spot}\hfill(2)\\
\t1\mathbf{ a\mapsto agentSpice(a)}\\
\mathbf{\t2+spice(position'(a))\}}\\
\mathbf{spice'=spice\oplus \{loc:POSITION | loc \in \ran position'\spot loc\mapsto 0\}}\hfill(3)\\
\end{schema}

\begin{enumerate}
\item This is a copy of the original proposition with the $welfare$ function now replacing the previous sugar level check;
\item Agents consume spice at their new locations;
\item Locations with agents present now have no remaining spice.
\end{enumerate}

\subsection{Pollution Formation $P_{\Pi,\chi}$}
The $Movement_{spicePollution}$ schema has the same alterations as the $Movement_{basicSpice}$
\begin{schema}{Movement_{pollutionSpice}}
\Delta SpiceScape\\
 \where
 step'=step\\
population'=population\\
 maxSugar'=maxSugar\\
 sex'=sex\\
 pollution'=pollution\\
 vision'=vision\\
 age'=age\\
 maxAge'=maxAge\\
 agentCulture'=agentCulture\\
 loanBook'=loanBook\\
 children'=children\\
 agentImmunity'=agentImmunity\\
  initialSugar'=initialSugar\\
   \mathbf{initialSpice'=initialSpice}\\
 diseases'=diseases\\
 metabolism'=metabolism\\
 \mathbf{spiceMetabolism'=spiceMetabolism}\\
  \mathbf{spiceLoanBook'=spiceLoanBook}\\
   \mathbf{maxSpice'=maxSpice}\\
\forall a:AGENT; l:POSITION | a\in \dom(position') \spot\\
\t1a\in population\implies distance(position'(a),position(a))\leq vision(a)\\
\t1 (distance(position(a),l)\leq vision(a) \land (l\notin \ran position'))\\
\t2 \implies \mathbf{welfare(agentSugar(a),metabolism(a),}\\
\t3\mathbf{agentSpice(a),spiceMetabolism(a),sugar(l),spice(l))}\\
\t3\mathbf{\div(1+pollution(l))}\\
\t2\mathbf{ < }\\
\t2\mathbf{welfare(agentSugar(a),metabolism(a),}\\
\t3\mathbf{agentSpice(a),spiceMetabolism(a),sugar(position'(a)),}\\
\t4\mathbf{spice(position'(a)))\div(1+pollution(position'(a))) }\\

agentSugar'=\{a:AGENT| a\in population \spot\\
\t1 a\mapsto agentSugar(a)+sugar(position'(a)) \}\\
 sugar'=sugar\oplus \{l:POSITION | l \in \ran position' \spot l\mapsto 0 \}\\

\mathbf{agentSpice'=\{a:AGENT| a\in population \spot}\\
 \t1\mathbf{ a\mapsto agentSpice(a)+spice(position'(a)) \}}\\
 \mathbf{spice'=spice\oplus \{l:POSITION | l \in \ran position' \spot l\mapsto 0 \}}\\

pollution'=pollution\oplus\\
\t1\{l:POSITION; x:AGENT| position'(x)=l\spot\\
\t1 l\mapsto (PRODUCTION*sugar(l)+CONSUMPTION*metabolism(x) ) \}\\

\end{schema}

\subsection{Pollution Diffusion}

\begin{schema}{PollutionDiffusion_{spice}}
\Delta SpiceLattice\\
\Xi Step
 \where
\\
maxSugar'=maxSugar\\
sugar'=sugar\\
\mathbf{maxSpice'=maxSpice}\\
\mathbf{spice'=spice}\\
(step \mod POLLUTIONRATE \neq 0) \implies pollution'=pollution\\
 (step \mod POLLUTIONRATE = 0) \implies  pollution'=\\
 \t1\{l:POSITION  \spot l\mapsto (pollution(north(l))+pollution(south(l))\\
 \t2+pollution(east(l))+pollution(west(l)))\div 4\}\\\\
 
\end{schema}

\subsection{Replacement}
 
 \begin{schema}{Death_{spice}}
\Delta SpiceAgents\\
 \where
population'=population\setminus\{a:AGENT  |age(a)=maxAge(a)\\
\t1\lor agentSugar(a)=0\mathbf{\lor agentSpice(a)=0}\spot a\}\\
loanBook'=population' \dres loanBook\rres\\
\t1\{x:AGENT\cross(\nat\cross\nat)|first(x)\in population'\} \\
\mathbf{spiceLoanBook'=population' \dres spiceLoanBook\rres}\\
\t1\mathbf{\{x:AGENT\cross(\nat\cross\nat)|first(x)\in population'\} }\\
\forall a:AGENT\spot\\
\t1 a \in population' \implies \\
\t2 (sex(a)=sex'(a)\land vision(a)=vision'(a)\\
\t2\land maxAge(a)=maxAge'(a)\land agentCulture(a)=agentCulture'(a)\\
\t2\land position(a)=position'(a)\land age(a)=age'(a)\\
\t2\land agentSugar(a)=agentSugar'(a)\\
\t2\land metabolism'(a)=metabolism(a)\\
\t2\land diseases'(a)=diseases(a)\\
\t2\land agentImmunity'(a)=agentImmunity(a)\\
\t2\land children'(a)=children(a)\\
 \t2\land initialSugar'(a)=initialSugar(a)\\
\t2\mathbf{\land agentSpice'(a)=agentSpice(a)}\\
\t2 \mathbf{\land initialSpice'(a)=initialSpice(a)}\\
\t2\mathbf{\land spiceMetabolism'(a)=spiceMetabolism(a)})\\
\end{schema}

\begin{schema}{Replacement_{spice}}
\Delta SpiceAgents\\
 \where
\#population'=INITIALPOPULATIONSIZE\\
loanBook'= loanBook\\
spiceLoanBook'= spiceLoanBook\\
\forall a:AGENT\spot\\
 a \in (population) \implies \\
\t1 (a\in population'\\
\t1 \land sex(a)=sex'(a)\land vision(a)=vision'(a)\\
\t1\land maxAge(a)=maxAge'(a)\land agentCulture(a)=agentCulture'(a)\\
\t1\land position(a)=position'(a)\land age(a)=age'(a)\\
\t1\land agentSugar'(a)=agentSugar(a)\\
\t1\land metabolism'(a)=metabolism(a)\\
 \t1\land initialSugar'(a)=initialSugar(a)\\
  \t1\mathbf{\land initialSpice'(a)=initialSpice(a)}\\
\t1\mathbf{\land agentSpice'(a)=agentSpice(a)}\\
\t1\mathbf{\land spiceMetabolism'(a)=spiceMetabolism(a)}\\
\t1\land diseases'(a)=diseases(a)\\
\t1\land agentImmunity'(a)=agentImmunity(a)\\
\t1\land children'(a)=children(a))\\
\forall a:AGENT\spot\\
\t1 a\in population'\setminus population\implies (age'(a) = 0\\
\t1\land INITIALSUGARMIN \leq agentSugar'(a) \leq INITIALSUGARMAX\\
\t1\land  initialSugar'(a)=agentSugar'(a)\\
\t1\mathbf{\land INITIALSPICEMIN\leq agentSpice'(a)\leq INITIALSPICEMAX}\\
\t1\mathbf{\land  initialSpice'(a)=agentSpice'(a)}\\
\t1\land diseases'(a)=\emptyset\land children'(a)=\emptyset)\\

\end{schema}

\subsection{Agent Mating}
\begin{schema}{AgentMating}
\Xi Lattice\\
\Delta Agents\\
 \where
 loanBook'=loanBook\land \mathbf{spiceLoanBook'= spiceLoanBook}\\
 
\exists potentialMatingPairs: \power (AGENT\cross AGENT)|\hfill(1)\\
potentialMatingPairs=\{(a:AGENT,b:AGENT)| sex(a)\neq sex(b)\\
\t1 \land isFertile(age(a),sex(a)) \land isFertile(age(head),sex(head))\\
\t1\land adjacent(position(a),position(head))\}\\
(population',position',vision',agentSugar',agentCulture',metabolism'\\
\t1,children',diseases',agentImmunity',age',sex',initialSugar',\\
\mathbf{spiceMetabolism',agentSpice',initialSpice'})=\hfill(2)\\
concurrentMating(getConfictFreePairs(potentialMatingPairs),population,position,vision,\\
\t1agentSugar,agentCulture,metabolism,children,\\
\t1diseases,agentImmunity,age,maxAge, sex,initialSugar,\\
\mathbf{spiceMetabolism,agentSpice,initialSpice})
\end{schema}

\begin{enumerate}
\item Generate the set of all possible mating pairs;
\item Recursively proceed with concurrent mating within the conflict free subsets.
\end{enumerate}

\begin{axdef}
concurrentMating: \seq \power(AGENT\cross AGENT)\\
 \cross \power AGENT\\
 \cross  AGENT  \inj POSITION\\
 \cross AGENT \pfun \nat_1\\
 \cross AGENT \pfun \nat\\
\cross AGENT \pfun \seq BIT\\
 \cross AGENT\pfun\nat\\
 \cross AGENT \pfun\power AGENT\\
  \cross AGENT\pfun\power \seq BIT\\
\cross AGENT\pfun \seq BIT\\
 \cross AGENT \pfun \nat\\
 \cross AGENT \pfun \nat_1\\
  \cross AGENT \pfun SEX\\
  \cross AGENT \pfun \nat\\
\mathbf{    \cross AGENT \pfun \nat}\\
\mathbf{       \cross AGENT \pfun \nat}\\
\mathbf{         \cross AGENT \pfun \nat}\\
\rel\\ 
\power AGENT\\
\cross  AGENT  \inj POSITION\\
 \cross AGENT \pfun \nat_1\\
\cross AGENT \pfun \nat\\
\cross AGENT \pfun \seq BIT \\
\cross AGENT\pfun\nat \\
\cross AGENT \pfun\power AGENT\\
 \cross AGENT\pfun\power \seq BIT\\
\cross AGENT\pfun \seq BIT\\
 \cross AGENT \pfun \nat\\
 \cross AGENT \pfun \nat_1\\
  \cross AGENT \pfun SEX\\
  \cross AGENT \pfun \nat\\
\mathbf{    \cross AGENT \pfun \nat}\\
\mathbf{       \cross AGENT \pfun \nat}\\
\mathbf{         \cross AGENT \pfun \nat}\\
\where
\forall tail :\seq \power(AGENT\cross AGENT);\\
head: \power(AGENT\cross AGENT);\\
 population: \power AGENT;\\
 position:  AGENT  \inj POSITION;\\
 vision: AGENT \pfun \nat_1;\\
 agentSugar,\mathbf{agentSpice}: AGENT \pfun \nat;\\
agentCulture: AGENT \pfun \seq BIT;\\
 metabolism,\mathbf{spiceMetabolism}: AGENT\pfun\nat;\\
 children: AGENT \pfun\power AGENT;\\
 diseases: AGENT\pfun\power \seq BIT;\\
agentImmunity: AGENT\pfun \seq BIT;\\
age: AGENT \pfun \nat;\\
maxAge: AGENT \pfun \nat_1;\\
sex: AGENT \pfun SEX;\\
 initialSugar,\mathbf{initialSpice}: AGENT \pfun \nat;\\
 \zbreak
  \exists newpopulation: \power AGENT;\\
 newposition:  AGENT  \inj POSITION;\\
 newvision: AGENT \pfun \nat_1;\\
 newagentSugar,\mathbf{newagentSpice}: AGENT \pfun \nat;\\
newagentCulture: AGENT \pfun \seq BIT;\\
 newmetabolism,\mathbf{newspiceMetabolism}: AGENT\pfun\nat;\\
 newchildren: AGENT \pfun\power AGENT;\\
 newdiseases: AGENT\pfun\power \seq BIT;\\
newagentImmunity: AGENT\pfun \seq BIT;\\
newage: AGENT \pfun \nat;\\
newmaxAge: AGENT \pfun \nat_1;\\
newsex: AGENT \pfun SEX;\\
 newinitialSugar,\mathbf{newinitialSpice}: AGENT \pfun \nat;|\\
(newpopulation,newposition,newvision,newagentSugar,newagentCulture,\\
\t1newmetabolism,newchildren,newdiseases,newagentImmunity,newage,newmaxAge,\\
\t1 newsex,newinitialSugar,\mathbf{newspiceMetabolism,newagentSpice,newinitialspice})=\\
applyMating(asSeq(head),population,position,vision,\\
\t1agentSugar,agentCulture,metabolism,children,\\
\t1diseases,agentImmunity,age,maxAge, sex,initialSugar,\\
\t1\mathbf{spiceMetabolism,agentSpice,initialspice})\spot\\

concurrentMating(\langle\rangle, population,position,vision,\\
\t1agentSugar,agentCulture,metabolism,children,\\
\t1diseases,agentImmunity,age,maxAge, sex,initialSugar)=\\
(population,position,vision,agentSugar,agentCulture,metabolism,\\
\t1children,diseases,agentImmunity,age,maxAge, sex,initialSugar,\\
\mathbf{spiceMetabolism,agentSpice,initialspice})\\

concurrentMating(\langle head\rangle\cat tail,population,position,vision,\\
\t1agentSugar,agentCulture,metabolism,children,\\
\t1diseases,agentImmunity,age,maxAge, sex,initialSugar)=\\
concurrentMating(tail,newpopulation,newposition,newvision,\\
\t1newagentSugar,newagentCulture,newmetabolism,newchildren,\\
\t1newdiseases,newagentImmunity,newage,newmaxAge, newsex,newinitialSugar,\\
\mathbf{newspiceMetabolism,newagentSpice,newinitialspice})\\

\end{axdef}

\begin{axdef}
applyMating:\seq (AGENT\cross AGENT)\\
 \cross \power AGENT\\
 \cross  AGENT  \inj POSITION\\
 \cross AGENT \pfun \nat_1\\
 \cross AGENT \pfun \nat\\
\cross AGENT \pfun \seq BIT\\
 \cross AGENT\pfun\nat\\
 \cross AGENT \pfun\power AGENT\\
  \cross AGENT\pfun\power \seq BIT\\
\cross AGENT\pfun \seq BIT\\
 \cross AGENT \pfun \nat\\
 \cross AGENT \pfun \nat_1\\
  \cross AGENT \pfun SEX\\
  \cross AGENT \pfun \nat\\
  \mathbf{    \cross AGENT \pfun \nat}\\
\mathbf{       \cross AGENT \pfun \nat}\\
\mathbf{         \cross AGENT \pfun \nat}\\
\rel\\ 
\power AGENT\\
\cross  AGENT  \inj POSITION\\
 \cross AGENT \pfun \nat_1\\
\cross AGENT \pfun \nat\\
\cross AGENT \pfun \seq BIT \\
\cross AGENT\pfun\nat \\
\cross AGENT \pfun\power AGENT\\
 \cross AGENT\pfun\power \seq BIT\\
\cross AGENT\pfun \seq BIT\\
 \cross AGENT \pfun \nat\\
 \cross AGENT \pfun \nat_1\\
  \cross AGENT \pfun SEX\\
  \cross AGENT \pfun \nat\\
  \mathbf{    \cross AGENT \pfun \nat}\\
\mathbf{       \cross AGENT \pfun \nat}\\
\mathbf{         \cross AGENT \pfun \nat}\\

\where
\forall population : \power AGENT; \\
position : AGENT  \inj POSITION;\\
sex: AGENT \pfun SEX;\\
vision: AGENT \pfun \nat_1;\\
age: AGENT \pfun \nat;\\
initialSugar,\mathbf{initialSpice}: AGENT \pfun \nat;\\
maxAge: AGENT \pfun \nat_1;\\
metabolism,\mathbf{spiceMetabolism}:AGENT\pfun\nat;\\
agentSugar,\mathbf{agentSpice}: AGENT \pfun \nat;\\
agentCulture:AGENT \pfun \seq BIT;\\
children: AGENT \pfun\power AGENT;\\
agentImmunity: AGENT\pfun \seq BIT;\\
diseases: AGENT\pfun\power \seq BIT;\\
head : AGENT\cross AGENT;\\
tail :\seq (AGENT\cross AGENT);\spot\\
\zbreak
\exists offspring,a,b:AGENT;\\
newsex: AGENT \pfun SEX;\\
newvision: AGENT \pfun \nat_1;\\
newmetabolism,newagentSugar,newinitialSugar: AGENT \pfun \nat;\\
\mathbf{newspiceMetabolism,newagentSpice,newinitialSpice: AGENT \pfun \nat;}\\
newmaxAge: AGENT \pfun \nat_1;\\
newagentCulture:AGENT \pfun \seq BIT;\\
newchildren: AGENT \pfun\power AGENT;\\
newagentImmunity: AGENT\pfun \seq BIT;\\
inheritedImmunity: \seq BIT;\\
inheritedCulture: \seq BIT;\\
| offspring\notin population\\
a=first(head)\land b=second(head)\\

newchildren: children\cup\{offspring\mapsto \emptyset\,a\mapsto children(a)\cup\{ offspring\},\\
\t3b\mapsto children(b)\cup\{ offspring\}\}\\

newsex = sex\cup\{offspring\mapsto male\}\\
 \t1\lor newsex=sex\cup\{offspring\mapsto female\}\\
newvision = vision\cup\{offspring\mapsto vision(a)\}\\
\t1 \lor newvision = vision\cup\{offspring\mapsto vision(b)\}\\

newmaxAge = maxAge\cup\{offspring\mapsto maxAge(a)\}\\
\t1 \lor newmaxAge = maxAge\cup\{offspring\mapsto maxAge(b)\}\\
newmetabolism =metabolism\cup\{offspring\mapsto metabolism(a)\}\\
\t1 \lor newmetabolism=metabolism\cup\{offspring\mapsto metabolism(b)\}\\

\mathbf{newspiceMetabolism =spiceMetabolism\cup\{offspring\mapsto spiceMetabolism(a)\}}\\
\t1\mathbf{ \lor newspiceMetabolism=spiceMetabolism\cup\{offspring\mapsto spiceMetabolism(b)\}}\\

newinitialSugar =initialSugar\oplus\\
\t1\{offspring\mapsto initialSugar(a)/2+initialSugar(b)/2\}\\
newagentSugar = agentSugar\oplus\{offspring\mapsto initialSugar(offspring),a\mapsto initialSugar(a)/2,b\mapsto initialSugar(b)/2\}\\

\mathbf{newinitialSpice =initialSpice\oplus}\\
\mathbf{\t1\{offspring\mapsto initialSpice(a)/2+initialSpice(b)/2\}}\\
\mathbf{newagentSpice = agentSpice\cup\{offspring\mapsto initialSpice(offspring),a\mapsto initialSpice(a)/2,b\mapsto initialSpice(b)/2\}}\\

\t1\land \forall n:1\upto IMMUNITYLENGTH\spot\\
\t2 (inheritedImmunity(n)=agentImmunity(a)(n)\\
\t2\lor inheritedImmunity(n)=agentImmunity(b)(n))\\
newagentImmunity: agentImmunity\cup\{offspring\mapsto inheritedImmunity\}\\

\t1\land \forall n:1\upto CULTURECOUNT\spot\\
\t2 (inheritedCulture(n)=agentCulture(a)(n)\\
\t2\lor inheritedCulture(n)=agentCulture(b)(n))\\
newagentCulture:agentCulture\cup\{offspring\mapsto inheritedCulture\}\\
\zbreak
applyMating(\langle\rangle, population,position,vision,agentSugar,agentCulture,\\
\t1metabolism,children,diseases, agentImmunity,age,maxAge, sex,initialSugar,\\
\t1\mathbf{spiceMetabolism,agentSpice,initialSpice})=\\
 (population,position,vision,agentSugar,agentCulture,metabolism,\\
\t1children,diseases, agentImmunity,age,maxAge, sex,initialSugar,\\
\t1\mathbf{spiceMetabolism,agentSpice,initialSpice})\\
~\\
applyMating(\langle head\rangle\cat tail, population,position,vision,agentSugar,agentCulture,\\
\t1metabolism,children,diseases, agentImmunity,age,maxAge, sex,initialSugar,\\
\t1\mathbf{spiceMetabolism,agentSpice,initialSpice})=\\
\IF ((\exists loc:POSITION|  (adjacent(loc, position(ag)))\\
\t2\lor adjacent(loc, position(head))\land loc\notin \dom position)\\
\t2\mathbf{\land (agentSpice(head)>initialSpice(head)) \land (agentSpice(ag)>initialSpice(ag))}
\t2\land (agentSugar(head)>initialSugar(head)) \land (agentSugar(ag)>initialSugar(ag))) \THEN\hfill(2a)\\

\t1applyMating(tail, population\cup\{offspring\},position\cup\{offspring\mapsto loc\},\\
\t1newvision,newagentSugar,newagentCulture,newmetabolism,\\
\t1 newchildren,diseases\cup\{offspring\mapsto \emptyset\}, newagentImmunity,\\
\t1 age\cup\{offspring\mapsto 0\},newmaxAge, newsex,initialSugar,
\t1\mathbf{newspiceMetabolism,newagentSpice,newinitialSpice})\\
\ELSE\hfill(2b)\\
\t1applyMating(tail, population,position,vision,agentSugar,agentCulture,\\
 \t2metabolism,children,diseases, agentImmunity,age,maxAge, sex,initialSugar,\\
 \t2\mathbf{spiceMetabolism,agentSpice,initialSpice})\\

\end{axdef}

\subsection{Culture}

\begin{schema}{Culture_{spice}}
\Delta SpiceAgents\\
 \where
 population'=population\\
 sex'=sex\\
 position'=position\\
 vision'=vision\\
 age'=age\\
 maxAge'=maxAge\\
 agentSugar'=agentSugar\\
 children'=children\\
 loanBook'=loanBook\\
 diseases'=diseases\\
  initialSugar'=initialSugar\\
 metabolism'=metabolism\\
 agentImmunity'=agentImmunity\\
  \mathbf{agentSpice'=agentSpice}\\
 \mathbf{spiceMetabolism'=spiceMetabolism}\\
 \mathbf{spiceLoanBook'=spiceLoanBook}\\
 \mathbf{initialSpice'=initialSpice}\\
\forall a:AGENT \spot a \in population\implies \\
\t1agentCulture'(a)= flipTags(agentCulture(a),\\
\t2asSeq(\{b:AGENT|adjacent(position(a),position(b))\}),agentCulture)\\
\end{schema}

\subsection{Disease}

\begin{schema}{ImmuneResponse_{spice}}
\Xi SpiceLattice\\
\Delta SpiceAgents\\
\Xi Step
 \where

loanBook'=loanBook\\
 \mathbf{spiceLoanBook'=spiceLoanBook}\\
 population'=population\\
 sex'=sex\\
 position'=position\\
 vision'=vision\\
 age'=age\\
 maxAge'=maxAge\\
 agentCulture'=agentCulture\\
 diseases'=diseases\\
  children'=children\\
  agentSugar'=agentSugar\\
   initialSugar'=initialSugar\\
    \mathbf{initialSpice'=initialSpice}\\
  metabolism'=metabolism\\
  \mathbf{agentSpice'=agentSpice}\\
 \mathbf{spiceMetabolism'=spiceMetabolism}\\
 agentImmunity'=\{a:AGENT|a\in population\spot\\
 \t1 a \mapsto applyDiseases(agentImmunity(a),asSeq(diseases(a)))\}\\
 
  \forall x:AGENT\spot x\in population\implies agentSugar'(x)=agentSugar(x)-\\
 \t1\#\{d:\seq BIT| d\in diseases(a)\land \lnot subseq(d,agentImmunity(a)\}\\
 
\forall x:AGENT\spot x\in population\implies agentSpice'(x)=agentSpice(x)-\\
 \t1\#\{d:\seq BIT| d\in diseases(a)\land \lnot subseq(d,agentImmunity(a)\}\\
\end{schema}

\begin{schema}{Transmission_{spice}}
\Delta SpiceAgents\\
 \where
 loanBook'=loanBook\\
  initialSugar'=initialSugar\\
 \mathbf{spiceLoanBook'=spiceLoanBook}\\
 population'=population\\
 sex'=sex\\
 position'=position\\
 vision'=vision\\
 age'=age\\
 maxAge'=maxAge\\
 agentCulture'=agentCulture\\
 agentImmunity'=agentImmunity\\
 children'=children\\
 agentSugar'=agentSugar\\
  \mathbf{agentSpice'=agentSpice}\\
 \mathbf{spiceMetabolism'=spiceMetabolism}\\
  \mathbf{initialSpice'=initialSpice}\\
 metabolism'=metabolism\\
\forall a: AGENT\spot a\in population\implies\\
diseases'(a)=diseases(a)\cup\\
\t1 newDiseases(asSeq(visibleAgents(a,position,1)),diseases)
\end{schema}

\subsection{Inheritance}

\begin{schema}{Inheritance_{spice}}
\Delta SpiceAgents\\
 \where
 population'=population\land sex'=sex\\
 position'=position\land vision'=vision\\
 age'=age\land maxAge'=maxAge\\
 agentCulture'=agentCulture\land children'=children\\
 metabolism'=metabolism\land \mathbf{spiceMetabolism'=spiceMetabolism}\\
 diseases'=diseases\land agentImmunity'=agentImmunity\\
  initialSugar'=initialSugar\\
   \mathbf{initialSpice'=initialSpice}\\
\exists dying: \power AGENT;\\
 inheritFromFemale,inheritFromMale:AGENT\pfun \mathbf{(\nat\cross\nat)} \spot\hfill(1)\\
\dom inheritFromFemale=\dom inheritFromMale=population\setminus dying\\
 \t1dying=\{x:AGENT | x\in population \land (age(x)=maxAge(x)\lor\\
 \t2 agentSugar(x)=0\mathbf{\lor agentSpice(x)=0})\}\hfill(2)\\

\forall x:AGENT; n,m:\nat |x \in population\setminus dying\spot\\

\t1getMother(x,children,sex) \notin dying\implies\\
 \t2 inheritFromFemale(x)=(0,0)\\
 \t1getFather(x,children,sex) \notin dying\implies\\
 \t2 inheritFromMale(x)=(0,0)\\

 \t1\exists m:AGENT | m=getMother(x,children,sex) \land m\in dying\implies\\
 \t2 inheritFromFemale(x)=\\
 \t3(agentSugar(m)\div\#(population\cap children(m)\setminus dying))\\
  \t3\mathbf{(agentSpice(m)\div\#(population\cap children(m)\setminus dying))}\hfill(3)\\
  \t1\exists d:AGENT | d=getFather(x,children,sex) \land d\in dying\implies\\
 \t2 inheritFromMale(x)=\\
 \t3agentSugar(d)\div\#(population\cap children(d)\setminus dying))\\
   \t3\mathbf{(agentSpice(m)\div\#(population\cap children(m)\setminus dying))}\\
   
x\in dying\\
\t1 \implies (agentSugar'(x)=0\mathbf{\land agentSpice'(x)=0})\\\\
x\notin dying\\
\t1 \implies (agentSugar'(x)=agentSugar(x)\\
\t2+first(inheritFromMale(x))+first(inheritFromFemale(x))\\
\t2\mathbf{\land agentSpice'(x)=agentSpice(x)}\\
\t3\mathbf{+second(inheritFromFemale(x))+second(inheritFromMale(x)))}\\
 loanBook'=disperseLoans(loanBook,asSeq(dying),children)\\
 \mathbf{spiceLoanBook'=}\\
 \mathbf{\t1disperseLoans(spiceLoanBook,asSeq(dying),children)}\hfill(4)\\

\end{schema}

\begin{enumerate}
\item Agents can now inherit two amounts, a $sugar$ inheritance and a $spice$ inheritance;
\item Death now occurs if either resource reaches zero;
\item The individual $spice$ inheritance is calculated in the same way as the $sugar$ inheritance;
\item Spice loans are dispersed amongst children.
\end{enumerate}

\subsection{Combat}
Combat is defined only in terms of sugar.  We can either accept this and assume combat is based only on sugar levels or we can extend combat by defining new versions of wealth and reward.  We note that no simulations combining combat with more than one resource are presented in the book.

We can extend the combat rule with a few simple assumptions. First the wealth of an agent is used to determine if we can attack that agent or if an agent can retaliate against us. In the single resource scenario we simply used the sugar that an agent carried.  With two resources we need to combine both sugar and spice.  The simplest approach is to add these two amounts together and in the absence of any guidelines this seems the sensible option.

$availMoves$ requires only minor changes to return the set of all safe moves that an agent can make. 

\begin{axdef}
availMoves_{spice}: AGENT\cross (AGENT\pinj POSITION)\cross (POSITION\pinj\nat)\cross (AGENT\pinj\nat)\\
\t1\cross(POSITION\pinj\nat)\cross (AGENT\pinj\nat)\cross (AGENT\pinj \seq BIT)\cross\nat\\
\t1\pfun\power POSITION
 \where
  \forall x, agent:AGENT;positions:AGENT\pinj POSITION;vision:\nat;\\
 sugar,spice:POSITION\pinj\nat;agentSpice,agentSugar:AGENT\pinj\nat;\\
  culture: AGENT\pinj \seq BIT\spot\\
~\\
availMoves_{spice}(agent,positions,sugar,agentSugar,spice,agentSpice,culture,vision)=\\
\{l:POSITION;x:AGENT|l\in distance(l,positions(agent))\leq vision\\
\land positions(x)=l\implies (agentSugar(x)\mathbf{+agentSpice(x)}<\\
\t4agentSugar(agent)+\mathbf{agentSpice(ag)}\\
\t1\land tribe(culture(x))\neq tribe(culture(agent)))\\
\land ((distance(positions(x),l)\leq vision)\hfill(3)\\
\t1\land tribe(culture(x))\neq tribe(culture(agent)))\implies \\
\t2agentSugar(x)\mathbf{+agentSpice(x)}<agentSugar(agent)\mathbf{+agentSpice(agent)}\\
\t2+reward(l,sugar,positions,agentSugar,COMBATLIMIT))\\
\t2\mathbf{+reward(l,spice,positions,agentSpice,SPICECOMBATLIMIT)}\spot l\}\\

 \end{axdef}

\begin{schema}{Combat_{spice}}
\Delta SugarScape\\
 \where
 step'=step\\
maxSugar'=maxSugar\\
 \mathbf{maxSpice'=maxSpice}\\
 pollution'=pollution\\
 loanBook'= population'\dres loanBook\rres (population'\dres (\ran loanBook))\\
  \mathbf{spiceLoanBook'=}\\
\mathbf{\t1 population'\dres spiceLoanBook\rres(population'\dres (\ran spiceLoanBook))}\\
  \forall ag:AGENT; l:POSITION  \spot\\
  ag\in population' \implies\hfill(3)\\
  \t1(sex'(ag)=sex(ag)\\
  \t1\land vision'(ag)=vision(ag)\\
  \t1\land age'(ag)=age(ag)\\
  \t1\land maxAge'(ag)=maxAge(ag)\\
  \t1\land children'(ag)=children(ag)\\
  \t1\land agentCulture'(ag)=agentCulture(ag)\\
  \t1\land agentImmunity'(ag)=agentImmunity(ag)\\
  \t1\land metabolism'(ag)=metabolism(ag)\\
  \t1\land  initialSugar'(ag)=initialSugar(ag)\\
   \t1\mathbf{\land initialSpice'(ag)=initialSpice(ag)}\\
   \t1\mathbf{\land spiceMetabolism'(ag)=spiceMetabolism(ag)}\\
  \t1\land diseases'(ag)=diseases(ag))\\
   (population',position',sugar',agentSugar',agentSpice')=\\
   \t1applyAllCombat_{spice}(asSeq(population),population,position,sugar,\\
   \t4agentSugar,agentSpice,vision,agentCulture)\\
\end{schema}

\begin{axdef}
applyAllCombat_{spice}: \seq AGENT\cross \power AGENT\cross (AGENT\pinj POSITION)\\
\t1\cross (POSITION\pinj\nat)\cross(AGENT\pinj\nat)\cross(AGENT\pinj\nat)\\
\t1\cross (AGENT\pfun\nat) \cross (AGENT\pfun \seq BIT)\\
 \fun\\
 (\power AGENT\cross (AGENT\pinj POSITION)\cross (POSITION\pinj\nat)\\
 \t1\cross(AGENT\pinj\nat)\cross(AGENT\pinj\nat))
\where
\forall head: AGENT; tail:\seq AGENT; pop:\power AGENT; \\
positions:AGENT\pinj POSITION; sugar: POSITION\pinj\nat; \\
agSugar,agSpice: AGENT\pinj\nat; \\
vision:AGENT\pfun\nat; culture: AGENT\pfun \seq BIT\spot\\

applyAllCombat(\langle\rangle,pop,positions,sugar,agSugar,agSpice,vision,culture) =\\
\t1(pop,positions,sugar,agSugar,agSpice)\\
~\\
applyAllCombat(\langle head\rangle\cat tail,pop,positions,sugar,agSugar,agSpice,vision,culture) =\\

\IF  (head\in pop) \THEN\\
\t1 applyAllCombat_{spice}(tail,\\
\t2 singleFight_{spice}(head,pop,positions,sugar,agSugar,agSpice,vision,culture))\\
 \ELSE\\
\t1 applyAllCombat_{spice}(tail,pop,positions,sugar,agSugar,agSpice,vision,culture)
\end{axdef}

\begin{axdef}
singleFight_{spice}:AGENT\cross \power AGENT\cross (AGENT\pinj POSITION)\cross (POSITION\pinj\nat)\\
\t1\cross(AGENT\pinj\nat)\cross(AGENT\pinj\nat)\cross (AGENT\pfun\nat) \cross (AGENT\pfun \seq BIT)\\
 \fun\\
 (\power AGENT\cross (AGENT\pinj POSITION)\cross (POSITION\pinj\nat)\\
 \t1\cross(AGENT\pinj\nat)\cross(AGENT\pinj\nat))
\where
\forall ag: AGENT; population:\power AGENT; positions:AGENT\pinj POSITION;\\
sugar: POSITION\pinj\nat; agSugar,agSpice: AGENT\pinj\nat; \\
vision:AGENT\pfun\nat; culture: AGENT\pfun \seq BIT\spot\\
~\\
singleFight_{spice}(ag,population,positions,sugar,agSugar,agSpice,vision,culture)=\\
\IF~(availMoves(ag,positions,sugar,agSugar,agSpice,culture,vision(ag))=\emptyset)~\THEN\\
\t1 (population,positions,sugar,agSugar,agSpice)\\
\ELSE\\
\t1\exists loc:POSITION; available:\power POSITION;\\
\t1\forall otherLoc:POSITION|\\
\t1 loc,otherLoc\in\\
\t2 availMoves_{spice}(ag,positions,sugar,agSugar,agSpice,culture,vision(ag))\\
 \t1 \land otherLocation\neq location\\
 \t1 reward(loc,sugar,position,agSugar,COMBATLIMIT)\\
 \t2\mathbf{+ reward(loc,spice,position,agSpice,SPICECOMBATLIMIT)}\\
 \t1 \geq reward(otherLoc,sugar,position,agSugar,COMBATLIMIT)\\
 \t2\mathbf{+ reward(otherLoc,spice,position,agSpice, SPICECOMBATLIMIT)}\\
 \t1(distance(position(ag),loc) > distance(position(otherLoc),position'(ag))\implies\\
 \t2reward(loc,sugar,position,agSugar,COMBATLIMIT)\\
 \t3\mathbf{+ reward(loc,sugar,position,agSpice,SPICECOMBATLIMIT)}\\
 \t2 > reward(otherLoc,sugar,position,agSugar,COMBATLIMIT)\\
 \t3\mathbf{+ reward(otherLoc,spice,position,agSpice,SPICECOMBATLIMIT)})\\
 \t1\spot\\
\t1(population\setminus\{positions\inv(loc)\},(positions\nrres \{loc\})\oplus\{ag\mapsto loc\},sugar\oplus\{loc\mapsto 0\},\\
\t1agSugar\oplus\{ag\mapsto agSugar(ag)+\\
\t2reward(position'(ag),sugar,position,agSugar,COMBATLIMIT)\},\\
\t1\mathbf{agSpice\oplus\{ag\mapsto agSpice(ag)+}\\
\t2\mathbf{reward(position'(ag),spice,position,agSpice,SPICECOMBATLIMIT)\}})\\
\end{axdef}

\begin{schema}{SynchronousCombat_{spice}}
\Delta SugarScape\\
 \where
 step'=step\\
 maxSugar'= maxSugar\\
 \mathbf{maxSpice'=maxSpice}\\
 pollution'=pollution\\
  loanBook'= population'\dres loanBook\rres (population'\dres (\ran loanBook))\\
  \mathbf{spiceLoanBook'=}\\
\mathbf{\t1 population'\dres spiceLoanBook\rres(population'\dres (\ran spiceLoanBook))}\\
    population'\subseteq population\\
    
   sugar'=sugar\oplus\{p:POSITION | p\in\ran position'\spot p\mapsto 0\}\\

  \forall ag:AGENT; l:POSITION  \spot\\

  ag\in population' \implies\\
  \t1(sex'(ag)=sex(ag)\\
  \t1\land vision'(ag)=vision(ag)\\
  \t1\land age'(ag)=age(ag)\\
  \t1\land maxAge'(ag)=maxAge(ag)\\
  \t1\land children'(ag)=children(ag)\\
  \t1\land agentCulture'(ag)=agentCulture(ag)\\
  \t1\land agentImmunity'(ag)=agentImmunity(ag)\\
  \t1\land metabolism'(ag)=metabolism(ag)\\
  \t1\land initialSugar'(ag)=initialSugar(ag)\\
  \t1\mathbf{\land initialSpice'(ag)=initialSpice(ag)}\\
  \mathbf{\land spiceMetabolism(ag)=spiceMetabolism(ag)}\\
  \t1\land diseases'(ag)=diseases(ag)\\
  \t1\land agentSugar'(ag)=agentSugar(ag)\\
\t2+reward(position'(ag),sugar,position,agentSugar,COMBATLIMIT)\\
\t1\land agentSpice'(ag)=agentSpice(ag)\\
\t2\mathbf{+reward(position'(ag),spice,position,agentSpice,}\\
\t4\mathbf{SPICECOMBATLIMIT)}\\
\t1\land position'(ag)\in\\
 \t2 availMoves_{spice}(ag,position,sugar,agentSugar,\\
 \t3agentSpice,agentCulture,vision(ag)))\\
  \t1)\\
 
ag\in population \setminus population'\implies \\
\t1 \exists x:AGENT \spot position'(x)=position(ag)\land tribe(culture(x))\neq tribe(culture(ag))\\

 ~\\
 (l\in availMoves_{spice}(ag,position,sugar,agentSugar,agentSpice,agentCulture,vision(ag))\\
 \t1\land reward(l,sugar,position,agentSugar,COMBATLIMIT)\\
  \t2\mathbf{+ reward(l,spice,position,agentSpice,SPICECOMBATLIMIT)}\\
 \t1 \geq reward(position'(ag),sugar,position,agentSugar,COMBATLIMIT)\\
 \t2\mathbf{+ reward(position'(ag),spice,position,agentSpice,}\\
\t4\mathbf{ SPICECOMBATLIMIT)}\\
 \t1\land distance(position(ag),l) < distance(position(ag),position'(ag)))\\
 \t1\implies \exists x:AGENT \spot position'\inv(l)=x\land position(x)\neq l\\
\end{schema}

\subsection{Credit}
Credit is defined with one resource.  It is incorporated into a dual resource simulation but no extra information is given as to what changes were made, if any.  The most logical approach is to assume that $spice$ loans are administered the the exact same way as $sugar$. We create a separate system of loans for $spice$ that is dealt with in the exact same manner as $sugar$ loans.  Our specification is now split into two parts, the parts dealing with $sugar$, already specified, and the parts dealing with $spice$, which are copies of their counterparts. The amount of a resource available to be borrowed now depends on which resource we are talking about.

\begin{axdef}
amtAvail_{new}:\nat\cross SEX\cross \nat \cross \nat \fun \nat\\
\where
\forall age,resource,baseAmt:\nat\spot\\
amtAvail_{new}(age,male,resource,baseAmt)=\\
\t1 resource\div 2\iff age> MALEFERTILITYEND\\

amtAvail_{new}(age,male,resource,baseAmt)= resource-baseAmt\iff \\
\t2 (age\leq MALEFERTILITYEND \land\\
\t2 isFertile(age,male)\land resource>baseAmt)\\

amtAvail_{new}(age,male,resource,baseAmt)= 0\iff \\
\t2 (age\leq MALEFERTILITYEND \land\\
\t2 \lnot(isFertile(age,male)\lor resource>baseAmt))\\

amtAvail_{new}(age,female,resource,baseAmt)=\\
\t1 resource\div 2\iff age> FEMALEFERTILITYEND\\

amtAvail_{new}(age,female,resource,baseAmt)= resource-baseAmt\iff \\
\t2 (age\leq FEMALEFERTILITYEND \land\\
\t2 isFertile(age,female)\land resource>baseAmt)\\

amtAvail_{new}(age,female,resource,baseAmt)= 0\iff \\
\t2 (age\leq FEMALEFERTILITYEND\\
\t2 \land \lnot(isFertile(age,female)\lor resource>baseAmt))\\

\end{axdef}

\begin{schema}{PaySugarLoans}
\Delta Agents\\
\Xi Step
 \where
 population'=population\\
 sex'=sex\\
 position'=position\\
 vision'=vision\\
 age'=age\\
 maxAge'=maxAge\\
 agentCulture'=agentCulture\\
 agentImmunity'=agentImmunity\\
 children'=children\\
 diseases'=diseases\\
 metabolism'=metabolism\\
 initialSugar'=initialSugar\\
  \mathbf{initialSpice'=initialSpice}\\
\mathbf{spiceMetabolism'=spiceMetabolism}\\
\mathbf{spiceLoanBook'=spiceLoanBook}\\
\mathbf{agentSpice'=agentSpice}\\

\exists dueLoans,newLoans:(AGENT\rel(AGENT\cross(\nat\cross\nat)))\spot\\ 
 dueLoans= loanBook\rres (\ran(loanBook)\rres\{a:(\nat\cross\nat)| second(a)=step\})\\
 (loanBook',agentSugar')=\\
 \t1 payExclusiveLoans(chooseConflictFreeSets(dueLoans),agentSugar,loanBook)\\
\end{schema}

\begin{schema}{PaySpiceLoans}
\Delta Agents\\
\Xi Step
 \where
 population'=population\\
 sex'=sex\\
 position'=position\\
 vision'=vision\\
 age'=age\\
 maxAge'=maxAge\\
 agentCulture'=agentCulture\\
 agentImmunity'=agentImmunity\\
 children'=children\\
 diseases'=diseases\\
 metabolism'=metabolism\\
 initialSugar'=initialSugar\\
  \mathbf{initialSpice'=initialSpice}\\
\mathbf{spiceMetabolism'=spiceMetabolism}\\
loanBook'=loanBook\\
agentSugar'=agentSugar\\

\exists dueLoans,newLoans:(AGENT\rel(AGENT\cross(\nat\cross\nat)))\spot\\ 
 dueLoans= loanBook\rres (\ran(spiceLoanBook)\rres\{a:(\nat\cross\nat)| second(a)=step\})\\
 (spiceLoanBook',agentSpice')=\\
 \t1 payExclusiveLoans(chooseConflictFreeSets(dueLoans),agentSpice,spiceLoanBook)\\
\end{schema}

\begin{schema}{MakeLoans_{Sugar}}
\Delta Agents\\
\Xi Step
 \where
  population'=population\land sex'=sex\\
 position'=position\land vision'=vision\\
 age'=age\land maxAge'=maxAge\\
 agentCulture'=agentCulture\land agentImmunity'=agentImmunity\\
  diseases'=diseases\land children'=children\\
 metabolism'=metabolism\\
 initialSugar'=initialSugar\\
 \mathbf{initialSpice'=initialSpice}\\
 \mathbf{spiceMetabolism'=spiceMetabolism}\\
\mathbf{agentSpice'=agentSpice}\\
\mathbf{spiceLoanBook'=spiceLoanBook}\\
\exists newLoans: \power(AGENT\cross(AGENT\cross(\nat \cross\nat)));\\
\forall ag,lender,borrower:AGENT; amt,due:\nat\spot\\
loanBook'=loanBook\cup newLoans\\
 ag\in\dom newLoans\implies\\
 \t1 agentSugar'(ag)=agentSugar(ag)-totalLoaned(ag,newLoans)\\
ag\in\dom(\ran newLoans)\implies\\ 
\t1agentSugar'(ag)=agentSugar(ag)+totalOwed(ag,newLoans)\\
ag\notin\dom (newLoans)\cup \dom(\ran newLoans)\implies\\
\t1 agentSugar'(ag)=agentSugar(ag)\\
 willBorrow(age(ag),sex(ag),agentSugar'(ag),\\
 \t1\ran (loanBook'\cap\{a:AGENT\cross(AGENT\cross(\nat \cross\nat))\\
\t3|borrower(a)=borrower(loan)\}))\implies\\
 \t2\lnot\exists ag2:AGENT\spot canLend(age(ag2),sex(ag2),agentSugar'(ag2))\\
 \t3\land adjacent(position(ag2),position(ag))\\
 
 totalLoaned(ag,newLoans)\leq \\
 \t1amtAvail_{new}(age(ag),sex(ag),agentSugar(ag),CHILDAMT)\\
totalOwed(ag,newLoans)\leq CHILDAMT-agentSugar(ag)\\

(lender,(borrower,(amt,due)))\in newLoans \implies\\
\t1 (canLend(age(lender),sex(lender),agentSugar(lender))\\
\t1\land willBorrow(age(borrower),sex(borrower),agentSugar(borrower),\\
\t2\{borrower\}\dres(\ran loanBook))\\

\t1\land amt\leq \\
\t2min(\{amtAvail_{new}(age(lender),sex(lender),agentSugar(lender),\\
\t6CHILDAMT),CHILDAMT-agentSugar(borrower))\})  \\
\t1\land due=step+DURATION\\
 \t1\land adjacent(position(lender),position(borrower)))\\
 
\end{schema}

\begin{schema}{MakeSpiceLoans}
\Xi SpiceLattice\\
\Delta SpiceAgents\\
\Xi Step
 \where
  population'=population\land sex'=sex\\
 position'=position\land vision'=vision\\
 age'=age\land maxAge'=maxAge\\
 agentCulture'=agentCulture\land agentImmunity'=agentImmunity\\
  diseases'=diseases\land children'=children\\
 metabolism'=metabolism\\
spiceMetabolism'=spiceMetabolism\\
agentSugar'=agentSugar\\
loanBook'=loanBook\\
initialSugar'=initialSugar\\
\mathbf{initialSpice'=initialSpice}\\
\exists newLoans: \power(AGENT\cross(AGENT\cross(\nat \cross\nat)));\\
\forall ag,lender,borrower:AGENT; amt,due:\nat\spot\\
spiceLoanBook'=spiceLoanBook\cup newLoans\\
 ag\in\dom newLoans\implies\\
 \t1 agentSpice'(ag)=agentSpice(ag)-totalLoaned(ag,newLoans)\\
ag\in\dom(\ran newLoans)\implies\\ 
\t1agentSpice'(ag)=agentSpice(ag)+totalOwed(ag,newLoans)\\
ag\notin\dom (newLoans)\cup \dom(\ran newLoans)\implies\\
\t1 agentSpice'(ag)=agentSpice(ag)\\
 willBorrow(age(ag),sex(ag),agentSpice'(ag),\\
 \t1\ran (loanBook'\cap\{a:AGENT\cross(AGENT\cross(\nat \cross\nat))\\
\t3|borrower(a)=borrower(loan)\}))\implies\\
 \t2\lnot\exists ag2:AGENT\spot canLend(age(ag2),sex(ag2),agentSpice'(ag2))\\
 \t3\land adjacent(position(ag2),position(ag))\\
 
 totalLoaned(ag,newLoans)\leq \\
 \t1amtAvail_{new}(age(ag),sex(ag),agentSpice(ag),SPICECHILDAMT)\\
totalOwed(ag,newLoans)\leq SPICECHILDAMT-agentSpice(ag)\\

(lender,(borrower,(amt,due)))\in newLoans \implies\\
\t1 (canLend(age(lender),sex(lender),agentSpice(lender))\\
\t1\land willBorrow(age(borrower),sex(borrower),agentSpice(borrower),\\
\t2\{borrower\}\dres(\ran spiceLoanBook))\\

\t1\land amt\leq min(\{amtAvail_{new}(age(lender),sex(lender),\\
\t4agentSpice(lender),SPICECHILDAMT)\\
\t3SPICECHILDAMT-agentSpice(borrower))\})  \\
\t1\land due=step+DURATION\\
 \t1\land adjacent(position(lender),position(borrower)))\\

\end{schema}

\subsection{Rule Application Sequence}
 \begin{zed}
Tick_{spice}\\
[\semi Growback_{spice}| \semi SeasonalGrowback_{spice}]\\
[\semi Movement_{basicSpice}| \semi (Movement_{pollutionSpice}\semi PollutionDiffusion_{spice})| \semi Combat_{spice}]\\
\{\semi Inheritance_{spice}\}\{\semi Death_{spice}[\semi Replacement_{spice} | \semi AgentMating_{spice}]\}\\
\{\semi Culture\}\{\semi PaySugarLoans\semi PaySpiceLoans\semi MakeSugarLoans\semi MakeSpiceLoans\}\\
\{\semi Transmission_{spice}\semi ImmuneResponse_{spice}\}\{\semi Trade\}
\end{zed}

 \section{Conclusions}
 We have shown that it is possible to apply formal methods fruitfully in the area of ABSS and  produced a full formal specification of the Sugarscape family of simulations.  It is, to the best of our knowledge, the first formal specification of the entire Sugarscape simulation family.  The purpose of the specification is to provide a clear, unambiguous and precise definition of Sugarscape.  The specification has identified many ambiguities and/or missing bits of information in the original rule definitions.  Where there is an obvious way of removing these ambiguities we have done so. If there is more than one possible solution we have identified them and chosen the most likely one. 

The issues with the model definition that we have encountered can broadly be grouped into three main types: Lack of Clarity, Missing Information and Sequential biases.
 \begin{description}
\item[ Lack of Clarity]
The rules, although simply stated in the appendix, lack clarity in their definition.  Only one version of each rule is presented even when many variations are referred to in the text. The variations presented cannot always be used together, for example the Movement rule defined in the appendix is not the variant required if the pollution rule is also used.  Our specification brings them all together in one place for ease of reference.
\item[ Missing Information] 
Missing or incomplete information is the biggest cause for concern. In many cases we can work out the most likely answer based on context but in some cases there is not one definitive correct answer.  If there was more than one arguably correct solution we chose the simplest. How we fill in these blanks can have a big effect on how the simulation proceeds.  These effects may be important if we are trying to compare different implementations of Sugarscape. If we take the disease transmission rule, for example, questions that are unanswered include:
\begin{enumerate}
\item Once an agent gains immunity from a particular disease, do we remove that particular disease from the set of diseases that the agent is carrying, or is the agent still a carrier?
\item When we transmit a disease do we only transmit diseases that we carry and have no immunity for, or, can any disease we carry be transmitted?
\item The Mating rule omits important information about parents contributing half of their resources to their offspring.  This has a huge effect on how mating works in a simulation.
\end{enumerate}

By replacing each ambiguous interpretation with one simple and precise interpretation we allow different developers to replicate their results and benchmark them against each other. All hidden assumptions that could serve to advantage one implementation over another are excised.
 \item[ Sequential Biases]
Sugarscape is based on the assumption that it will be implemented sequentially. While this may have been a good assumption at the time it was written it is not now necessarily the case. Improvements in processing speed have recently been attained mainly through the introduction of concurrency. Simulations are now almost always run on multicore or even multiprocessor machines. 

The Z specification is free from assumptions about implementation.  It achieves this without having to specify or constrain in any way what conflict resolution or avoidance strategies are employed. This leaves developers the freedom to try out different approaches as suits their implementation platform.
\end{description}

Because the specification is high level and only defines the before and after state of each rule it makes few assumptions as to how any rule will be implemented. All inherent biases towards a sequential implementation are removed. Implementers have complete freedom as to what programming model they employ (Object-oriented, imperative, functional, or any concurrent approach).  Any simulations that adhere to the standard can be properly compared in terms of performance or patterns of behaviour. This will put on a firmer foundation any claims made by researchers about their implementations. 
 
\subsection{Further Work}
Further work remains to be done in getting agreement from the ABM community on the decisions made in producing this interpretation of Sugarscape.  Any incorrect assumptions made in removing ambiguities need to be identified and agreed upon.  This provides a route to address the issues of replication of experimental results in ABSS.

Sugarscape can now be used as a benchmark (or rather set of benchmarks) for ABM implementers. This is particularly useful for those proposing new approaches to concurrency that promise performance improvements. Current trends, for example, include the use of Graphics Processor Units (GPUs)\cite{Deissenberg1,lysenko2008,Richmond3}, containing hundreds to thousands of individual processors. These approaches tend not to be tested on the more complex rules in Sugarscape (such as Combat, Inheritance and Trade) as they are not easily parallelized. By providing a precise and full set of these rules it is now possible for researchers to properly compare how different models cope with more complex and more realistic ABMs.

Z itself is rather verbose and can be hard to parse when reading.  The lack of modularity made function definition signatures overly long.  The available tools such as CZT \cite{Malik05czt:a} make the process of writing the specification easier but I have altered the specifications to remove bracketing where I thought it made the specification easier to read even if this was flagged as an error in the type checker.  These issues could be overcome through the use of a variant of Z such as Object-Z or Alloy. The issue of whether ABM modellers would be willing to use formal specifications remains unknown.

There are differences between the outcomes of the synchronous and asynchronous approaches. Sugarscape assumes an asynchronous approach and this affects the style of specification that we use.  We have shown in the case of combat the differences in a synchronous and asynchronous specification.  While we regard the synchronous specification as somewhat simpler to produce but others may disagree. We tackle the question as to which approach is the more correct elsewhere. 

\bibliographystyle{apalike}
\bibliography{papers}

\end{document}